\documentclass[12pt]{article}
\usepackage{amsmath,amssymb,theorem,cite,epsfig,url,psfrag,eepic,mathtools,amsmath,mathtools,graphicx,xcolor}
\mathtoolsset{showonlyrefs}
\usepackage{hyperref}
\advance \topmargin by -\headheight
\advance \topmargin by -\headsep     
\evensidemargin \oddsidemargin
\def\Maketitle{{\def\newpage{}\maketitle}}
\makeatletter
\def\Appendix{\appendix
	\def\@seccntformat##1{Appendix~\csname the##1\endcsname.~~}}
\makeatother
\makeatletter
\@addtoreset{equation}{section}

\makeatother
\makeatletter
\newcommand{\zerounderset}[3][\mathord]{%
	#1{\vtop{
			\let\\\cr
			\baselineskip\z@skip\lineskip.25ex
			\ialign{\hidewidth$##$\hidewidth\crcr
				\omit$#3$\cr
				#2\crcr
			}%
	}}%
}
\makeatother
\begin{document}
\rightline{\texttt{\today}}
\title{\textbf{Liouville reflection operator,\\affine Yangian and Bethe ansatz}\vspace*{.3cm}}
\date{}
\author{Alexey Litvinov$^{1,2}$ and Ilya Vilkoviskiy$^{2,3}$\\[\medskipamount]
\parbox[t]{0.85\textwidth}{\normalsize\it\centerline{1. Landau Institute for Theoretical Physics, 142432 Chernogolovka, Russia}}\\
\parbox[t]{0.85\textwidth}{\normalsize\it\centerline{2. Center for Advanced Studies, Skolkovo Institute of Science and Technology,		 143026 Moscow, Russia}}\\
\parbox[t]{0.85\textwidth}{\normalsize\it\centerline{3. National Research University Higher School of Economics, 119048 Moscow, Russia}}}
\Maketitle
\begin{abstract}
  In these notes we study integrable structure of conformal field theory by means of Liouville reflection operator/Maulik-Okounkov $R$-matrix. We discuss  the relation between $RLL$ and current realization of the affine Yangian of $\mathfrak{gl}(1)$. We construct the family of commuting transfer matrices related to the Intermediate Long Wave hierarchy  and derive Bethe ansatz equations for their spectra discovered by Nekrasov and Okounkov and independently by one of the authors. Our derivation mostly follows the one by Feigin, Jimbo, Miwa and Mukhin, but is adapted to the conformal case.
\end{abstract}
\section{Introduction}
There is a large class of $2D$ QFT's defined by Toda action
\begin{equation}\label{Toda-action}
S_{0}=\int\left(\frac{1}{8\pi}\bigl(\partial_{\mu}\boldsymbol{\varphi}\cdot \partial_{\mu}\boldsymbol{\varphi}\bigr)+
\Lambda\sum_{r=1}^{N}e^{\bigl(\boldsymbol{\alpha}_{r}\cdot\boldsymbol{\varphi}\bigr)}\right)\,d^{2}x,
\end{equation}
where $\boldsymbol{\varphi}=(\varphi_{1},\dots,\varphi_{N})$ is the $N-$component bosonic field and $(\boldsymbol{\alpha}_{1},\dots,\boldsymbol{\alpha}_{N})$ is the set of linearly independent vectors. The theory \eqref{Toda-action}, properly coupled to a background metric, defines a conformal field theory. However, it is well known, that under some conditions on the set   $(\boldsymbol{\alpha}_{1},\dots,\boldsymbol{\alpha}_{N})$ it also enjoys enlarged conformal symmetry usually referred as $W-$algebra \cite{Zamolodchikov:1985wn}. There is a class of such distinguishable sets $(\boldsymbol{\alpha}_{1},\dots,\boldsymbol{\alpha}_{N})$ with semi-classical behavior 
\begin{equation}
\boldsymbol{\alpha}_{r}=b\boldsymbol{e}_{r}\quad\text{for all}\quad r=1,\dots,N,
\end{equation}
where $\boldsymbol{e}_{r}$ are finite in the limit $b\rightarrow0$. The vectors $\boldsymbol{e}_{r}$ have to be simple roots of a semi-simple Lie algebra $\mathfrak{g}$ of rank $N$.
	
An interesting question arises if one perturbs the theory \eqref{Toda-action} by an additional exponential field
\begin{equation}\label{integrable-perturbation}
S_{0}\rightarrow S_{0}+\lambda\int e^{\bigl(\boldsymbol{\alpha}_{N+1}\cdot\boldsymbol{\varphi}\bigr)}\,d^{2}x.
\end{equation}
Typically this perturbation breaks down all the $W-$algebra symmetry down to Poincar\'e symmetry. However, there is a special class of perturbations, called the integrable ones, which survive an infinite symmetry of the original theory in a very non-trivial way \cite{Zamolodchikov:1989zs}. Namely, one can argue that there are infinitely many mutually commuting local Integrals of Motion $\mathbf{I}_{s}^{\lambda}$ and $\bar{\mathbf{I}}_{s}^{\lambda}$ which are perturbative in $\lambda$
\begin{equation}
\mathbf{I}_{s}^{\lambda}=\mathbf{I}_{s}+O(\lambda),\qquad
\bar{\mathbf{I}}_{s}^{\lambda}=\bar{\mathbf{I}}_{s}+O(\lambda),
\end{equation}
where $(\mathbf{I}_s,\bar{\mathbf{I}}_s)$ are defined in CFT.
	
Thus any integrable perturbation \eqref{integrable-perturbation} inherits a distinguishable set of local IM's $\mathbf{I}_{s}$  in conformal field theory.  The seminal program devoted to calculation of simultaneous   spectra of $\mathbf{I}_{s}$ has been initiated by Bazhanov, Lukyanov and Zamolodchikov in \cite{Bazhanov:1994ft,Bazhanov:1996dr,Bazhanov:1998dq} for $\mathfrak{sl}(2)/\textrm{KdV}$ case. The culmination was the discovery \cite{Bazhanov:2004fk} of Gaudin-like equations for the spectrum. In current notes we use an alternative approach,  based on affine Yangian symmetry. We consider the case of $\mathfrak{sl}(n)$ symmetry. Actually, it will be convenient for us to extend the theory by adding an auxiliary non-interacting bosonic field, leading to the action
\begin{equation}\label{Toda-action-gln}
 S=\int\left(\frac{1}{8\pi}\bigl(\partial_{\mu}\boldsymbol{\varphi}\cdot \partial_{\mu}\boldsymbol{\varphi}\bigr)+
 \Lambda\sum_{k=1}^{n-1}e^{b(\varphi_{k+1}-\varphi_{k})}+\Lambda e^{b(\varphi_{1}-\varphi_{n})}\right)\,d^{2}x,
\end{equation}
where the last term, corresponding to the affine root of $\mathfrak{sl}(n)$, is known to lead to an integrable perturbation.
With the last term dropped, the theory \eqref{Toda-action-gln} defines the conformal field theory, whose symmetry algebra can be described by quantum Miura-Gelfand-Dikii transformation \cite{Fateev:1987zh,Lukyanov_1988}
\begin{equation}\label{Miura-transformation}
 \bigl(Q\partial-\partial\varphi_{n}\bigr)\bigl(Q\partial-\partial\varphi_{n-1}\bigr)\dots\bigl(Q\partial-\partial\varphi_{2}\bigr)\bigl(Q\partial-\partial\varphi_{1}\bigr)=
 (Q\partial)^{n}+\sum_{k=1}^{n}W^{(k)}(z)(Q\partial)^{n-k},
\end{equation}
where $Q=b+b^{-1}$. In fact, one can drop any other exponent in \eqref{Toda-action-gln}, leading to different, but isomorphic $W-$algebra. For example, dropping the term $e^{b(\varphi_2-\varphi_1)}$, one has different  formula 
\begin{equation}\label{Miura-transformation-different}
 \bigl(Q\partial-\partial\varphi_{1}\bigr)\bigl(Q\partial-\partial\varphi_{n}\bigr)\dots\bigl(Q\partial-\partial\varphi_{3}\bigr)\bigl(Q\partial-\partial\varphi_{2}\bigr)=
 (Q\partial)^{n}+\sum_{k=1}^{n}\tilde{W}^{(k)}(z)(Q\partial)^{n-k}.
\end{equation}
By symmetry arguments, it is clear that local Integrals of Motion $\boldsymbol{I}_s$ should belong to the intersection of these two $W-$algebras. In particular, one can check that (for $n$ large enough)
\begin{equation}\label{local-IMs}
 \begin{aligned}
	&\mathbf{I}_1=-\frac{1}{2\pi}\int\left[\sum_{i<j}^{n}(\boldsymbol{h}_i\cdot\partial\boldsymbol{\varphi})(\boldsymbol{h}_j\cdot\partial\boldsymbol{\varphi})\right]dx,\\
	&\mathbf{I}_2=\frac{1}{2\pi}\int\left[\sum_{i<j<k}^{n}(\boldsymbol{h}_i\cdot\partial\boldsymbol{\varphi})(\boldsymbol{h}_j\cdot\partial\boldsymbol{\varphi})(\boldsymbol{h}_k\cdot\partial\boldsymbol{\varphi})+Q\sum_{i<j}(\boldsymbol{h}_i\cdot\partial\boldsymbol{\varphi})(\boldsymbol{h}_j\cdot\partial^2\boldsymbol{\varphi})\right]dx,\\
	&\mathbf{I}_3=\frac{1}{2\pi}\int\left[\sum_{i<j<k<l}^{n}(\boldsymbol{h}_i\cdot\partial\boldsymbol{\varphi})(\boldsymbol{h}_j\cdot\partial\boldsymbol{\varphi})(\boldsymbol{h}_k\cdot\partial\boldsymbol{\varphi})(\boldsymbol{h}_l\cdot\partial\boldsymbol{\varphi})+\dots\right]dx,\\
	&\dots\dots\dots\dots\dots\dots\dots\dots\dots
 \end{aligned}
\end{equation}
where 
\begin{equation}
 \boldsymbol{h}_i=\boldsymbol{e}_i-\frac{1}{n}\sum_{k=1}\boldsymbol{e}_k,\qquad
(\boldsymbol{h}_i\cdot\boldsymbol{h}_j)=\delta_{ij}-\frac{1}{n}.
\end{equation}
indeed satisfy this requirement. We note that in \eqref{local-IMs} we excluded trivial IM's build out of $U(1)$ field
\begin{equation}\label{U(1)-field}
 J=\frac{1}{n}\sum_k\partial\varphi_k.
\end{equation}
In general, one expects an existence of local Integrals of Motion for all $s\neq0(\textrm{mod}\,n)$.
	
This point of view that IM's should belong to intersection of two $W$-algebras given by \eqref{Miura-transformation} and \eqref{Miura-transformation-different} automatically implies that the intertwining operator $T_1$
\begin{equation}\label{O-operator}
 T_1\tilde{W}^{(k)}(z)=W^{(k)}(z)T_1,
\end{equation}
will be itself an Integral of Motion. The operator $T_1$ will be primarily important for us. We call it Knizhnik-Zamolodchikov operator (see section \ref{IM}). Actually it is natural to define more  operators, which will map between different $W-$algebras corresponding to different permutations of factors in \eqref{Miura-transformation}. The Maulik-Okounkov $R-$matrix \cite{Maulik:2012wi} corresponds to elementary transposition
\begin{equation}\label{R-definition}
	\mathcal{R}_{i,j}\bigl(Q\partial-\partial\varphi_{i}\bigr)\bigl(Q\partial-\partial\varphi_{j}\bigr)=\bigl(Q\partial-\partial\varphi_{j}\bigr)\bigl(Q\partial-\partial\varphi_{i}\bigr)\mathcal{R}_{i,j},
\end{equation}
while the operator $T_1$ introduced in \eqref{O-operator} corresponds to the long cycle permutation
\begin{equation}
 T_1=\mathcal{R}_{1,2}\mathcal{R}_{1,3}\dots\mathcal{R}_{1,n-1}\mathcal{R}_{1,n}.
\end{equation}
	
The operator $\mathcal{R}_{i,j}$ acts in the tensor product of two Fock representations of Heisenberg algebra with the highest weight parameters $u_{i}$ and $u_{j}$
\begin{equation}
 \mathcal{F}_{u_{i}}\otimes\mathcal{F}_{u_{j}}\overset{\mathcal{R}_{i,j}}{\longrightarrow}\mathcal{F}_{u_{i}}\otimes\mathcal{F}_{u_{j}}
\end{equation}
and its matrix  depends on the difference $u_{i}-u_{j}$. Then it follows immediately from the definition \eqref{R-definition} that $\mathcal{R}_{i,j}(u_{i}-u_{j})$ satisfies the Yang-Baxter equation
\begin{equation}\label{YB-equation}
 \mathcal{R}_{1,2}(u_{1}-u_{2})\mathcal{R}_{1,3}(u_{1}-u_{3})\mathcal{R}_{2,3}(u_{2}-u_{3})=\mathcal{R}_{2,3}(u_{2}-u_{3})\mathcal{R}_{1,3}(u_{1}-u_{3})\mathcal{R}_{1,2}(u_{1}-u_{2}),
\end{equation}
and hence the whole machinery of quantum inverse scattering method  can be applied. In particular, one can construct a family of commuting transfer-matrices on $n-$sites
\begin{equation}\label{Transfer-matrix}
 \mathbf{T}(u)=\textrm{Tr}'\bigl(\mathcal{R}_{0,1}(u-u_{1})\mathcal{R}_{0,2}(u-u_{2})\dots\mathcal{R}_{0,n-1}(u-u_{n-1})\mathcal{R}_{0,n}(u-u_{n})\bigr)\Bigl|_{\mathcal{F}_{u}}.
\end{equation}
At $u=u_{1}$ one has $\mathcal{R}_{0,1}=\mathcal{P}_{0,1}$ a permutation operator and hence
\begin{equation}\label{KZ-operator-introduction}
 \mathbf{T}(u_{1})=\mathcal{R}_{1,2}\mathcal{R}_{1,3}\dots\mathcal{R}_{1,n-1}\mathcal{R}_{1,n}=T_1,
\end{equation}
which implies that $\mathbf{T}(u)$ commutes with local Integrals of Motion $\mathbf{I}_{s}$ and can be taken as a generating function. 
	
In \eqref{Transfer-matrix} the notation $\textrm{Tr}'$ corresponds to certain regularization of the trace, which goes through the introduction of the twist parameter $q$
\begin{equation}
 \textrm{Tr}'(\dots)\overset{\text{def}}{=}\lim\limits_{q\rightarrow 1}\,\frac{1}{\chi(q)}\,\textrm{Tr}\left(q^{L_0^{(0)}}\dots\right),\quad\text{where}\quad\chi(q)=\prod_{k=1}^{\infty}\frac{1}{1-q^k}
\end{equation}
and $L_0^{(0)}=\sum_{k>0}a_{-k}^{(0)}a_{k}^{(0)}$ is the level operator in auxiliary space $\mathcal{F}_u$. Remarkably, the introduction of the twist parameter does not spoil the integrability, that is the twist deformed transfer-matrices 
\begin{equation}\label{Transfer-matrix-q-deformed}
 \mathbf{T}_q(u)=\textrm{Tr}\bigl(q^{L_0^{(0)}}\mathcal{R}_{0,1}(u-u_{1})\mathcal{R}_{0,2}(u-u_{2})\dots\mathcal{R}_{0,n-1}(u-u_{n-1})\mathcal{R}_{0,n}(u-u_{n})\bigr)\Bigl|_{\mathcal{F}_{u}},
\end{equation}
still commute. On the level of local Integrals of Motion \eqref{local-IMs} this deformation corresponds to the non-local deformation $\mathbf{I}_{s}\rightarrow\mathbf{I}_{s}(q)$ called quantum $\textrm{ILW}_n$ (Intermediate Long Wave) integrable system \cite{Litvinov:2013zda}. In particular
\begin{equation}\label{ILW-local-integrals}
 \begin{aligned}
	&\mathbf{I}_1(q)=\frac{1}{2\pi}\int\left[\frac{1}{2}\sum_{k=1}^{n}(\partial\varphi_k)^2\right]dx,\\
	&\mathbf{I}_2(q)=\frac{1}{2\pi}\int\left[\frac{1}{3}\sum_{k=1}^{n}(\partial\varphi_k)^3+Q\left(\frac{i}{2}\sum_{i,j}\partial\varphi_iD\partial\varphi_j+\sum_{i<j}\partial\varphi_i\partial^2\varphi_j\right)\right]dx,\\
	&\mathbf{I}_3(q)=\frac{1}{2\pi}\int\left[\frac{1}{4}\sum_{k=1}^{n}(\partial\varphi_k)^4+\dots\right]dx,\\
	&\dots\dots\dots\dots\dots\dots\dots\dots\dots
 \end{aligned}
\end{equation}
where $D$ is the non-locality operator whose Fourier image is
\begin{equation}
 D(k)=k\frac{1+q^k}{1-q^k}.
\end{equation}
We note that the limit $q\rightarrow1$ is a little subtle since the  operator $D$ is singular at $q\rightarrow1$ and hence some eigenvalues of $\mathbf{I}_{s}(q)$ become infinite.  However, one can show that on a subspace spanned by eigenfunctions with \emph{finite} eigenvalues the modes of the $J(z)$ field \eqref{U(1)-field} are not exited. It implies in particular that
\begin{equation}
 \mathbf{I}_2=\mathbf{I}_{2}(q)\Bigr|_{J\rightarrow0}.
\end{equation}
	
The spectrum of $\textrm{ILW}_n$ integrable system is governed by finite type Bethe ansatz equations which have been conjectured by Nekrasov and Okounkov\footnote{See Okounkov's \href{http://scgp.stonybrook.edu/video_portal/video.php?id=524}{talk} at Facets of Integrability conference, SCGP January 2013.} and independently by one of the authors in \cite{Litvinov:2013zda}
\begin{equation}\label{Bethe-ansatz-equations-introduction}
 q\prod_{j\neq i}\frac{(x_i-x_j-b)(x_i-x_j-b^{-1})(x_i-x_j+Q)}{(x_i-x_j+b)(x_i-x_j+b^{-1})(x_i-x_j-Q)}\prod_{k=1}^n\frac{x_i-u_k-\frac{Q}{2}}{x_i-u_k+\frac{Q}{2}}=1\quad\text{for all}\quad i=1,\dots,N,
\end{equation}
such that the eigenvalues of  $\mathbf{I}_{s}(q)$ are symmetric polynomials in Bethe roots
\begin{equation}\label{Bethe-ansatz-eigenvalues-introduction}
 \mathbf{I}_1(q)\sim-\frac{1}{2}\sum_{k=1}^nu_k^2+N,\qquad
 \mathbf{I}_2(q)\sim\frac{1}{3}\sum_{k=1}^nu_k^3-2i\sum_{j=1}^Nx_j,\qquad\dots
\end{equation}
Equations \eqref{Bethe-ansatz-equations-introduction}-\eqref{Bethe-ansatz-eigenvalues-introduction} have been checked in \cite{Litvinov:2013zda} by explicit calculations on lower levels. A proof has been suggested in \cite{Feigin:2015raa} in the trigonometric ($\mathfrak{q}$-deformed)  case. 
	
We note that Bethe ansatz equations \eqref{Bethe-ansatz-eigenvalues-introduction} are simplified drastically for $q^{\pm1}\rightarrow0$, which is equivalent to $D(k)\rightarrow\pm|k|$. The limit of $\textrm{ILW}_n$ system at $q^{\pm1}\rightarrow0$ is known as $\textrm{BO}_n$ integrable system (Benjamin-Ono). The basis of its eigenfunctions stands behind AGT correspondence \cite{Alday:2009aq}. Namely, it has been shown in \cite{Alba:2010qc,Fateev:2011hq} that the matrix elements of semi-degenerate $W_n$-primary fields, dressed by suitably chosen $U(1)$ vertex operators, sandwiched between the $\textrm{BO}_n$ eigenfunctions coincide with bi-fundamental contribution to the Nekrasov partition function \cite{Nekrasov:2002qd} for corresponding quiver gauge theory.

The Maulik-Okounkov $R-$matrix defines in a standard way the  Yang-Baxter algebra (RLL algebra). We note that  $R_{i,j}$ intertwines  two representations of Heisenberg algebra \eqref{R-definition} ($\widehat{\mathfrak{gl}}(1)$ current algebra). Since the matrix elements of $R_{i,j}$ are rational functions of the highest weight/spectral parameter (see below), it is natural to call the corresponding  Yang-Baxter algebra the Yangian of $\widehat{\mathfrak{gl}}(1)$, or affine Yangian of $\mathfrak{gl}(1)$. The algebra under the same name has been introduced by  Tsymbaliuk in \cite{Tsymbaliuk:2014fvq}. It has been given by explicit commutation relations (the so called current  realization).  We will show that both algebras do not literally coincide, but rather Tsymbaliuk's algebra $\textrm{Y}\bigl(\widehat{\mathfrak{gl}}(1)\bigr)$ is obtained from the Yang-Baxter algebra $\textrm{YB}\bigl(\widehat{\mathfrak{gl}}(1)\bigr)$ by factorization over infinite-dimensional center. 
	
As we already mentioned, the Yangian $\textrm{Y}\bigl(\widehat{\mathfrak{gl}}(1)\bigr)$ is the rational counterpart of the trigonometric algebra called Ding-Iohara-Miki algebra or quantum toroidal $\mathfrak{gl}(1)$ algebra \cite{Tsymbaliuk:2014fvq}. This algebra has been extensively studied by Feigin and collaborators in \cite{Awata:2011fk,Feigin_2012,Feigin:2015raa,Feigin:2017gcv}. Another but equivalent approach through the methods of geometric representation theory was developed by Okounkov and collaborators \cite{Okounkov:2016sya,Aganagic:2017gsx}. We borrow many ideas developed in \cite{Awata:2011fk,Feigin_2012,Feigin:2015raa,Feigin:2017gcv} and \cite{Okounkov:2016sya,Aganagic:2017gsx} for our study. In particular, Bethe anzatz equations as well as Bethe vectors can be found in \cite{Feigin:2015raa} in the $\mathfrak{q}-$deformed case and in \cite{Aganagic:2017gsx}. 
	
This paper is organized as follows. In section \ref{R-matrix} we define the main actor of our study -- Liouville reflection operator/Maulik-Okounkov $R-$matrix and discuss its general properties and various representations. In section \ref{Affine-Yangian and BO integrable system} we study corresponding $RLL$ algebra and discuss its relation to affine Yangian of $\mathfrak{gl}(1)$. In section \ref{IM} we introduce quantum Integrals of Motion corresponding to ILW system and prove Bethe ansatz equations for the spectrum. In section \ref{Conclusions}  we give some conclusions an emphasize future possible directions of study. In appendices we present some explicit formulae and calculations used in the main text. 
\section{Maulik-Okounkov \texorpdfstring{$R$}{R}-matrix as Liouville reflection operator}\label{R-matrix}
It is clear from the definition \eqref{R-definition} that $\mathcal{R}_{i,j}$ trivially commutes with the center of mass field $\varphi_{i}+\varphi_{j}$, that is
\begin{equation}\label{R-as-R}
 \mathcal{R}_{i,j}=\mathcal{R}\Bigl|_{J\rightarrow\frac{\partial\varphi_{i}-\partial\varphi_{j}}{2}},
\end{equation}
where $\mathcal{R}$ is the Liouville reflection operator for the $U(1)$ current algebra
\begin{equation}\label{U(1)-algebra}
 J(z)J(w)=\frac{1}{2(z-w)^{2}}+\dots
\end{equation}
which is defined as
\begin{equation}\label{RJ-relation}
 \mathcal{R}(-J^{2}+Q\partial J)=(-J^{2}-Q\partial J)\mathcal{R}.
\end{equation}
We note that here the $U(1)$ current has different normalization compared to one used before. In particular the commutation relations for modes of $J(x)$ are differ by the factor of $2$
\begin{equation}
 [a_{m},a_{n}]=\frac{m}{2}\delta_{m,-n}.
\end{equation}
	
The relation \eqref{RJ-relation}  can be used for calculation of $\mathcal{R}$. Consider highest weight representation of the $U(1)$ algebra \eqref{U(1)-algebra}. It is generated by the negative mode operators $a_{-k}$  from the vacuum state $|u\rangle$:
\begin{equation}
  a_{0}|u\rangle=u|u\rangle,\qquad a_{n}|u\rangle=0\quad\text{for}\quad n>0.
\end{equation}
Then  \eqref{RJ-relation} is equivalent to the infinite set of relations 
\begin{equation}\label{R-action-on-L}
 \mathcal{R}L^{(+)}_{-\lambda_{1}}\dots L^{(+)}_{-\lambda_{n}}|u\rangle=\mathcal{R}^{\textrm{vac}}(u)L^{(-)}_{-\lambda_{1}}\dots L^{(-)}_{-\lambda_{n}}|u\rangle,
\end{equation}
where $L^{(\pm)}_{n}$ are the components of $T^{(\pm)}=-J^{2}\pm Q\partial J$ 
\begin{equation}\label{Bosonization}
 L^{(\pm)}_{n}=\sum_{k\neq0,n}a_{k}a_{n-k}+(2a_{0} \pm inQ)a_{n},\quad L^{(+)}_{0}=L^{(-)}_{0}=\frac{Q^{2}}{4}+a_{0}^{2}+2\sum_{k>0}a_{-k}a_{k}.
\end{equation}
and $\mathcal{R}^{\textrm{vac}}(u)$ is an eigenvalue for the vacuum state. In the following we will usually take 
\begin{equation}\label{R-normalization}
	\mathcal{R}^{\textrm{vac}}(u)=1.
\end{equation}
Using \eqref{R-action-on-L} as a set of equations one can compute the matrix of $\mathcal{R}$. For example at the level $1$ one has
\begin{equation}\label{R-level-1}
 \mathcal{R}L^{(+)}_{-1}|u\rangle=L^{(-)}_{-1}|u\rangle\implies
 \mathcal{R}a_{-1}|u\rangle=\frac{2u+iQ}{2u-iQ}a_{-1}|u\rangle.
\end{equation}
Similarly, at the level $2$ one obtains
\begin{equation}\label{R-level-2}
 \begin{aligned}
  &\mathcal{R}a_{-2}|u\rangle=\frac{\left(\bigl(8u^{3}+2u(3Q^{2}-1)-iQ(2Q^{2}+1)\bigr)a_{-2}-8iQua_{-1}^{2}\right)|u\rangle}{(2u-iQ)(2u-iQ-ib)(2u-iQ-ib^{-1})},\\
  &\mathcal{R}a_{-1}^{2}|u\rangle=\frac{\left(-4iQua_{-2}+\bigl(8u^{3}+2u(3Q^{2}-1)+iQ(2Q^{2}+1)\bigr)a_{-1}^{2}\right)|u\rangle}{(2u-iQ)(2u-iQ-ib)(2u-iQ-ib^{-1})}.
 \end{aligned}
\end{equation}
	
Apart from explicit expressions on lower levels the reflection operator is not known in a closed form. However it shares several properties allowing to judge about its structure:
\paragraph{Poles.} It is clear that apart from the normalization factor the operator $\mathcal{R}$ is a meromorphic functions of the momentum $u$. In fact, it can be argued that it has only simple poles located at the Kac points
\begin{equation}\label{Kac-points}
 u=u_{m,n}=i\left(\frac{mb}{2}+\frac{n}{2b}\right),\quad m,n>0,
\end{equation}
i.e. $\mathcal{R}(u)$ can be written in the form
\begin{equation}\label{R-meromorphic-form}
 \mathcal{R}(u)=1+\sum_{m,n>0}\frac{R_{m,n}}{u-u_{m,n}}.
\end{equation}
Indeed it is well known that for the values \eqref{Kac-points} the map from the Fock module $\mathcal{F}_u$ to Verma module $\mathcal{V}_{\Delta}$ given by the $L^{+}$ generators has a kernel. More precisely all the states of the form  
\begin{equation}\label{co-singular-vector}
 L^{(+)}_{-\lambda_{1}}\dots L^{(+)}_{-\lambda_{n}}\Bigl((L_{-1}^{+})^{mn}+\dots\Bigr)|u\rangle,
\end{equation} 
where $|\chi_{m,n}\rangle\overset{\text{def}}{=}\Bigl((L_{-1}^{+})^{mn}+\dots\Bigr)|u\rangle$ is a special state called co-singular vector, vanish at $u=u_{m,n}$. Explicitly, one has
\begin{equation}
 |\chi_{1,1}\rangle=L_{-1}^{+}|u\rangle,\quad
 |\chi_{2,1}\rangle=\Bigl((L_{-1}^{+})^2-b^2L_{-2}^{+}\Bigr)|u\rangle,\quad
 |\chi_{1,2}\rangle=\Bigl((L_{-1}^{+})^2-b^{-2}L_{-2}^{+}\Bigr)|u\rangle\quad\text{etc}
\end{equation}
At the same time the reflected states
\begin{equation}\label{singular-vector}
 L^{(-)}_{-\lambda_{1}}\dots L^{(-)}_{-\lambda_{n}}\Bigl((L_{-1}^{-})^{mn}+\dots\Bigr)|u\rangle,
\end{equation}
do not vanish for $u=u_{m,n}$ which implies that $\mathcal{R}$ should exhibit a singularity at \eqref{singular-vector}, namely a simple pole, which implies \eqref{R-meromorphic-form}. 

We note that the formula \eqref{R-meromorphic-form} is a reminiscent of the Alyosha Zamolodchikov's recurrence formula for conformal block \cite{Zamolodchikov:1985ie}.  In particular, one can use  \eqref{R-meromorphic-form} as a tool for calculation of the matrix of $\mathcal{R}(u)$.  
\paragraph{Relation to Liouville $S-$matrix}
The Liouville reflection operator $\mathcal{R}$ is closely related to the Liouville $S-$matrix introduced in Zamolodchikov's paper \cite{Zamolodchikov:1995aa}. Namely, they differ by the sign change operator $\boldsymbol{\pi}J(x)=-J(x)\boldsymbol{\pi}$ as 
\begin{equation}
 \mathcal{R}(u)=\boldsymbol{\pi}\hat{S}(u).
\end{equation}
According to \eqref{R-action-on-L} the $S$-matrix $\hat{S}(u)$ acts between different Fock modules $\mathcal{F}_u\xrightarrow{\hat{S}(u)}\mathcal{F}_{-u}$ as follows
\begin{equation}\label{S-action-Virasoro}
 \hat{S}(u)L^{(+)}_{-\lambda_{1}}\dots L^{(+)}_{-\lambda_{n}}|u\rangle=L^{(+)}_{-\lambda_{1}}\dots L^{(+)}_{-\lambda_{n}}|-u\rangle.
\end{equation}
\paragraph{Expression through Screening operators.}Given the stress energy tensor $T^{+}=-J^2+Q\partial J$  with $J=\partial\varphi$, one finds that the exponential fields  $e^{2b^{\pm1}\varphi(z)}$ satisfy 
\begin{equation}\label{screening-vanishing-of-integral}
 \oint_{\mathcal{C}_{\xi}} e^{2b^{\pm1}\varphi(z)}T^{+}(\xi)dz=0.
\end{equation}  
Then suppose that $u=-u_{m,n}$ for $m,n\geq0$. In this case one can define a \emph{closed} contour $\mathcal{C}$ such that the normalized operator $\mathcal{F}_{-u_{m.n}}\xrightarrow{\mathcal{Q}_{m,n}}\mathcal{F}_{u_{m.n}}$
\begin{equation}
 \mathcal{Q}_{m,n}=\Omega_{m,n}\oint_\mathcal{C}e^{2b\varphi(z_1)}\dots e^{2b\varphi(z_m)}e^{2b^{-1}\varphi(z_{m+1})}\dots e^{2b^{-1}\varphi(z_{m+n})}dz_1\dots dz_{m+n}:\qquad \mathcal{Q}_{m,n}|-u_{m,n}\rangle=|u_{m,n}\rangle
\end{equation} 
called the screening operator, is well defined. Then the formula \eqref{screening-vanishing-of-integral} implies that
\begin{equation}\label{Q-action-Virasoro}
 \mathcal{Q}_{m,n}L^{(+)}_{-\lambda_{1}}\dots L^{(+)}_{-\lambda_{n}}|-u_{m,n}\rangle=L^{(+)}_{-\lambda_{1}}\dots L^{(+)}_{-\lambda_{n}}|u_{m,n}\rangle.
\end{equation}
Comparing \eqref{S-action-Virasoro} and \eqref{Q-action-Virasoro} one finds
\begin{equation}\label{S-as-screening}
 \hat{S}(-u_{m,n})=\mathcal{Q}_{m,n}\implies\mathcal{R}(-u_{m,n})=\boldsymbol{\pi}\mathcal{Q}_{m,n}.
\end{equation}
\paragraph{Large momenta expansion.}
We note that $\mathcal{R}_{1,2}$ coincides with the KZ operator \eqref{KZ-operator-introduction} for $n=2$ and hence $\mathcal{R}$ commutes with the system of local Integrals of Motion of quantum KdV (mKdV) system
\begin{equation}\label{IM's-local}
 \mathbf{I}_{2n-1}=\frac{1}{2\pi}\int\Bigl(J^{2n}+\text{higher derivatives}\Bigr)dx.
\end{equation}  
It can be shown that $\mathcal{R}$ is an exponent of semi-local (non-polynomial) Integral of Motion\footnote{The existence of such semi-local Integrals of Motion for KdV (mKdV) equation has been noticed by Boris Dubrovin \cite{Dubrovin_2006}.}
\begin{equation}\label{R-semilocal-IM}
 \mathcal{R}=\exp\left[\frac{iQ}{2\pi}\int\Bigl(2J\log J+\frac{1-2Q^2}{24}\frac{J_x^2}{J^3}+\text{higher derivatives}\Bigr)dx\right].
\end{equation}
The formula  \eqref{R-semilocal-IM} is rather symbolic and requires a regularization prescription to make sense.  It can be defined as a large $u$ expansion. Namely, if one splits $J$ into constant and zero-mean parts $J=u+\tilde{J}$, then the expansion coefficients 
\begin{equation}
 J\log J=u\log u+\tilde{J}(\log u+1)+\sum_{k=1}^{\infty}\frac{(-1)^{k+1}}{k(k+1)u^k}\,\tilde{J}^{k+1},\quad
	\frac{J_x^2}{J^3}=\frac{\tilde{J}_x^2}{2}\sum_{k>1}\frac{(-1)^{k}(k-2)(k-3)}{2u^{k-1}}\tilde{J}^{k-4},\quad\dots
\end{equation} 
are zeta-valued regularized (similar regularization is used in the definition of local IM's \eqref{IM's-local}). So that \eqref{R-semilocal-IM}  leads to large $u$ expansion 
\begin{equation}\label{R-large-P-expansion}
 \mathcal{R}(u)=\exp\left[\frac{iQ}{2\pi}\int\left(2u\log u+\frac{\tilde{J}^2}{u}-\frac{\tilde{J}^3}{3u^2}+O\Bigl(u^{-3}\Bigr)\right)dx\right].
\end{equation}
We note that in \eqref{R-semilocal-IM} and \eqref{R-large-P-expansion} the normalization is different from the one used before, i.e. $\mathcal{R}_{\text{vac}}(u)\neq1$.
\paragraph{Free-fermion point.} One can show that $\mathcal{R}$ admits simple representation at the free-fermion point $c=-2$. Namely, if one uses boson-fermion correspondence to represent
\begin{equation}
 J(x)=u+\frac{1}{\sqrt{2}}:\psi^{+}(x)\psi(x):,
\end{equation}
where $(\psi(x),\psi^+(x))$ is the chiral part of Dirac fermion, then
up to normalization factor one has an explicit formula (see appendix  \ref{SUSY-ILW})
\begin{equation}\label{R-free-fermion}
 \mathcal{R}(u)\Bigl|_{c=-2}\sim
 \exp\left(\frac{1}{2\pi}\int_{0}^{2\pi}:\psi^{+}(x)\log\left(1+\frac{i}{u\sqrt{2}}\partial\right)\psi(x):\,dx\right).
\end{equation}
For $c\neq-2$ formula is more complicated and \eqref{R-free-fermion} will include multiple fermion terms.
\paragraph{Smirnov's fermion formula.} There is also Smirnov's formula for Maulik-Okounkov $R-$matrix involving an infinite product of fermionic operators \cite{Smirnov:2013hh}.  Unfortunately, we do not known any practical use of it for our purposes. 

Given the Liouville reflection operator $\mathcal{R}$ the Maulik-Okounkov matrix $\mathcal{R}_{i,j}$ is given by substitution \eqref{R-as-R}. We also found it convenient to take highest weight parameters to be purely imaginary. More precisely in \eqref{R-as-R} we replace
\begin{equation}
u\rightarrow-\frac{i}{2}(u_i-u_j).
\end{equation} 
\section{Yang-Baxter algebra}\label{Affine-Yangian and BO integrable system}
The Maulik-Okounkov $R$-matrix defines the Yang-Baxter algebra in a standard way
\begin{equation}\label{YB-algebra}
 \mathcal{R}_{ij}(u-v)\mathcal{L}_{i}(u)\mathcal{L}_{j}(v)=\mathcal{L}_{j}(v)\mathcal{L}_{i}(u)\mathcal{R}_{ij}(u-v).
\end{equation}
Here $\mathcal{L}_i(u)$ is treated as an operator in some quantum space, a tensor product of $n$ Fock spaces in our case, and as a matrix in auxiliary Fock space $\mathcal{F}_u$. The algebra \eqref{YB-algebra} becomes an infinite set of quadratic relations between the matrix elements labeled by two partitions
\begin{equation}
 \mathcal{L}_{\scriptscriptstyle{\boldsymbol{\lambda},\boldsymbol{\mu}}}(u)\overset{\text{def}}{=}\langle u|a_{\boldsymbol{\lambda}}\mathcal{L}(u)a_{-\boldsymbol{\mu}}|u\rangle\quad\text{where}\quad
 a_{-\boldsymbol{\mu}}|u\rangle=a_{-\mu_1}a_{-\mu_2}\dots|u\rangle.
\end{equation}
Since \eqref{YB-algebra} is defined by the rational $R$-matrix which intertwines two representations of the Heisenberg algebra, which is the same as $\widehat{\mathfrak{gl}}(1)$, it looks natural to associate it to the Yangian algebra of $\widehat{\mathfrak{gl}}(1)$ introduced by Tsymbaliuk in \cite{Tsymbaliuk:2014fvq}. In fact the algebras do not literally coincide.   We will show that \eqref{YB-algebra} is related to the  Yangian of $\widehat{\mathfrak{gl}}(1)$ by factorization over its center. This is similar to the well known fact that the Yangians of $\mathfrak{gl}(n)$ and of $\mathfrak{sl}(n)$ are  differ by central elements \cite{Kulish:1981bi}. We note that, compared to the non-affine case,  the center of \eqref{YB-algebra} is infinite dimensional.  We will denote the Yang-Baxter algebra as $\textrm{YB}\bigl(\widehat{\mathfrak{gl}}(1)\bigr)$, reserving the notation $\textrm{Y}\bigl(\widehat{\mathfrak{gl}}(1)\bigr)$ for Tsymbaliuk's algebra.
	
In discussions below we will mainly follow the analysis of the relation between the $RLL$  and current realizations performed in \cite{ding1993} for quantum groups. We introduce three basic currents of degree $0$, $1$ and $-1$ (see appendix \ref{Yangian-relations} for more details)  
\begin{equation}\label{efh-def}
 h(u)\overset{\text{def}}{=}\mathcal{L}_{\scriptscriptstyle{\varnothing,\varnothing}}(u),\qquad
 e(u)\overset{\text{def}}{=}h^{-1}(u)\cdot\mathcal{L}_{\scriptscriptstyle{\varnothing,\Box}}(u)\quad\text{and}\quad
 f(u)\overset{\text{def}}{=}\mathcal{L}_{\scriptscriptstyle{\Box,\varnothing}}(u)\cdot h^{-1}(u),
\end{equation}
as well as an auxiliary current (as we will see \eqref{hh-relation} it  also belongs to the Cartan sub-algebra of $\textrm{YB}\bigl(\widehat{\mathfrak{gl}}(1)\bigr)$)
\begin{equation}\label{psi-definition}
 \psi(u)\overset{\text{def}}{=}\Bigl(\mathcal{L}_{\scriptscriptstyle{\Box,\Box}}(u-Q)-\mathcal{L}_{\scriptscriptstyle{\varnothing,\Box}}(u-Q)h^{-1}(u-Q)
 \mathcal{L}_{\scriptscriptstyle{\Box,\varnothing}}(u-Q)\Bigr)h^{-1}(u-Q)
\end{equation}
As follows from definition of the $R$-matrix these currents admit large $u$ expansion
\begin{equation}\label{currents-large-u-expansion}
	h(u)=1+\frac{h_{0}}{u}+\frac{h_{1}}{u^{2}}+\dots,\quad e(u)=\frac{e_{0}}{u}+\frac{e_{1}}{u^{2}}+\dots,\quad
	f(u)=\frac{f_{0}}{u}+\frac{f_{1}}{u^{2}}+\dots,\quad \psi(u)=1+\frac{\psi_{0}}{u}+\frac{\psi_{1}}{u^{2}}+\dots
\end{equation}
As we will see below, it proves convenient to introduce higher currents labeled by $3D$ partitions. In particular, on level $2$ one has three $e_{\boldsymbol{\lambda}}(u)$ currents
\begin{equation}\label{level2-currents}
 \begin{gathered}
	e_{\includegraphics[scale=0.035]{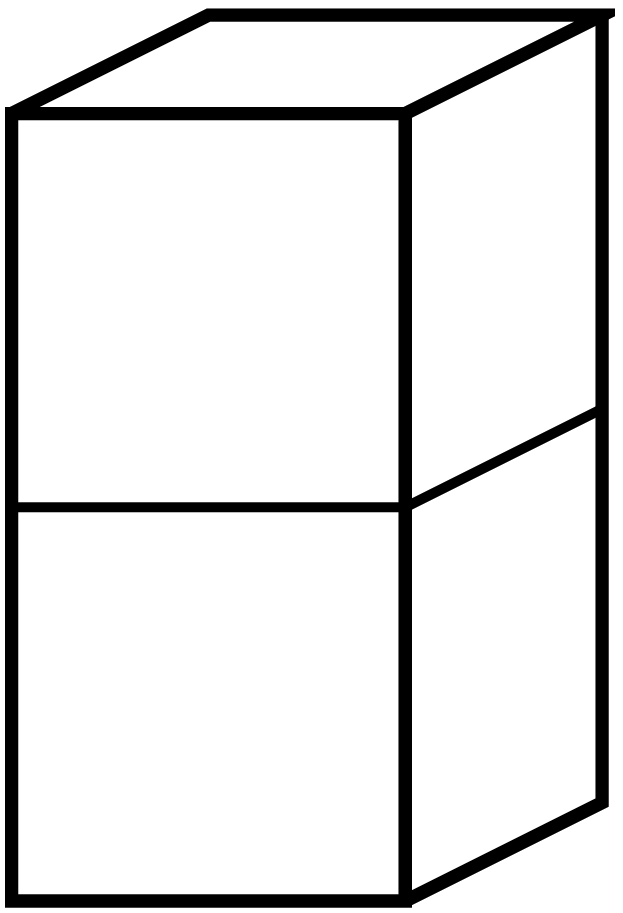}}(u)=\frac{ibQ}{(b^{2}-1)(b^{2}+2)}h^{-1}(u)\left(\mathcal{L}_{\scriptscriptstyle{\varnothing,\Box\hspace*{-1.2pt}\vspace*{2pt}\Box}}(u)-
	ib\mathcal{L}_{\scriptscriptstyle{\varnothing},\zerounderset{\scriptscriptstyle{\Box}}{\vspace*{-1.55pt}\scriptscriptstyle{\Box}}}(u)\right),\\
	e_{\includegraphics[scale=0.035]{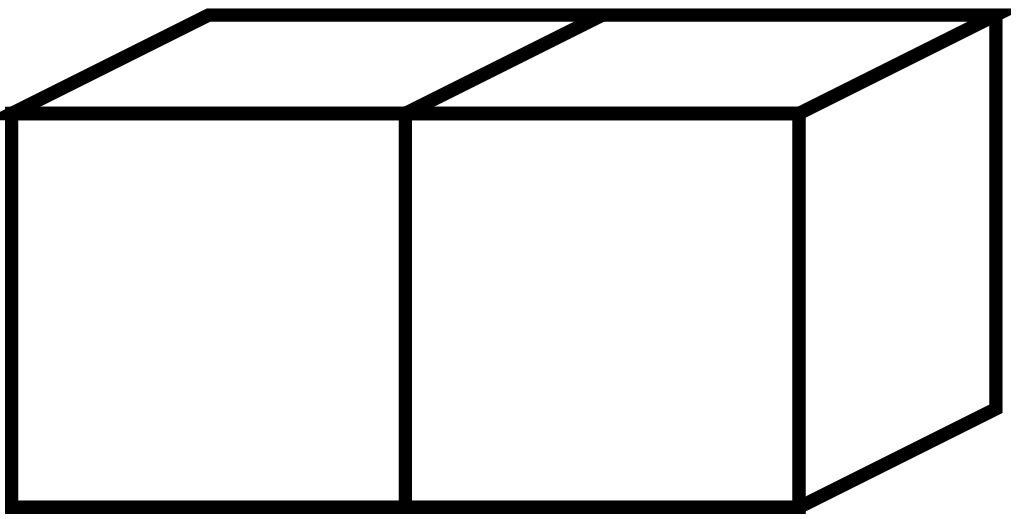}}(u)=\frac{ib^{-1}Q}{(b^{-2}-1)(b^{-2}+2)}h^{-1}(u)
	\left(\mathcal{L}_{\scriptscriptstyle{\varnothing,\Box\hspace*{-1.2pt}\vspace*{2pt}\Box}}(u)-
	ib^{-1}\mathcal{L}_{\scriptscriptstyle{\varnothing},\zerounderset{\scriptscriptstyle{\Box}}{\vspace*{-1.55pt}\scriptscriptstyle{\Box}}}(u)\right),\quad
	e_{\includegraphics[scale=0.035]{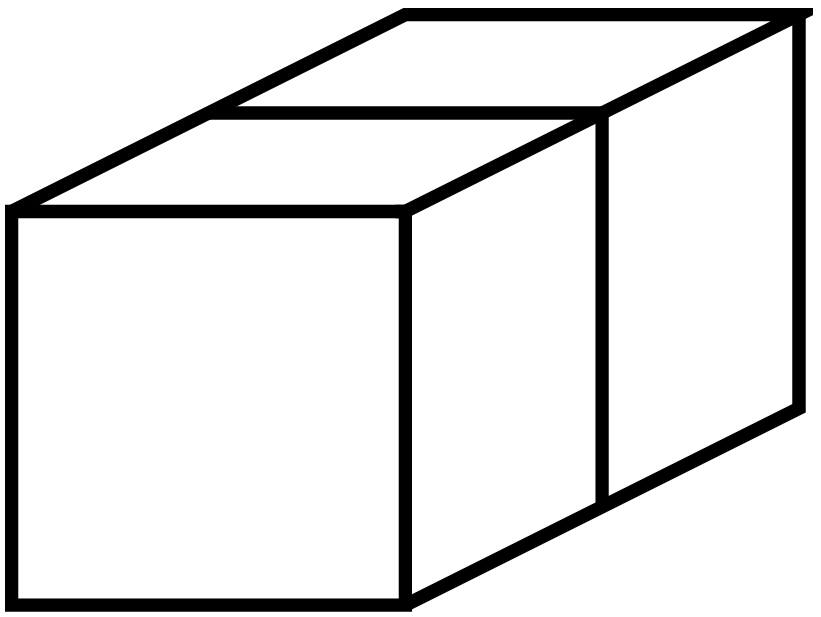}}(u)=Q\left[be_{\includegraphics[scale=0.035]{partition3.eps}}(u)+b^{-1}e_{\includegraphics[scale=0.035]{partition2.eps}}(u)-e^{2}(u)\right].
 \end{gathered}
\end{equation} 
and similarly
\begin{equation}\label{level2-currents-f}
 \begin{gathered}
	f_{\includegraphics[scale=0.035]{partition3.eps}}(u)=\frac{ibQ}{(b^{2}-1)(b^{2}+2)}h^{-1}(u)\left(\mathcal{L}_{\scriptscriptstyle{\Box\hspace*{-1.2pt}\vspace*{2pt}\Box,\varnothing}}(u)-
	ib\mathcal{L}_{\zerounderset{\scriptscriptstyle{\Box}}{\vspace*{-1.55pt}\scriptscriptstyle{\Box}},\scriptscriptstyle{\varnothing}}(u)\right),\\
	f_{\includegraphics[scale=0.035]{partition2.eps}}(u)=\frac{ib^{-1}Q}{(b^{-2}-1)(b^{-2}+2)}h^{-1}(u)
	\left(\mathcal{L}_{\scriptscriptstyle{\Box\hspace*{-1.2pt}\vspace*{2pt}\Box,\varnothing}}(u)-
	ib^{-1}\mathcal{L}_{\zerounderset{\scriptscriptstyle{\Box}}{\vspace*{-1.55pt}\scriptscriptstyle{\Box}},\scriptscriptstyle{\varnothing}}(u)\right),\quad
	f_{\includegraphics[scale=0.035]{partition1.eps}}(u)=Q\left[bf_{\includegraphics[scale=0.035]{partition3.eps}}(u)+b^{-1}f_{\includegraphics[scale=0.035]{partition2.eps}}(u)-f^{2}(u)\right].
 \end{gathered}
\end{equation} 
As we will see below these currents are algebraically depending on the basic ones \eqref{efh-def}.
	
It will be more convenient to use Nekrasov epsilon notations rather than Liouville notations. Formally, they are  obtained by replacing central charge parameter
\begin{equation}\label{epsilonNotation}
 b\rightarrow \frac{\epsilon_{1}}{\sqrt{\epsilon_1 \epsilon_2}},\quad b^{-1}\rightarrow \frac{\epsilon_{2}}{\sqrt{\epsilon_1 \epsilon_2}},\quad Q\rightarrow-\frac{\epsilon_{3}}{\sqrt{\epsilon_1 \epsilon_2}}\implies\epsilon_{1}+\epsilon_{2}+\epsilon_{3}=0,
\end{equation}
together with the normalization of bosonic fields:
\begin{equation}\label{uepsilonNotation}
 \varphi(x)\to -i \frac{\varphi(x)}{\sqrt{\epsilon_1\epsilon_2}} . 
\end{equation}
Altogether, this leads to the following Miura transformation
\begin{eqnarray}
 \mathcal{W}^{(2)}(z)=(i\epsilon_3\partial -\partial \phi_1)(i\epsilon_3 \partial -\partial \phi_2)
\end{eqnarray}
We also have to scale our basic current $e(u)$ and $f(u)$ as
\begin{equation}\label{currents-scaling}
 e(u)\rightarrow\sqrt{\epsilon_3}e(u),\qquad
 f(u)\rightarrow\sqrt{\epsilon_3}f(u).
\end{equation}
\subsection{Current realisation of the Yang-Baxter algebra \texorpdfstring{$\textrm{YB}\bigl(\widehat{\mathfrak{gl}}(1)\bigr)$}{YB(gl(1))}}\label{comrel}
Using the definition \eqref{efh-def} and \eqref{psi-definition} and explicit expression for the $R$-matrix on first three levels one finds (see appendix \ref{Yangian-relations} for details)
\begin{subequations}\label{Yangian-relation-main}
\begin{gather}
		[h(u),\psi(v)]=0,\quad[\psi(u),\psi(v)]=0,\quad[h(u),h(v)]=0,\label{hh-relation}\\
		(u-v-\epsilon_{3})h(u)e(v)=(u-v)e(v)h(u)\textcolor{blue}{-\epsilon_3 h(u) e(u)}, \label{he-relation} \\ (u-v-\epsilon_{3})f(v)h(u)=(u-v)h(u)f(v)\textcolor{blue}{-\epsilon_3 f(u)h(u)},\label{hf-relation} \\
		[e(u),f(v)]=\frac{\psi(u)-\psi(v)}{u-v}\label{ef-relation}, 
\end{gather}
as well as $ee$, $ff$ relations
\begin{multline}\label{ee-exact-relation}
		g(u-v)\Bigl[e(u)e(v)\textcolor{blue}{-\frac{e_{\includegraphics[scale=0.035]{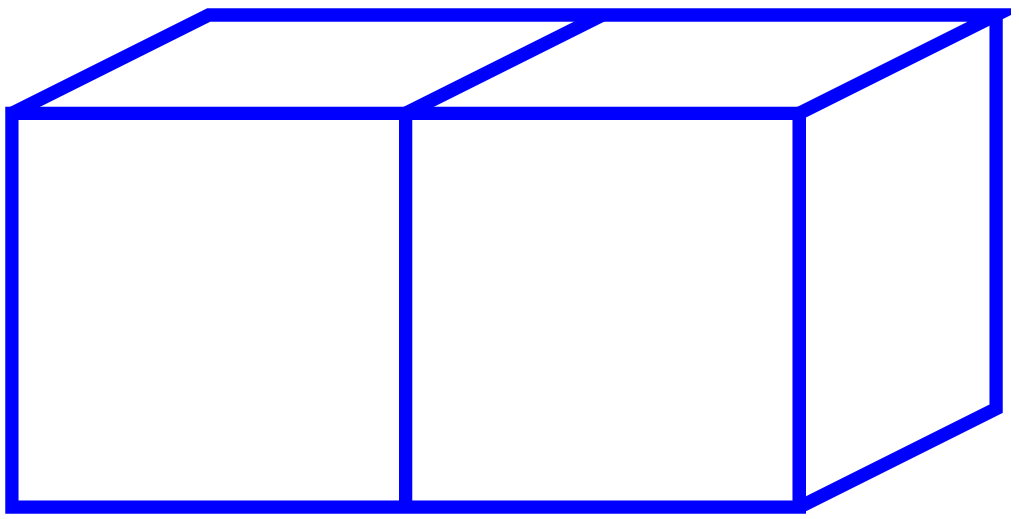}}(v)}{u-v+\epsilon_{1}}-
			\frac{e_{\includegraphics[scale=0.035]{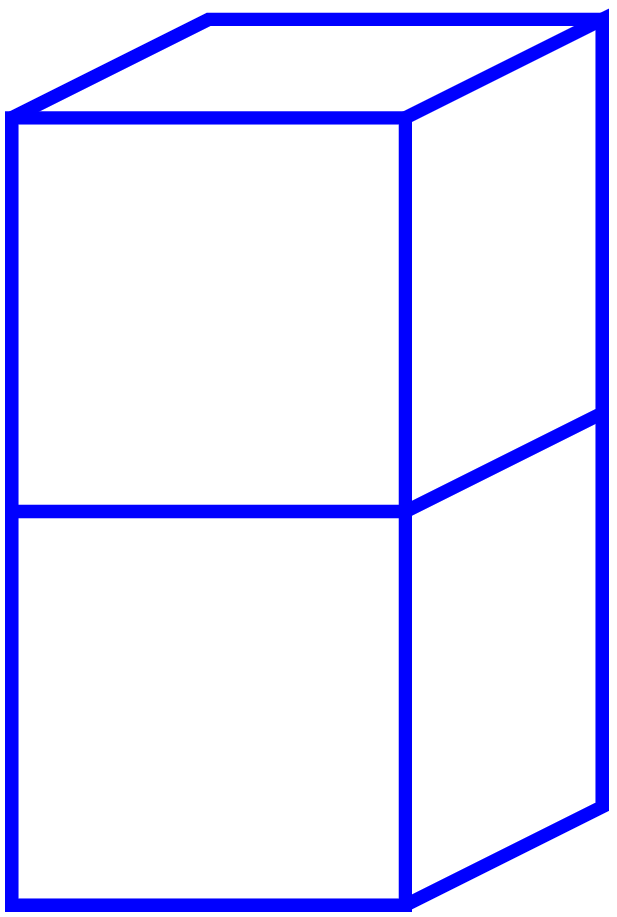}}(v)}{u-v+\epsilon_{2}}-\frac{e_{\includegraphics[scale=0.035]{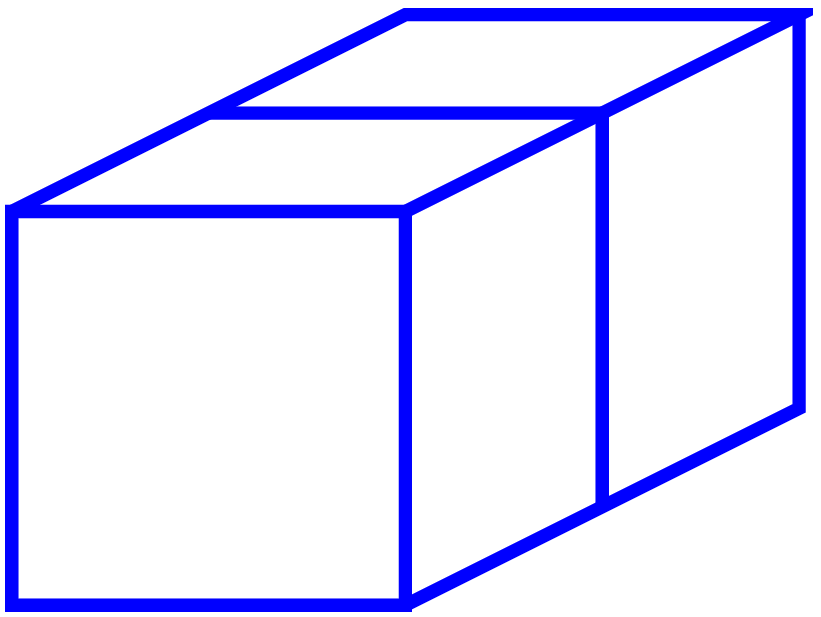}}(v)}{u-v+\epsilon_{3}} }\Bigr]=\\=
		\bar{g}(u-v)\Bigl[e(v)e(u)\textcolor{blue}{-\frac{e_{\includegraphics[scale=0.035]{partition2-blue.eps}}(u)}{u-v-\epsilon_{1}}-
			\frac{e_{\includegraphics[scale=0.035]{partition3-blue.eps}}(u)}{u-v-\epsilon_{2}}-\frac{e_{\includegraphics[scale=0.035]{partition1-blue.eps}}(u)}{u-v-\epsilon_{3}}}\Bigr], 
\end{multline}
\begin{multline}\label{ff-exact-relation}
		\bar{g}(u-v)\Bigl[f(u)f(v)\textcolor{blue}{-\frac{f_{\includegraphics[scale=0.035]{partition2-blue.eps}}(v)}{u-v-\epsilon_{1}}-
			\frac{f_{\includegraphics[scale=0.035]{partition3-blue.eps}}(v)}{u-v-\epsilon_{2}}-\frac{f_{\includegraphics[scale=0.035]{partition1-blue.eps}}(v)}{u-v-\epsilon_{3}}}\Bigr]=\\=
		g(u-v)\Bigl[f(v)f(u)\textcolor{blue}{-\frac{f_{\includegraphics[scale=0.035]{partition2-blue.eps}}(u)}{u-v+\epsilon_{1}}-
			\frac{f_{\includegraphics[scale=0.035]{partition3-blue.eps}}(u)}{u-v+\epsilon_{2}}-\frac{f_{\includegraphics[scale=0.035]{partition1-blue.eps}}(u)}{u-v+\epsilon_{3}}}\Bigr],
\end{multline}
and Serre relations
\begin{gather}    
	\sum_{\sigma\in\mathbb{S}_{3}}(u_{\sigma_{1}}-2u_{\sigma_{2}}+u_{\sigma_{3}})e(u_{\sigma_{1}})e(u_{\sigma_{2}})e(u_{\sigma_{3}})+\sum_{\sigma\in\mathbb{S}_{3}}
	[e(u_{\sigma_1}),e_{\includegraphics[scale=0.035]{partition1.eps}}(u_{\sigma_2})+e_{\includegraphics[scale=0.035]{partition2.eps}}(u_{\sigma_2})+e_{\includegraphics[scale=0.035]{partition3.eps}}(u_{\sigma_2})]=0,\label{Serre-relations-e} \\ 
	\sum_{\sigma\in\mathbb{S}_{3}}(u_{\sigma_{1}}-2u_{\sigma_{2}}+u_{\sigma_{3}})f(u_{\sigma_{1}})f(u_{\sigma_{2}})f(u_{\sigma_{3}})+\sum_{\sigma\in\mathbb{S}_{3}}
	[f(u_{\sigma_1}),f_{\includegraphics[scale=0.035]{partition1.eps}}(u_{\sigma_2})+f_{\includegraphics[scale=0.035]{partition2.eps}}(u_{\sigma_2})+f_{\includegraphics[scale=0.035]{partition3.eps}}(u_{\sigma_2})]=0. 
	\label{Serre-relations}
\end{gather}
\end{subequations}
In the relations above we have used the following notations
\begin{equation}
 g(x)\overset{\text{def}}{=}(x+\epsilon_{1})(x+\epsilon_{2})(x+\epsilon_{3}),\quad
 \bar{g}(x)\overset{\text{def}}{=}(x-\epsilon_{1})(x-\epsilon_{2})(x-\epsilon_{3}).
\end{equation}
The higher currents $e_{\boldsymbol{\lambda}}$ and $f_{\boldsymbol{\lambda}}$ in \eqref{ee-exact-relation}-\eqref{Serre-relations} are related to \eqref{level2-currents} and \eqref{level2-currents-f} by change of notations \eqref{epsilonNotation}-\eqref{uepsilonNotation} and by certain scaling factors. 
	
We note that the terms shown by blue in \eqref{he-relation}-\eqref{ff-exact-relation} depend only on one parameter either $u$ or $v$.  We call such terms \emph{local}. If one writes the commutation relations for the modes of the currents \eqref{currents-large-u-expansion}, \emph{local} terms affect only few of them. Indeed if we apply
\begin{equation}\label{ij-integral-application}
 \frac{1}{(2\pi i)^{2}}\oint_{\mathcal{C}_{\infty}}\oint_{\mathcal{C}_{\infty}}u^{i}v^{j}du\,dv
\end{equation}
to \eqref{he-relation} for $i,j\geq 0$ the \emph{local} term  $h(u)e(u)$ does not contribute and we obtain
\begin{equation}
 [h_{i+1},e_j]-[h_i,e_{j+1}]=\epsilon_3 h_i e_j.
\end{equation}
The local term appears if we apply \eqref{ij-integral-application} with $j=-1$
\begin{equation}\label{first-local-relation}
 [h(u),e_0]=\epsilon_3h(u)e(u)\overset{\eqref{efh-def}}{\implies}\mathcal{L}_{\scriptscriptstyle{\varnothing,\Box}}(u)=[e_0,h(u)].
\end{equation}
Similarly, applying 
\begin{equation}
 \frac{1}{(2\pi i)^{2}}\oint_{\mathcal{C}_{\infty}}\oint_{\mathcal{C}_{\infty}}u^{i}v^{j}du\,dv\cdot
\end{equation}
for $i,j\geq0$ to  \eqref{ee-exact-relation}, local terms represented by $e_{\boldsymbol{\lambda}}(u)$ do not contribute and we obtain
\begin{equation}\label{ee-ff-in-modes}
	[e_{i+3},e_{j}]-3[e_{i+2},e_{j+1}]+3[e_{i+1},e_{j+2}]-[e_{i},e_{j+3}]+\sigma_{2}\left([e_{i+1},e_{j}]-[e_{i},e_{j+1}]\right)=\sigma_{3}\{e_{i},e_{j}\},
\end{equation}
where $\sigma_{k}$ are elementary symmetric polynomials in $\epsilon_{j}$.  However, taking either $i$ or $j$ negative allow to express  the higher currents  $e_{\boldsymbol{\lambda}}$ in terms of commutators 
\begin{equation}\label{higher-currents-as-commutators}
 \begin{aligned}
	&e_{\includegraphics[scale=0.035]{partition2.eps}}(u)=\frac{1}{(\epsilon_{1}-\epsilon_{2})(\epsilon_{1}-\epsilon_{3})}
	\Bigl((u-\epsilon_{2})(u-\epsilon_{3})[e(u),e_{0}]-(2u+\epsilon_{1})[e(u),e_{1}]+[e(u),e_{2}]-3[e_{1},e_{0}]\Bigr),\\
	&e_{\includegraphics[scale=0.035]{partition3.eps}}(u)=\frac{1}{(\epsilon_{2}-\epsilon_{1})(\epsilon_{2}-\epsilon_{3})}
	\Bigl((u-\epsilon_{1})(u-\epsilon_{3})[e(u),e_{0}]-(2u+\epsilon_{2})[e(u),e_{1}]+[e(u),e_{2}]-3[e_{1},e_{0}]\Bigr),\\
	&e_{\includegraphics[scale=0.035]{partition1.eps}}(u)=\frac{1}{(\epsilon_{3}-\epsilon_{1})(\epsilon_{3}-\epsilon_{2})}
	\Bigl((u-\epsilon_{1})(u-\epsilon_{2})[e(u),e_{0}]-(2u+\epsilon_{3})[e(u),e_{1}]+[e(u),e_{2}]-3[e_{1},e_{0}]\Bigr),
 \end{aligned}
\end{equation}
and similar expressions for $f_{\boldsymbol{\lambda}}(u)$. 
	
Using the relations \eqref{higher-currents-as-commutators} one can express generators of the Yangian $\mathcal{L}_{\scriptscriptstyle{\varnothing,\boldsymbol{\lambda}}}(u)$ with $|\boldsymbol{\lambda}|=2$ as 
\begin{equation}\label{second-local-relation}
 \mathcal{L}_{\scriptscriptstyle{\varnothing},\zerounderset{\scriptscriptstyle{\Box}}{\vspace*{-1.55pt}\scriptscriptstyle{\Box}}}(u)=[e_0,[e_0,h(u)]],\qquad
 \mathcal{L}_{\scriptscriptstyle{\varnothing,\Box\hspace*{-1.2pt}\vspace*{2pt}\Box}}(u)= [[e_1,e_0],h(u)]
\end{equation}
and similarly for $\mathcal{L}_{\scriptscriptstyle{\boldsymbol{\lambda},\varnothing}}(u)$ with $e_k$ being replaced by $f_k$.  These equations as well as \eqref{first-local-relation} suggest that generic generator $\mathcal{L}_{\scriptscriptstyle{\boldsymbol{\lambda},\boldsymbol{\mu}}}(u)$ can be obtained as an adjoint action of $e_k$ and $f_k$ generators on $h(u)$. Using the $RLL$ relations \eqref{YB-algebra} at level $3$ one can find
\begin{equation}
 \mathcal{L}_{\scriptscriptstyle{\varnothing},\zerounderset{\scriptscriptstyle{\Box}}{\vspace*{-1.55pt}\zerounderset{\scriptscriptstyle{\Box}}{\vspace*{-1.55pt}\scriptscriptstyle{\Box}}}}(u)=[e_0,[e_0,[e_0,h(u)]]],\quad \mathcal{L}_{\scriptscriptstyle{\varnothing},\hspace*{4pt}\zerounderset{\scriptscriptstyle{\Box}\hspace*{-1.3pt}\scriptscriptstyle{\Box}}{\vspace*{-1.5pt}\hspace*{-4.07pt}\scriptscriptstyle{\Box}}}\hspace*{3pt}(u)=[e_0,[[e_1,e_0],h(u)],\quad \mathcal{L}_{\scriptscriptstyle{\varnothing,\Box\hspace*{-1.2pt}\vspace*{2pt}\Box\hspace*{-1.2pt}\vspace*{2pt}\Box}}(u)=[[e_1,[e_1,e_0]],h(u)]
\end{equation}
In general, we have found nice representation for the generating function
\begin{equation}
 \langle u|e^{\sum\limits_n t_{-n} a_n}\mathcal{L}(u)e^{\sum\limits_n t_n a_{-n}}|u\rangle=e^{\sum\limits_n t_{-n}t_n} e^{\sum\limits_nJ_{-n}t_n}e^{-\sum\limits_nJ_{n}t_{-n}}h(u)e^{\sum\limits_nJ_{n}t_{-n}}e^{-\sum\limits_nJ_{-n}t_n}.
\end{equation}
Here $J_n$ are the modes of the Heisenberg $W^{(1)}$ current \eqref{Jn-}, which can be expressed in terms of $e$ and $f$ currents \eqref{Jef}
\begin{equation}
 \begin{aligned}
   &J_{-1}=e_0,\quad &&J_{-2} = [e_1,e_0],\quad &&J_{-3}=[e_1,[e_1,e_0]],\quad&&\dots\\
   &J_{1}=f_0,\quad &&J_{2} = [f_1,f_0],\quad &&J_{3}=[f_1,[f_1,f_0]],\quad&&\dots
 \end{aligned}
\end{equation}
\subsection{Center of \texorpdfstring{$\textrm{YB}\left(\widehat{\mathfrak{gl}}(1)\right)$}{YB(gl(1))}}
In this section we will show that the algebra $YB\left(\widehat{\mathfrak{gl}}(1)\right)$ contains a huge center. Namely for any singular vector $|s\rangle$ of $W_n$ algebra in the space of $n$ bosons we assign a central element $D_s$ \eqref{Ds}.
First representative of this series is related to the operator $\psi(u)$ as
\begin{equation}
  D_{1,1}(u)=
	\psi(u) \frac{h(u)h(u+\epsilon_3)}{h(u-\epsilon_1)h(u-\epsilon_2)}.
\end{equation}
In  representation of $YB\left(\widehat{\mathfrak{gl}}(1)\right)$ in the space of $n$ bosons $\mathfrak{F}_{u_1}\otimes\dots\mathfrak{F}_{u_n}$ the element $D_{1,1}(u)$ acts by the function:
\begin{equation}
	D_{1,1}(u)|\boldsymbol{\varnothing}\rangle=V(u)|\boldsymbol{\varnothing}\rangle\quad\text{where}\quad V(u)=\prod_{k=1}^{n}\frac{u-u_{k}+\epsilon_{3}}{u-u_{k}},
\end{equation}
which we call the weight of representation.
	
In order to see it we note that the algebra \eqref{Yangian-relation-main} contains additional Hamiltonian $\psi(u)$ which commutes with $h(v)$. One can derive that
\begin{equation}\label{psi-e-relation}
 \psi(u)e(v)=\prod_{\alpha=1}^{3}\frac{(u-v-\epsilon_{\alpha})}{(u-v+\epsilon_{\alpha})}e(v)\psi(u) \textcolor{blue}{+locals}
\end{equation}
Using the relation
\begin{equation}
 e(v+\epsilon_3)h^{-1}(v+\epsilon_3)=h^{-1}(v+\epsilon_3)e(v),
\end{equation}
which immediately follows from \eqref{he-relation} at $u=v+\epsilon_3$, we may transform the operator $\psi(u)$ to the more convenient form
\begin{equation}
 \psi(u)=-h^{-1}(u+\epsilon_3)\,_{u+\epsilon_3}\langle\varnothing|\otimes\,_{u}\langle\varnothing| a_{1}^{(2)}|\mathcal{L}^1(u+\epsilon_3)\mathcal{L}^2(u)\bigl(a_{-1}^{(1)}-a_{-1}^{(2)}\bigr)|\varnothing\rangle_{u}\otimes |\varnothing\rangle_{u+\epsilon_3} h^{-1}(u).
\end{equation}
Using another identity
\begin{equation}
 \,_{u+\epsilon_3}\langle\varnothing|\otimes\,_{u}\langle\varnothing| (a_1^{(1)}+a_{1}^{(2)})\mathcal{L}^1(u)\mathcal{L}^2(u+\epsilon_3)(a_{-1}^{(1)}-a_{-1}^{(2)})|\varnothing\rangle_{u}\otimes |\varnothing\rangle_{u+\epsilon_3}=0,
\end{equation}
we find
\begin{equation}
 \psi(u)=\frac{\langle s_{1,1}|\mathcal{L}^1(u)\mathcal{L}^2(u+\epsilon_3)|s_{1,1}\rangle}{h(u)h(u+\epsilon_3)}, \label{singular-Psi}
\end{equation}
where
\begin{equation}
 |s_{1,1}\rangle_u\overset{\text{def}}{=}\bigl(a_{-1}^{(1)}-a_{-1}^{(2)}\bigr)|\varnothing\rangle_{u}\otimes |\varnothing\rangle_{u+\epsilon_3}
\end{equation}
is a singular vector of a $W$-algebra which appears in the tensor product of two Fock spaces $\mathcal{F}_{u_1}\otimes\mathcal{F}_{u_2}$ at $u_2=u_1+\epsilon_3$. Indeed it can be checked that under the resonance condition $u_2=u_1+\epsilon_3$ the vector $|s_{1,1}\rangle$ is annihilated by positive modes of the $W-$currents defined by Miura formula:
\begin{equation}
 -\epsilon_3^2\partial^2-i \epsilon_3 \mathcal{W}^{(1)}(z)\partial +\mathcal{W}^{(2)}(z)=(i\epsilon_3 \partial -\partial \phi_1)(i\epsilon_3 \partial -\partial \phi_2)
\end{equation}
Due to the property that singular vector is annihilated by all positive modes of $W-$currents it follows that the $R$-matrix acts trivially on the tensor product of the vacuum and the singular vector. In our particular case we have
\begin{equation}\label{Intertvining-vac-S}
	R_{0,1}(u-v)R_{0,2}(u-v+\epsilon_3)|\varnothing\rangle_u \otimes |s_{1,1}\rangle_v=\frac{u-v+\epsilon_3}{u-v}|\varnothing\rangle_u \otimes |s_{1,1}\rangle_v 
\end{equation}
Relation \eqref{Intertvining-vac-S} implies the commutativity of $h(v)$ and $\psi(u)$ and ensures that the Hamiltonian $\psi(u)$ acts on the vacuum $|\varnothing\rangle_v$ by the highest weight
\begin{equation}
 \psi(u)|\varnothing\rangle_v=\frac{u-v+\epsilon_3}{u-v}|\varnothing\rangle_v \label{HighestWeight}
\end{equation}
	
We also found by explicit calculation that $\psi(u) \frac{h(u)h(u+\epsilon_3)}{h(u-\epsilon_1)h(u-\epsilon_2)}$ commutes with $e(v)$ and $f(v)$ and so belongs to the center of RLL algebra \footnote{This fact is an analog of similar relation in $\mathfrak{gl}(2)$ Yangian: operator $\psi(u)$ is a direct analog of $q-$determinant \cite{Kulish:1981bi}
\begin{equation}
	 qDet\bigl[L^{gl(2)}(u)\bigr]=\langle\uparrow\otimes\downarrow-\downarrow\otimes\uparrow|R^{(1)}(u)R^{(2)}(u+\epsilon_3)|\uparrow\otimes\downarrow-\downarrow\otimes\uparrow|\rangle
\end{equation}
and $qDet\bigl[L^{gl(2)}(u)\bigr]$ belongs to the center of $Y\bigl(\mathfrak{gl}(2)\bigr)$.}.
In order to understand this phenomenon, let us note that R-matrix between two vector spaces which are representations of $W_{\infty}$ algebra is completely (up to a normalization constant) fixed by the eigenvalues of zero modes $W_{0}$ of $W$-currents on vacuum and intertwining identity
\begin{multline}\label{IntertwiningIdentity}
 \mathcal{R}\Big(\sum\limits_{k\ge 0}\big(\mathcal{W}^{(k)}(z)\otimes 1\big) \big(i\epsilon_3\partial\big)^{n_1-k}\Big)\Big(\sum\limits_{k\ge 0}\big(\mathcal{W}^{(k)}(z)\otimes 1\big) \big(i\epsilon_3\partial\big)^{n_1-k}\Big)\Big(\sum\limits_{k\ge 0}\big(1\otimes \mathcal{W}^{(k)}(z) \big)\big(i\epsilon_3\partial\big)^{n_2-k}\Big)=\\=\Big(\sum\limits_{k\ge 0}\big(1\otimes \mathcal{W}^{(k)}(z) \big) \big(i\epsilon_3\partial\big)^{n_2-k}\Big)\Big(\sum\limits_{k\ge 0}\big(\mathcal{W}^{(k)}(z)\otimes 1\big) \big(i\epsilon_3\partial\big)^{n_1-k}\Big)\mathcal{R} 
\end{multline}
	
We will consider two representations of $W$-algebra, one in the space of one boson, and other in the space of finite number of bosons $n$. We take two different representations of $W_n$ algebra - one is the standard Fock representation and the other is the highest weight representation arising from the singular vector $|s\rangle_u$. 
Let us compute exchanging relation of higher Hamiltonian
\begin{equation}
 h_s(u)=\,_{u}\langle s|\mathcal{L} |s\rangle_u 
\end{equation}
and the current $e(v)$. On general grounds, it has the form
\begin{equation}\label{h_s-e-relation}
 h_s(u)e(v)=F_s(u-v) e(v)h_s(u)\textcolor{blue}{+locals}, 
\end{equation}
where $F_s(u-v)$ is some rational function. Let us concentrate on the first term of \eqref{h_s-e-relation}, because local terms are fixed by a demand that l.h.s of \eqref{h_s-e-relation} doesn't have poles\footnote{First term of \eqref{h_s-e-relation} obviously has poles because of the rational function $F_s(u-v)$. Its residues should be canceled by local terms which fixes them unambiguously.}.
According to the $RLL$ relation, the function $F_s(u-v)$ is equal to the matrix element 
\begin{equation}\label{F_s}
	F_s(u-v)=V^{-1}(u-v){}_u\langle s|\otimes{}_v\langle\varnothing|a_1 \mathcal{R}(u-v)a_{-1}|\varnothing\rangle_v|s\rangle_u,  
\end{equation}
where $V(u-v)$ is the weight of representation arising from the singular vector $|s\rangle$
\begin{equation}
	\mathcal{R}(u-v)|\varnothing\rangle_v|s\rangle_u =V(u-v)|\varnothing\rangle_v|s\rangle_u
\end{equation}

In order to calculate the matrix element \eqref{F_s} let us act by the minus first mode of intertwining identity \eqref{IntertwiningIdentity}, specified to the case $n_1$=1, $n_2=n$, $\Big(\sum\limits_{k\ge 0}\big(\mathcal{W}^{(k)}(z)\otimes 1\big) \big(i\epsilon_3\partial\big)^{n_1-k}\Big)\to i\epsilon_3\partial-\partial\phi(z)$, on vacuum
\begin{multline}\label{BoxS}
 \mathcal{R}\Big[ a_{-1}\sum_k \mathcal{W}_0^{(k)}\big(i\epsilon_3\partial\big)^{n-k}+\sum_k W^{(k)}_{-1}(-\epsilon_3+i\epsilon_3\partial+u)\big(i\epsilon_3\partial\big)^{n-k}\Big]|\varnothing\rangle_u\otimes|s\rangle_v=\\=V(u-v)\Big[ a_{-1}\sum_k\mathcal{W}^{(k)}_{0}\big(-\epsilon_3+i\epsilon_3\partial\big)^{n-k}+\sum_k\mathcal{W}^{(k)}_{-1}(i\epsilon_3\partial+u)\big(i\epsilon_3\partial\big)^{n-k}\Big]|\varnothing\rangle_u|\otimes|s\rangle_v 
\end{multline}
The desired matrix element can be found by  solving a linear system and excluding all $W_{-1}^{(k)}$ modes in the l.h.s of \eqref{BoxS}. However one can avoid this complicated calculation simply by substitution $i\epsilon_3\partial\to -u+\epsilon_3$:
	\begin{equation}
	{}_v\langle s|\otimes\,{}_u\langle \square| R(u-v)|\square\rangle_u\times|s\rangle_v=V(u-v)\frac{\sum_k W_0^{(k)}\big(-u+\epsilon_3\big)^{n-k}}{\sum_kW^{(k)}_{0}\big(-u\big)^{n-k}}.
	\end{equation}
Thus we find that the exchanging function in \eqref{h_s-e-relation} depends only on the polynomial $P_s(u)$:
\begin{equation}\label{Ps}
	P_s(u)=\sum_kW^{(k)}_{0}\big(-u\big)^{n-k}=\prod\limits_{k=1}^{n}(u-v_k)
\end{equation} 
as
\begin{equation}\label{Fs}
	F_s(u-v)=\frac{P_s(u+\epsilon_3)}{P_s(u)}. 
\end{equation}
For example, explicit calculation for singular vector on the first level $|s_{1,1}\rangle=(a_{-1}^{(1)}-a_{-1}^{(2)})|\varnothing\rangle_{v,v+\epsilon_3}$ provides
\begin{equation}
 P_{s_{1,1}}(u)=(u-v-\epsilon_1)(u-v-\epsilon_2)
\end{equation}
		
More generally for a singular vector in $W_2$ algebra $s_{m,n}$ at level $mn$
\begin{equation}
 P_{s_{m,n}}=(u-v-m \epsilon_1)(u-v-n\epsilon_2) \label{Pmn}
\end{equation}
Let us note that the same polynomial corresponds to a vacuum vector in two Fock spaces $F_{v-m\epsilon_1}\otimes F_{v-n\epsilon_2}$. This calculation immediately implies that current $D_{m,n}(u)=\frac{h_{s_{m,n}}(u)}{h(u-m\epsilon_1)h(u-n\epsilon_2)}$ commute with $e(v), f(v),h(v)$ and so belongs to the center of $YB(\widehat{\mathfrak{gl}}(1))$. Indeed:
\begin{equation}
 D_{m,n}(u)e(v)=e(v)D_{m,n}(u)\textcolor{blue}{+locals}
\end{equation}
However, as we have seen, all local terms came up with poles which should be canceled with residues of non local term. Since non-local terms do not have poles no local terms allowed. Thus, we proved
\begin{equation}
 D_{m,n}(u)e(v)=e(v)D_{m,n}(u)   
\end{equation}
Exchanging relation with $f(u)$ is similar, and hence we prove that $D_{m,n}(u)$ is indeed belongs to the center of $YB(\widehat{\mathfrak{gl}}(1))$.
	
In general, any singular vector of $W_n$ algebra in the space of $n$ Fock modules gives rise to central element of $YB(\widehat{\mathfrak{gl}}(1))$. As we explained exchanging relations of higher Hamiltonian $h_s$ with $e(v),f(v)$ currents are encoded in a single polynomial \eqref{Ps}. And the element:
\begin{equation}\label{Ds}
	D_s=\frac{h_s(u)}{\prod\limits_{i=1}^n h(u-v_i)}
\end{equation}
is central.
\subsection{Zero twist integrable system}
The Yang-Baxter algebra $\textrm{YB}(\widehat{\mathfrak{gl}}(1))$ contains commutative subalgebra spanned by modes of the current $h(u)$. If one consider a representation of  the $\textrm{YB}(\hat{\mathfrak{gl}}(1))$ algebra on $n$ sites, this integrable system is known to coincide with matrix generalization of quantum Benjamin-Ono integrable hierarchy. It attracted some attention because it is directly related to AGT representation for conformal blocks   \cite{Alday:2009aq}. Much is known about this integrable system. In particular, its spectra and eigenfunctions can be written rather explicitly.
	
Suppose, one has an eigenvector of $h(u)$
\begin{equation}
 h(u)|\Lambda\rangle=h_{\Lambda}(u)|\Lambda\rangle.
\end{equation}
Then one can try to create new states by repetitive  application of $e(v)$. Using \eqref{he-relation}, one finds that
\begin{equation}\label{e-application-trial}
 h(u)e(v)|\Lambda\rangle=\frac{u-v}{u-v-\epsilon_{3}}h_{\Lambda}(u)e(v)|\Lambda\rangle-\frac{\epsilon_{3}}{u-v-\epsilon_{3}}L_{\scriptscriptstyle{\varnothing,\Box}}(u)|\Lambda\rangle,
\end{equation}
and hence in general $e(v)|\Lambda\rangle$ is not an eigenvector of $h(u)$. However if $e(v)|\Lambda\rangle$ develops a singularity at some value $v=x$, typically a pole, then the second term in the r.h.s. of \eqref{e-application-trial} is negligible and we  have a new eigenvector
\begin{equation}\label{h-main-property}
 |\tilde{\Lambda}\rangle=\frac{1}{2\pi i}\oint\limits_{\mathcal{C}_{x}}e(v)|\Lambda\rangle dv,\qquad
 h(u)|\tilde{\Lambda}\rangle=\frac{(u-x)}{(u-x-\epsilon_{3})}h_{\Lambda}(u)|\tilde{\Lambda}\rangle
\end{equation}
Similar argument applies to the operator $\psi(u)$
\begin{equation}\label{psi-main-property}
 \psi(u)|\tilde{\Lambda}\rangle=\prod_{\alpha=1}^{3}\frac{(u-x-\epsilon_{\alpha})}{(u-x+\epsilon_{\alpha})}\psi_{\lambda}(u)|\tilde{\Lambda}\rangle
\end{equation}
and to any higher Hamiltonian  $h_s(u)$ from the previous section. 

Using \eqref{h-main-property}-\eqref{psi-main-property}, one can generate any eigenvector from the vacuum state by successive application of $e(u)$. We note that the operators $e(u)$ do not commute. However the structure of commutation relations \eqref{ee-exact-relation} implies the following property
\begin{equation}\label{ee-main-property}
 \oint\limits_{\mathcal{C}_{y}}dv \oint\limits_{\mathcal{C}_{x}}du\, e(u)e(v)|\Lambda\rangle=
	\prod_{\alpha=1}^{3}\frac{(x-y-\epsilon_{\alpha})}{(x-y+\epsilon_{\alpha})}
	\oint\limits_{\mathcal{C}_{y}}dv \oint\limits_{\mathcal{C}_{x}}du\, e(v)e(u)|\Lambda\rangle
\end{equation}
provided that $x$ and $y$ are \emph{simple} poles and that $y\neq x+\epsilon_{\alpha}$. 
	
The properties  \eqref{h-main-property}-\eqref{psi-main-property} and \eqref{ee-main-property} are used to show that the eigenstates are in correspondence with tuples of Young diagrams or more generally with 3D partitions. In order to demonstrate how it works, we take our quantum space to be the tensor product of $n$ Fock modules generated from the vacuum state $|\boldsymbol{\varnothing}  \rangle=|x_{1}\rangle\otimes|x_{2}\rangle\otimes\dots\otimes|x_{n}\rangle$
\begin{equation}
 \mathcal{F}_{x_{1}}\otimes\dots\otimes\mathcal{F}_{x_{n}}=\textrm{span}\{a_{-\boldsymbol{\lambda}^{(1)}}^{(1)}\dots a_{-\boldsymbol{\lambda}^{(n)}}^{(n)}|\boldsymbol{\varnothing}  \rangle:\boldsymbol{\lambda}^{(k)}=\lambda^{(k)}_{1}\geq\lambda^{(k)}_{2}\geq\dots\}.
\end{equation}
Our normalization of $h(u)$, which is inherited from our normalization of the $R-$matrix \eqref{R-normalization},  implies that $h(u)|\boldsymbol{\varnothing}\rangle=|\boldsymbol{\varnothing}\rangle$. Then it follows from the definition of $\psi(u)$ \eqref{psi-definition} that 
\begin{equation}\label{psi-eigenvalue}
	\psi(u)|\boldsymbol{\varnothing}\rangle=\prod_{k=1}^{n}\frac{u-x_{k}+\epsilon_{3}}{u-x_{k}}|\boldsymbol{\varnothing}\rangle. 
\end{equation}
Moreover the vacuum state is annihilated  by $f(u)$
\begin{equation}
	f(u)|\boldsymbol{\varnothing}\rangle=0,
\end{equation}
while the new states are generated by the modes of $e(u)$. In principle, one can rewrite a generic state in $\mathcal{F}_{x_{1}}\otimes\dots\otimes\mathcal{F}_{x_{n}}$ as an integral
\begin{equation}
	a_{-\boldsymbol{\lambda}^{(1)}}^{(1)}\dots a_{-\boldsymbol{\lambda}^{(n)}}^{(n)}|\boldsymbol{\varnothing}  \rangle=
	\int\dots\int \rho_{\vec{\boldsymbol{\lambda}}}(\boldsymbol{u})e(u_{N})\dots e(u_{1})|\boldsymbol{\varnothing}  \rangle\,du_{1}\dots du_{N}\quad\text{where}\quad
	N=\sum_{k=1}^{n}|\boldsymbol{\lambda}^{(k)}|,
\end{equation}
for some function $\rho_{\vec{\boldsymbol{\lambda}}}(\boldsymbol{u})$ (see \cite{Prochazka:2015deb,Gaberdiel:2017dbk} for explicit formulae on lowest levels). The eigenfunctions of  $h(u)$ provide another basis $|\vec{\boldsymbol{\lambda}}\rangle$ in $\mathcal{F}_{x_{1}}\otimes\dots\otimes\mathcal{F}_{x_{n}}$ which has very simple form in terms of $e(u)$ generators
\begin{equation}\label{lambda-contour-integral}
 |\vec{\boldsymbol{\lambda}}\rangle \sim 
	\oint\limits_{\mathcal{C}_{N}}du_{N}\dots\oint\limits_{\mathcal{C}_{1}}du_{1}\,e(u_{N})\dots e(u_{1})|\boldsymbol{\varnothing}  \rangle,\qquad
	N=|\vec{\boldsymbol{\lambda}}|=\sum_{k=1}^{n}|\boldsymbol{\lambda}^{(k)}|,
\end{equation}

We will specify the proportionality coefficient in \eqref{lambda-contour-integral} later. In fact it depends on the order in which we perform the integrations. The contours in \eqref{lambda-contour-integral} go counterclockwise around simple poles located at the contents of  Young diagrams in $\vec{\boldsymbol{\lambda}}$. By definition  a content of a cell with coordinates $(i,j)$ in Young diagram $\boldsymbol{\lambda}^{(k)}$ is
\begin{equation}
 c_{\Box}=x_{k}-(i-1)\epsilon_{1}-(j-1)\epsilon_{2}.
\end{equation}
The order of the contours $\mathcal{C}_{i}$ in \eqref{lambda-contour-integral} should follow the order of any standard Young tableaux  associated to $\boldsymbol{\lambda}^{(k)}$. Different choices of the ordering would lead to the same state which might differ by a factor, later we will provide a formula for eigenvector $|\vec{\boldsymbol{\lambda}}\rangle$ which is independent of the ordering (see \eqref{eigenvector}). 

The state defined by \eqref{lambda-contour-integral} is an eigenstate of $h(u)$ and $\psi(u)$ with eigenvalues 
\begin{equation}\label{psi-h-eigenvalues}
 h(u)|\vec{\boldsymbol{\lambda}}\rangle=\prod_{\Box\in\vec{\boldsymbol{\lambda}}}\frac{(u-c_{\Box})}{(u-c_{\Box}-\epsilon_{3})}|\vec{\boldsymbol{\lambda}}\rangle,\qquad
	\psi(u)|\vec{\boldsymbol{\lambda}}\rangle=\prod_{\alpha=1}^{3}\prod_{\Box\in\vec{\boldsymbol{\lambda}}}\frac{(u-c_{\Box}-\epsilon_{\alpha})}{(u-c_{\Box}+\epsilon_{\alpha})}
	\prod_{k=1}^{n}\frac{(u-x_{k}+\epsilon_{3})}{(u-x_{k})}|\vec{\boldsymbol{\lambda}}\rangle
\end{equation}
We note that \eqref{psi-h-eigenvalues} follows immediately from \eqref{he-relation}, \eqref{psi-e-relation} and \eqref{psi-eigenvalue} provided that the surrounded singularities of the integrand in \eqref{lambda-contour-integral} are all simple poles.  This statement can be proven by induction in level $N$:
\begin{itemize}
 \item The base of induction. Let us consider generic states at level one: $e(u)|\boldsymbol{\varnothing}\rangle$. In order to find it's poles we use \eqref{ef-relation}
	  \begin{equation}\label{f-simplest-action}
		f(v)e(u)|\boldsymbol{\varnothing}\rangle=-\frac{\psi(u)-\psi(v)}{u-v}|\boldsymbol{\varnothing}\rangle,
		\end{equation}
		which implies that poles of $e(u)|\boldsymbol{\varnothing}\rangle$ are located exactly at $u=x_{k}$ and hence 
		\begin{equation}
		|\Box_{k}\rangle\sim\oint\limits_{\mathcal{C}_{k}}du\,e(u)|\boldsymbol{\varnothing}\rangle
		\end{equation}
		are the eigenstates of $h(u)$.
	\item Let us assume that up to level $N$ the operators $e(u)$ and $f(u)$ act as follows
	\begin{equation}\label{induction-assumption}
	\begin{aligned}
	&e(u)|\vec{\boldsymbol{\lambda}}\rangle=\sum_{\Box\in\textrm{addable}(\vec{\boldsymbol{\lambda}})}\frac{E(\vec{\boldsymbol{\lambda}},\vec{\boldsymbol{\lambda}}+\Box)}{u-c_{\Box}}
	|\vec{\boldsymbol{\lambda}}+\Box\rangle&&\quad\text{for}\quad |\vec{\boldsymbol{\lambda}}|<N,\\
	&f(u)|\vec{\boldsymbol{\lambda}}\rangle=\sum_{\Box\in\textrm{removable}(\vec{\boldsymbol{\lambda}})}\frac{F(\vec{\boldsymbol{\lambda}},\vec{\boldsymbol{\lambda}}-\Box)}{u-c_{\Box}}
		|\vec{\boldsymbol{\lambda}}-\Box\rangle&&\quad\text{for}\quad |\vec{\boldsymbol{\lambda}}|\leq N,
		\end{aligned}
	\end{equation}
	where the amplitudes $E(\vec{\boldsymbol{\lambda}},\vec{\boldsymbol{\lambda}}+\Box)$ and $F(\vec{\boldsymbol{\lambda}},\vec{\boldsymbol{\lambda}}-\Box)$ are given by
	\begin{align}\label{E-amplitude-explicit}
	&E(\vec{\boldsymbol{\lambda}},\vec{\boldsymbol{\lambda}}+\Box)=\frac{\epsilon_{1}\epsilon_{2}}{\epsilon_{3}}\prod_{\Box'\in\vec{\boldsymbol{\lambda}}+\Box}
	S^{-1}(c_{\Box'}-c_{\Box})\prod_{k=1}^{n}\frac{(c_{\Box}-x_{k}+\epsilon_{3})}{(c_{\Box}-x_{k})},\\\label{F-amplitude-explicit}
	&F(\vec{\boldsymbol{\lambda}},\vec{\boldsymbol{\lambda}}-\Box)=\prod_{\Box'\in\vec{\boldsymbol{\lambda}}-\Box}S(c_{\Box}-c_{\Box'}),
	\end{align}
	with 
	 \begin{equation}\label{S-function-def}
		S(x)=\frac{(x+\epsilon_1)(x+\epsilon_2)}{x(x-\epsilon_3)}.
	 \end{equation}
	 In \eqref{induction-assumption} the sets $\textrm{addable}(\vec{\boldsymbol{\lambda}})$ and $\textrm{removable}(\vec{\boldsymbol{\lambda}})$ corresponds to the sets of all boxes which can be either added or removed from $\vec{\boldsymbol{\lambda}}$.
	\item We have to show that $e(u)|\vec{\boldsymbol{\lambda}}\rangle$ with $|\vec{\boldsymbol{\lambda}}|=N$ has poles at addable points. Consider $u$ poles of the following vector
		\begin{equation}\label{trial-vector-induction}
		f(v)e(u)|\vec{\boldsymbol{\lambda}}\rangle=-\frac{\psi(u)-\psi(v)}{u-v}|\vec{\boldsymbol{\lambda}}\rangle+e(u)f(v)|\vec{\boldsymbol{\lambda}}\rangle.
		\end{equation}
		There are two sources of poles in the r.h.s of  \eqref{trial-vector-induction}: the eigenvalue of $\psi(u)$ and the $e(u)f(v)|\vec{\boldsymbol{\lambda}}\rangle$ term. It is easy to show that both terms have poles only at  addable and removable points. Formula \eqref{F-amplitude-explicit} provides exact cancellation of poles at removable points, which implies the statement.
	\end{itemize}
Finally, we provide the normalized formula \eqref{lambda-contour-integral} for the eigenvector $|\vec{\boldsymbol{\lambda}}\rangle$ which agrees with formulas \eqref{induction-assumption}
\begin{equation}\label{eigenvector}
 |\vec{\boldsymbol{\lambda}}\rangle =\lim\limits_{u_i\to c_k}\prod\limits_{i,k} \frac{u_i-x_k}{u_i-x_k-\epsilon_3} \prod\limits_{i<j}S(u_i-u_j)e(u_N)...e(u_1)|0\rangle 
\end{equation}
\section{ILW Integrals of Motion and Bethe ansatz}\label{IM}
Consider the monodromy matrix on $n$ sites $\mathbf{T}_q(u)$ defined by \eqref{Transfer-matrix-q-deformed}. One can easily see that $\mathbf{T}_q(u)$ admits the following large $u$ expansion
\begin{equation}
 \mathbf{T}_{q}(u)=\Lambda(u,q)\exp\left(\frac{1}{u}\mathbf{I}_{1}+\frac{1}{u^{2}}\mathbf{I}_{2}+\dots\right),
\end{equation} 
where $\Lambda(u,q)$ is a normalization factor and $\mathbf{I}_{1}$ and $\mathbf{I}_{2}$ are the first $\textrm{ILW}_n$ Integrals of Motion \eqref{ILW-local-integrals}. As explained in Introduction among other Integrals of Motion there is a particular one called KZ integral\footnote{This KZ operator shares the same notation with its $q\rightarrow1$ limit introduced in Introduction \eqref{O-operator}. We hope this will not lead to confusion.}
\begin{equation}\label{First-KZ-integral}
	T_1\overset{\text{def}}{=} \mathbf{T}_q(u_1).
\end{equation} 
Using the fact that $\mathcal{R}_{0,1}(0)=\mathcal{P}_{0,1}$ a permutation operator, one finds
\begin{equation}
 T_1=q^{L_0^{(1)}}\mathcal{R}_{1,2}(u_1-u_{2})\mathcal{R}_{1,3}(u_1-u_{3})\dots\mathcal{R}_{1,n}(u_1-u_{n}).
\end{equation}
	
As announced in Introduction the simultaneous spectrum of  $\mathbf{T}_{q}(u)$ is governed by Bethe ansatz equations \eqref{Bethe-ansatz-eigenvalues-introduction}. In this section we will prove these equations. 
\subsection{Off-shell Bethe vector}\label{off-shell-section}
The basic ingredient of algebraic Bethe ansatz is the construction of the so-called off-shell Bethe vector. For the case of $Y(\widehat{\mathfrak{gl}}(1))$ algebra they have been introduced in \cite{Aganagic:2017gsx} by the methods of geometric representation theory, here we use an equivalent but more direct approach. We take the tensor product of $n+N$ Fock spaces, with $n$ ``quantum'' and $N$ ``auxiliary'' spaces
\begin{equation}\label{total-Fock}
 \underbrace{\mathcal{F}_{u_{1}}\otimes\dots\otimes\mathcal{F}_{u_{n}}}_{\text{quantum space}}\otimes
	\underbrace{\mathcal{F}_{x_{1}}\otimes\dots\otimes\mathcal{F}_{x_{N}}}_{\text{auxiliary space}}
\end{equation}
generated from the vacuum state
\begin{equation}
	|\boldsymbol{\varnothing}\rangle_{\boldsymbol{x}}\otimes|\boldsymbol{\varnothing}\rangle_{\boldsymbol{u}}=|x_{1}\rangle\otimes\dots\otimes|x_{N}\rangle\otimes|u_{1}\rangle
	\otimes\dots\otimes|u_{n}\rangle.
\end{equation}
	
Consider the special state in the auxiliary space
\begin{equation}
	|\chi\rangle_{\scriptscriptstyle{\boldsymbol{x}}}\overset{\text{def}}{=}|\underbrace{\Box,\dots,\Box}_N\rangle\sim\oint\limits_{\mathcal{C}_{N}}dz_{N}\dots\oint\limits_{\mathcal{C}_{1}}dz_{1}\,
	e(z_{N})\dots e(z_{1})|\boldsymbol{\varnothing}\rangle_{\boldsymbol{x}},
\end{equation}
where the contour $\mathcal{C}_k$ encircles the point $x_k$ in counter-clockwise direction. The vector $|\chi\rangle_{\scriptscriptstyle{\boldsymbol{x}}}$ is an eigenvector of zero twist integrable system 
\begin{equation}\label{chi-eigenvalue}
  h(u)|\chi\rangle_{\scriptscriptstyle{\boldsymbol{x}}}=\prod_{k=1}^N\frac{u-x_k}{u-x_k-\epsilon_3}|\chi\rangle_{\scriptscriptstyle{\boldsymbol{x}}}.
\end{equation}
We note also convenient formula
\begin{equation}\label{f-action-on-chi}
 _{\boldsymbol{x}}\hspace*{-1pt}\langle\boldsymbol{\varnothing}|f(z)\dots f(z_1)|\chi\rangle_{\boldsymbol{x}}=\text{Sym}_{\boldsymbol{x}}\left(\prod_{a=1}^N\frac{1}{z_a-x_a}\prod_{a<b}S(x_a-x_b)\right),
\end{equation}  
which is an immediate consequence of \eqref{induction-assumption} and \eqref{F-amplitude-explicit}.
	
Now we define the off-shell Bethe vector as \cite{Aganagic:2017gsx}
\begin{equation}\label{off-shell-bethe-vector-definition}
	|B(\boldsymbol{x})\rangle_{\boldsymbol{u}}\overset{\text{def}}{=} _{\boldsymbol{x}}\hspace*{-4pt}\langle\boldsymbol{\varnothing}|\mathcal{R}(\boldsymbol{x},\boldsymbol{u})
	|\chi\rangle_{\boldsymbol{x}}\otimes|\boldsymbol{\varnothing}\rangle_{\boldsymbol{u}}\quad\text{where}\quad
	\mathcal{R}(\boldsymbol{x},\boldsymbol{u})=\mathcal{R}_{x_{1}u_{1}}\dots\mathcal{R}_{x_{N}u_{1}}\dots\mathcal{R}_{x_{1}u_{n}}\dots\mathcal{R}_{x_{N}u_{n}}.
\end{equation}
The off-shell Bethe vector $|B(\boldsymbol{x})\rangle$ can be represented by the following picture
\begin{equation}\label{off-shell-bethe-vector-definition-picture}
 \psfrag{p}{$|B(\boldsymbol{x})\rangle=$} 
	\psfrag{v}{$\scriptscriptstyle{\varnothing}$} 
	\psfrag{c}{\resizebox{1.3cm}{!}{$|\chi\rangle_{\scriptscriptstyle{\boldsymbol{x}}}$}}
	\psfrag{u1}{$\scriptstyle{u_1}$}
	\psfrag{u2}{$\scriptstyle{u_2}$}
	\psfrag{u3}{$\scriptstyle{u_3}$}
	\psfrag{u4}{$\scriptstyle{u_4}$}
	\psfrag{u5}{$\scriptstyle{u_{n-3}}$}
	\psfrag{u6}{$\scriptstyle{u_{n-2}}$}
	\psfrag{u7}{$\scriptstyle{u_{n-1}}$}
	\psfrag{u8}{$\scriptstyle{u_n}$}
	\psfrag{x1}{$\scriptstyle{x_N}$}
	\psfrag{x2}{$\scriptstyle{x_{N-1}}$}
	\psfrag{x3}{$\scriptstyle{x_{2}}$}
	\psfrag{x4}{$\scriptstyle{x_1}$}
	\includegraphics[scale=0.35]{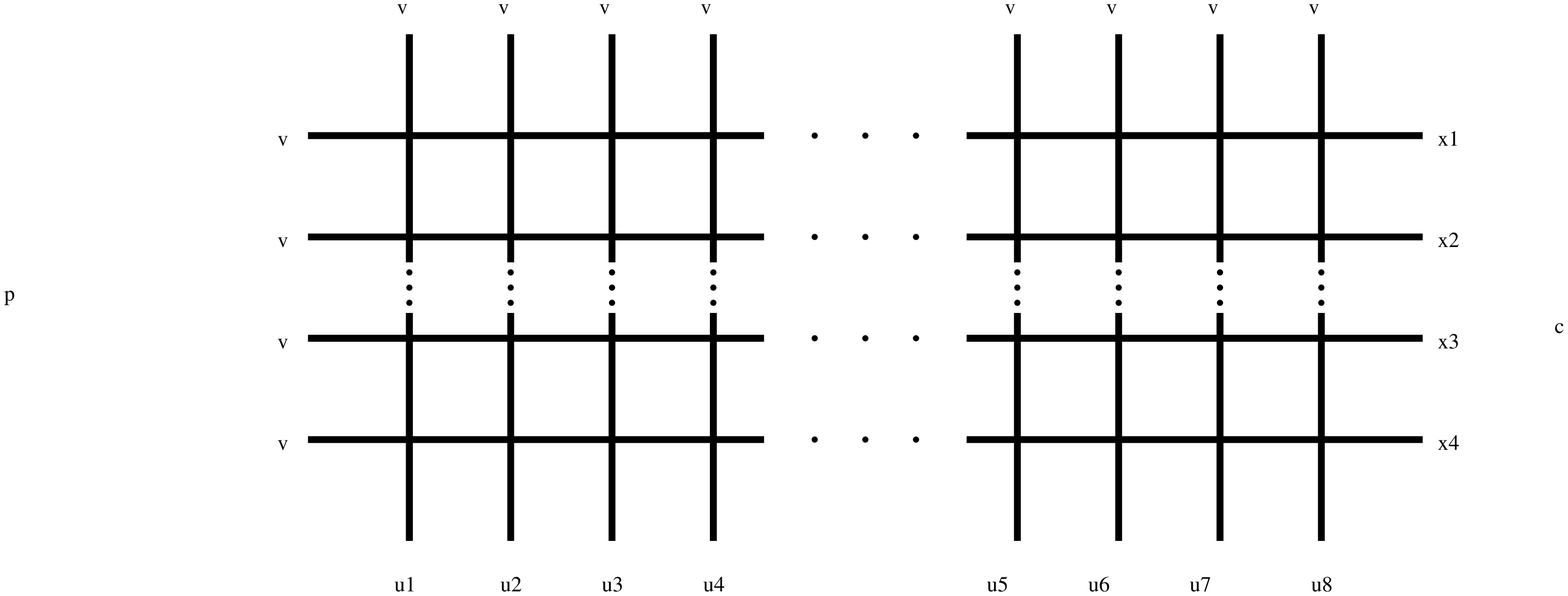}
\end{equation}
Consider the matrix element between $|B(\boldsymbol{x})\rangle_{\boldsymbol{u}}$ and generic state 
\begin{equation}\label{off-shell-projected}
	\omega_{\vec{\boldsymbol{\lambda}}}(\boldsymbol{x}|\boldsymbol{u})\overset{\text{def}}{=}\,_{\boldsymbol{u}}\hspace*{-1pt}\langle\boldsymbol{\varnothing}|a^{(1)}_{\boldsymbol{\lambda}^{(1)}}\dots a^{(n)}_{\boldsymbol{\lambda}^{(n)}}
	|B(\boldsymbol{x})\rangle_{\boldsymbol{u}}=_{\boldsymbol{x}}\hspace*{-4pt}\langle\boldsymbol{\varnothing}|
	\mathcal{L}_{\boldsymbol{\lambda}^{(1)},\varnothing}(u_{1})\dots\mathcal{L}_{\boldsymbol{\lambda}^{(n)},\varnothing}(u_{n})|\chi\rangle_{\boldsymbol{x}},
\end{equation}
which is non-zero only if 
\begin{equation}
	|\vec{\boldsymbol{\lambda}}|=\sum_{k=1}^n|\boldsymbol{\lambda}^{(k)}|=N.
\end{equation}
Following \cite{Tarasov:1993vs}, we call  $\omega_{\vec{\boldsymbol{\lambda}}}(\boldsymbol{x}|\boldsymbol{u})$   the weight function. It can be simplified by noting that the matrix element of Lax operator $\mathcal{L}_{\boldsymbol{\lambda},\varnothing}(u)$ can be expressed through $h(u)$ and $f(z)$ via contour integral\footnote{See Appendix \ref{VectorChi} for details.}
\begin{equation}\label{matrix-element-Lax-integral-representaion}
	\mathcal{L}_{\boldsymbol{\lambda},\varnothing}(u)=\frac{1}{(2\pi i)^{|\boldsymbol{\lambda}|}}\oint_{\mathcal{C}_1}\dots\oint_\mathcal{C_{|\boldsymbol{\lambda}|}} F_{\boldsymbol{\lambda}}(\boldsymbol{z}|u)\,h(u)f(z_{|\boldsymbol{\lambda}|})\dots f(z_1)dz_1\dots dz_{|\boldsymbol{\lambda}|},
\end{equation} 
where each contour $\mathcal{C}_k$ goes clockwise around $\infty$ and $u-\epsilon_{3}$. Using \eqref{matrix-element-Lax-integral-representaion} the weight function \eqref{off-shell-projected} can be rewritten as
\begin{multline}\label{off-shell-projected-rewritten}
\omega_{\vec{\boldsymbol{\lambda}}}(\boldsymbol{x}|\boldsymbol{u})=\frac{1}{(2\pi i)^{N}}\times\\\times\oint
F_{\vec{\boldsymbol{\lambda}}}(\vec{\boldsymbol{z}}|\boldsymbol{u})\,\,
_{\boldsymbol{x}}\hspace*{-1pt}\langle\boldsymbol{\varnothing}|
h(u_1)\underbrace{f(z_1^{(1)})f(z_2^{(1)})\dots }_{|\boldsymbol{\lambda}^{(1)}|}\,h(u_2)\underbrace{f(z_1^{(2)})f(z_2^{(2)})\dots}_{|\boldsymbol{\lambda}^{(2)}|}\,\,\dots\,\, h(u_n)\underbrace{f(z_1^{(n)})f(z_2^{(n)})\dots}_{|\boldsymbol{\lambda}^{(n)}|}|\chi\rangle_{\boldsymbol{x}}\,d\vec{\boldsymbol{z}},
\end{multline}
where
\begin{equation}\label{F-lambda-vec}
F_{\vec{\boldsymbol{\lambda}}}(\vec{\boldsymbol{z}}|\boldsymbol{u})=\prod_{k=1}^{n}F_{\boldsymbol{\lambda}^{(k)}}\left(z_1^{(k)},\dots,z^{(k)}_{|\boldsymbol{\lambda}^{(k)}|}\bigr|u_k\right).
\end{equation}
Then the matrix element in \eqref{off-shell-projected-rewritten} can be explicitly computed using \eqref{hf-relation} and \eqref{f-action-on-chi}. One obtains
\begin{equation}\label{off-shell-projected-rewritten-2}
\omega_{\vec{\boldsymbol{\lambda}}}(\boldsymbol{x}|\boldsymbol{u})=\frac{1}{(2\pi i)^{N}}\oint\dots\oint \Omega_{\vec{\boldsymbol{\lambda}}}(\vec{\boldsymbol{z}}|\boldsymbol{u})\,\,\text{Sym}_{\boldsymbol{x}}\left(\prod_{a=1}^N\frac{1}{z_a-x_a}\prod_{a<b}S(x_a-x_b)\right) d\vec{\boldsymbol{z}},
\end{equation}
where $(z_1,\dots,z_N)=(z_1^{(1)},\dots z_{\boldsymbol{\lambda}^{(1)}}^{(1)},z_1^{(2)},\dots z_{\boldsymbol{\lambda}^{(2)}}^{(2)},\dots,z_1^{(n)},\dots z_{\boldsymbol{\lambda}^{(n)}}^{(n)})$ and the  function
\begin{multline}\label{Integrand-function-transformed}
\Omega_{\vec{\boldsymbol{\lambda}}}(\vec{\boldsymbol{z}}|\boldsymbol{u})=F_{\vec{\boldsymbol{\lambda}}}(\vec{\boldsymbol{z}}|\boldsymbol{u})\left(\prod_{j=1}^{|\boldsymbol{\lambda}^{(1)}|}\frac{u_2-z_j^{(1)}}{u_2-z_j^{(1)}-\epsilon_3}\right)\left(\prod_{j=1}^{|\boldsymbol{\lambda}^{(2)}|}\frac{u_3-z_j^{(2)}}{u_3-z_j^{(2)}-\epsilon_3}\prod_{j=1}^{|\boldsymbol{\lambda}^{(1)}|}\frac{u_3-z_j^{(1)}}{u_3-z_j^{(1)}-\epsilon_3}\right)\dots\\\dots
\left(\prod_{j=1}^{|\boldsymbol{\lambda}^{(n-1)}|}\frac{u_n-z_j^{(n-1)}}{u_n-z_j^{(n-1)}-\epsilon_3}\prod_{j=1}^{|\boldsymbol{\lambda}^{(n-2)}|}\frac{u_n-z_j^{(n-2)}}{u_n-z_j^{(n-2)}-\epsilon_3}\dots\prod_{j=1}^{|\boldsymbol{\lambda}^{(1)}|}\frac{u_n-z_j^{(1)}}{u_n-z_j^{(1)}-\epsilon_3}\right)
\end{multline}
has been obtained from $F_{\vec{\boldsymbol{\lambda}}}(\vec{\boldsymbol{z}}|\boldsymbol{u})$ as a result of application of \eqref{hf-relation}. We note that as explained in appendix \ref{VectorChi} the local terms do not appear in \eqref{off-shell-projected-rewritten-2} if one extends the integration contour to include all new singularities \eqref{Integrand-function-transformed}. It implies that the integral shrinks to the points $\boldsymbol{x}$ and one obtains
\begin{equation}\label{off-shell-projected-computed}
\omega_{\vec{\boldsymbol{\lambda}}}(\boldsymbol{x}|\boldsymbol{u})=\text{Sym}_{\boldsymbol{x}}\left(\Omega_{\vec{\boldsymbol{\lambda}}}(\vec{\boldsymbol{x}}|\boldsymbol{u})\prod_{a<b}S(x_a-x_b)\right). 
\end{equation}
	
Let us note finally that this last equation implies the well known co-product property of weight function:
\begin{equation}\label{coproduct}
\omega^{V_1,V_2}_{\vec{\lambda}^{(1)},\vec{\lambda}^{(2)}}= \sum\limits_{I=I_1+I_2} \omega^{V_1}_{\vec{\lambda}_1}(x^{(1)})\omega^{V_2}_{\vec{\lambda}_2}(x^{(2)})\prod\limits_{x_i\in I_2} V_1(x_i) \prod_{i \in I_1, j \in I_2} S(x_{i}-x_j),
\end{equation}
Where the sum is over partitions of set of indices $I={1,2,...N}$ into two sets $(I_1,I_2)$ of lengths $(N_1,N_2)$, we also denote all $x_i$ variables from set $I_a$ $x^{(a)}$.
\subsection{Diagonalization of KZ Integral}
The action of the KZ Integral of Motion  on off-shell Bethe vector $|B(\boldsymbol{x})\rangle_{\boldsymbol{u}}$ is very simple and can be explained by the following  picture
\begin{equation}\label{off-shell-bethe-vector-action-on-KZ}
 \psfrag{v}{$\scriptscriptstyle{\varnothing}$} 
 \psfrag{c}{\resizebox{1.3cm}{!}{$|\chi\rangle_{\scriptscriptstyle{\boldsymbol{x}}}$}}
 \psfrag{u1}{$\scriptscriptstyle{u_1}$}
 \psfrag{u2}{$\scriptscriptstyle{u_2}$}
 \psfrag{u3}{$\scriptscriptstyle{u_3}$}
 \psfrag{u4}{$\scriptscriptstyle{u_4}$}
 \psfrag{u5}{$\scriptscriptstyle{u_{n-3}}$}
 \psfrag{u6}{$\scriptscriptstyle{u_{n-2}}$}
 \psfrag{u7}{$\scriptscriptstyle{u_{n-1}}$}
 \psfrag{u8}{$\scriptscriptstyle{u_n}$}
 \psfrag{x1}{$\scriptscriptstyle{x_{N}}$}
 \psfrag{x2}{$\scriptscriptstyle{x_{N-1}}$}
 \psfrag{x3}{$\scriptscriptstyle{x_{2}}$}
 \psfrag{x4}{$\scriptscriptstyle{x_1}$}
 \psfrag{q}{$q^{L_0}$}
 \psfrag{e}{$=$}
 \hspace*{-0.5cm}
 \includegraphics[scale=0.27]{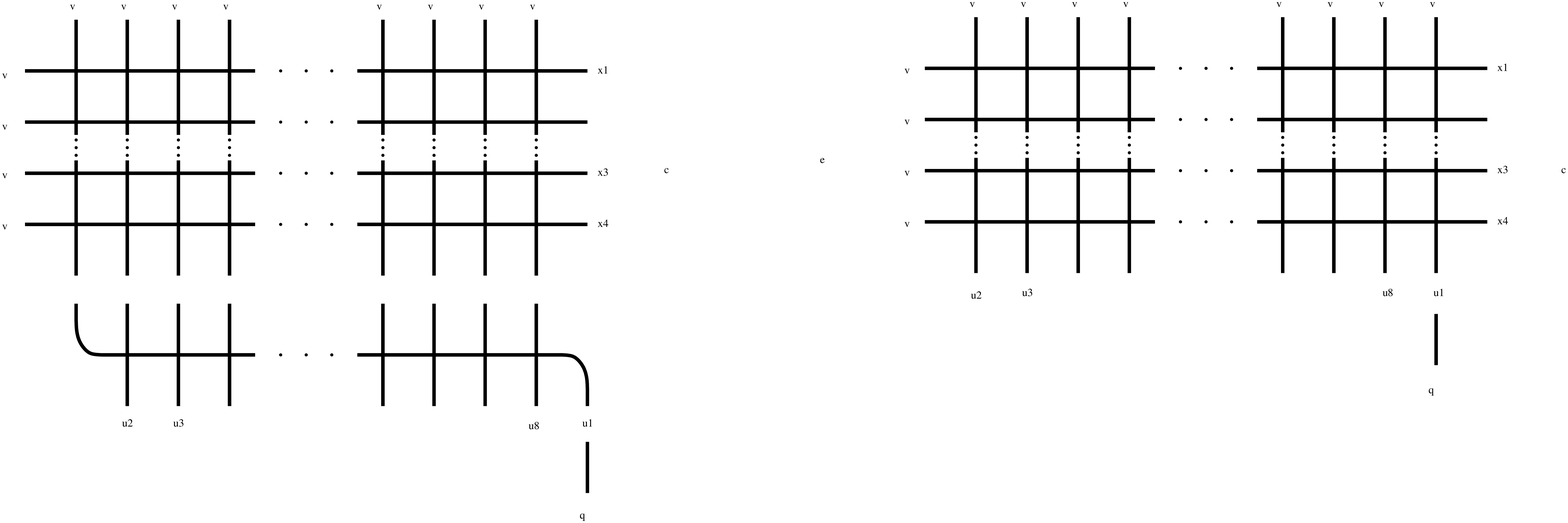}
\end{equation}
Projecting this equation on arbitrary state, one obtains
\begin{equation}\label{off-shell-bethe-vector-action-on-KZ-projected}
_{\boldsymbol{u}}\hspace*{-2pt}\langle\boldsymbol{\varnothing}|a^{(1)}_{\boldsymbol{\lambda}^{(1)}}\dots a^{(n)}_{\boldsymbol{\lambda}^{(n)}}|T_1
|B(\boldsymbol{x})\rangle_{\boldsymbol{u}}=q^{|\boldsymbol{\lambda}^{(1)}|}\hspace*{1pt}_{\boldsymbol{x}}\langle\boldsymbol{\varnothing}|
\mathcal{L}_{\boldsymbol{\lambda}^{(2)},\varnothing}(u_{2})\dots\mathcal{L}_{\boldsymbol{\lambda}^{(n)},\varnothing}(u_{n})\mathcal{L}_{\boldsymbol{\lambda}^{(1)},\varnothing}(u_{1})|\chi\rangle_{\boldsymbol{x}}
\end{equation}
If we require that $|B(\boldsymbol{x})\rangle_{\boldsymbol{u}}$ is an eigenstate for $T_1$ we have to demand
\begin{equation}\label{eigen-equation-on-off-shell-bethe-vector}
q^{|\boldsymbol{\lambda}^{(1)}|}\hspace*{1pt}_{\boldsymbol{x}}\langle\boldsymbol{\varnothing}|
\mathcal{L}_{\boldsymbol{\lambda}^{(2)},\varnothing}(u_{2})\dots\mathcal{L}_{\boldsymbol{\lambda}^{(n)},\varnothing}(u_{n})\mathcal{L}_{\boldsymbol{\lambda}^{(1)},\varnothing}(u_{1})|\chi\rangle_{\boldsymbol{x}}=T_1(\boldsymbol{u})\hspace*{1pt}_{\boldsymbol{x}}\langle\boldsymbol{\varnothing}|
\mathcal{L}_{\boldsymbol{\lambda}^{(1)},\varnothing}(u_{1})\dots\mathcal{L}_{\boldsymbol{\lambda}^{(n)},\varnothing}(u_{n})|\chi\rangle_{\boldsymbol{x}},
\end{equation}
which should hold for any set of partitions $\vec{\boldsymbol{\lambda}}$. The eigenvalue $T_1(\boldsymbol{u})$ can be found from \eqref{eigen-equation-on-off-shell-bethe-vector} by taking  $\boldsymbol{\lambda}^{(1)}=\varnothing$
\begin{equation}
 T_1(\boldsymbol{u})=\prod_{k=1}^N\frac{x_k-u_1}{x_k-u_1+\epsilon_3}.
\end{equation}
For generic $\vec{\boldsymbol{\lambda}}$ the eigenstate equation \eqref{eigen-equation-on-off-shell-bethe-vector}  implies the integral identity
\begin{multline}\label{eigen-equation-on-off-shell-bethe-vector-integral}
q^{|\boldsymbol{\lambda^{(1)}}|}\oint
F_{\vec{\boldsymbol{\lambda}}}(\vec{\boldsymbol{z}}|\boldsymbol{u})\,\,_{\boldsymbol{x}}\hspace*{-1pt}\langle\boldsymbol{\varnothing}|
h(u_2)\underbrace{f(z_1^{(2)})\dots}_{|\boldsymbol{\lambda}^{(2)}|}\,\,\dots\,\, h(u_n)\underbrace{f(z_1^{(n)})\dots}_{|\boldsymbol{\lambda}^{(n)}|}h(u_1)\underbrace{f(z_1^{(1)})\dots }_{|\boldsymbol{\lambda}^{(1)}|}|\chi\rangle_{\boldsymbol{x}}\,d\vec{\boldsymbol{z}}=\\=T_1(\boldsymbol{u})\oint
F_{\vec{\boldsymbol{\lambda}}}(\vec{\boldsymbol{z}}|\boldsymbol{u})\,\,_{\boldsymbol{x}}\hspace*{-1pt}\langle\boldsymbol{\varnothing}|\,h(u_1)\underbrace{f(z_1^{(1)})\dots}_{|\boldsymbol{\lambda}^{(1)}|}\,
h(u_2)\underbrace{f(z_1^{(2)})\dots}_{|\boldsymbol{\lambda}^{(2)}|}\,\,\dots\,\, h(u_n)\underbrace{f(z_1^{(n)})\dots}_{|\boldsymbol{\lambda}^{(n)}|}|\chi\rangle_{\boldsymbol{x}}\,d\vec{\boldsymbol{z}},
\end{multline}
which holds provided that $\boldsymbol{x}$ obeys Bethe ansatz equations
\begin{equation}\label{Bethe-ansatz-equations}
	q\prod_{j\neq i}\prod_{\alpha=1}^3\frac{x_i-x_j-\epsilon_{\alpha}}{x_i-x_j+\epsilon_{\alpha}}\prod_{k=1}^n\frac{x_i-u_k+\epsilon_3}{x_i-u_k}=1\quad\text{for all}\quad i=1,\dots,N.
\end{equation}
Indeed using \eqref{hf-relation} and \eqref{ff-exact-relation} one can drag all $f(z_j^{(1)})$'s to the left. As the integrals over $z_i^{(j)}$ go around simple poles located at the points $x_1,\dots,x_N$ (see \eqref{off-shell-projected-rewritten-2}), the local terms in  \eqref{hf-relation} and \eqref{ff-exact-relation} do not contribute and we will have 
\begin{multline}\label{eigen-equation-on-off-shell-bethe-vector-integral-2}
 q^{|\boldsymbol{\lambda^{(1)}}|}\oint
	F_{\vec{\boldsymbol{\lambda}}}(\vec{\boldsymbol{z}}|\boldsymbol{u})\,\,
	_{\boldsymbol{x}}\hspace*{-1pt}\langle\boldsymbol{\varnothing}|
	h(u_2)\underbrace{f(z_1^{(2)})\dots}_{|\boldsymbol{\lambda}^{(2)}|}\,\,\dots\,\, h(u_n)\underbrace{f(z_1^{(n)})\dots}_{|\boldsymbol{\lambda}^{(n)}|}h(u_1)\underbrace{f(z_1^{(1)})\dots}_{|\boldsymbol{\lambda}^{(1)}|}|\chi\rangle_{\boldsymbol{x}}\,d\vec{\boldsymbol{z}}=\\=\oint
	F_{\vec{\boldsymbol{\lambda}}}(\vec{\boldsymbol{z}}|\boldsymbol{u})\prod_{k=1}^{|\boldsymbol{\lambda}^{(1)}|}\mathcal{D}(z_k^{(1)}|\boldsymbol{z})\,\,
	_{\boldsymbol{x}}\hspace*{-1pt}\langle\boldsymbol{\varnothing}|\underbrace{f(z_1^{(1)})\dots}_{|\boldsymbol{\lambda}^{(1)}|}\,
	h(u_2)\underbrace{f(z_1^{(2)})\dots}_{|\boldsymbol{\lambda}^{(2)}|}\,\,\dots\,\, h(u_n)\underbrace{f(z_1^{(n)})\dots}_{|\boldsymbol{\lambda}^{(n)}|}\,h(u_1)|\chi\rangle_{\boldsymbol{x}}\,d\vec{\boldsymbol{z}}=\\=
	T_1(\boldsymbol{u})\oint
	F_{\vec{\boldsymbol{\lambda}}}(\vec{\boldsymbol{z}}|\boldsymbol{u})\prod_{k=1}^{|\boldsymbol{\lambda}^{(1)}|}\mathcal{D}(z_k^{(1)}|\boldsymbol{z})\,\,
	_{\boldsymbol{x}}\hspace*{-1pt}\langle\boldsymbol{\varnothing}|\underbrace{f(z_1^{(1)})\dots}_{|\boldsymbol{\lambda}^{(1)}|}\,
	h(u_2)\underbrace{f(z_1^{(2)})\dots}_{|\boldsymbol{\lambda}^{(2)}|}\,\,\dots\,\, h(u_n)\underbrace{f(z_1^{(n)})\dots}_{|\boldsymbol{\lambda}^{(n)}|}|\chi\rangle_{\boldsymbol{x}}\,d\vec{\boldsymbol{z}},
\end{multline}
where $\boldsymbol{z}$ denotes the set of all $z_i^{(j)}$ and
\begin{equation}\label{deffect-factor}
 \mathcal{D}(z|\boldsymbol{z})=q\prod_{z_j\neq z}\prod_{\alpha=1}^3\frac{z-z_j-\epsilon_{\alpha}}{z-z_j+\epsilon_{\alpha}}\prod_{k=1}^n\frac{z-u_k+\epsilon_3}{z-u_k}.
\end{equation}
One can easily show that under Bethe ansatz equations \eqref{Bethe-ansatz-equations} each additional factor $\mathcal{D}(z_k^{(1)}|\boldsymbol{z})$ in \eqref{eigen-equation-on-off-shell-bethe-vector-integral-2} equals to $1$, which implies the statement.
\subsection{KZ equation} \label{Section KZ}
Off-shell Bethe vectors  \eqref{off-shell-bethe-vector-definition}, are closely related to solutions of difference KZ equation. Namely, let us introduce auxiliary functions
\begin{equation}
	V^{(\hbar)}(x)=\frac{\Gamma(\frac{x}{h})}{\Gamma(\frac{x-\epsilon_3}{\hbar})},\qquad
	\Phi^{(\hbar)}(x)=\frac{\Gamma(\frac{x-\epsilon_{1}}{\hbar})\Gamma(\frac{x-\epsilon_{2}}{\hbar})}{\Gamma(\frac{x}{\hbar})\Gamma(\frac{x+\epsilon_{3}}{\hbar})},
\end{equation}
where $S(x)$ is given by \eqref{S-function-def}. The main property of these functions which will be used is the shift relation
\begin{equation}
	V^{(\hbar)}(x+\hbar)=V^{(\hbar)}(x)\frac{x}{x-\epsilon_3},\quad \Phi^{(\hbar)}(x+\hbar)=S(-x)\Phi^{(\hbar)}(x)
\end{equation}
Then the wave function
\begin{equation}
 |\psi(\boldsymbol{u})\rangle\overset{\text{def}}{=}\oint q^{\sum\frac{x_a}{\hbar}} \prod_{a=1}^N\prod_{j=1}^{n}V^{(\hbar)}(u_j-x_a)\prod\limits_{a\neq b}\Phi^{(\hbar)}(x_a-x_b)|B(\boldsymbol{x})\rangle_{\boldsymbol{u}}\,\frac{d^N\boldsymbol{x}}{(2\pi i)^N}, \label{WaveFunction}
\end{equation}
is a solution of difference KZ equation:
\begin{equation}\label{eqKZ}
 |\psi(u_1+\hbar,u_2,...,u_n)\rangle=T_1|\psi(u_1,...,u_n)\rangle,
\end{equation}
where $T_1$ is the first KZ operator \eqref{First-KZ-integral}.
	
The proof of \eqref{eqKZ} is simple. Let us pick a tuple of Young diagrams $\vec{\boldsymbol{\lambda}}=\{\boldsymbol{\lambda}^{(1)},...,\boldsymbol{\lambda}^{(n)}\}$ and consider the projection of the wave function $|\psi(\boldsymbol{u})\rangle$
\begin{equation}\label{KZ-equation-lhs}
 _{\boldsymbol{u}}\hspace*{-2pt}\langle\boldsymbol{\varnothing}|a^{(1)}_{\boldsymbol{\lambda}^{(1)}}\dots a^{(n)}_{\boldsymbol{\lambda}^{(n)}}|\psi(\boldsymbol{u})\rangle=\oint q^{\sum_a\frac{x_a}{\hbar}} \prod_{a=1}^N\prod_{j=1}^{n}V^{(\hbar)}(u_j-x_a)\prod\limits_{a\neq b}\Phi^{(\hbar)}(x_a-x_b)\omega_{\vec{\boldsymbol{\lambda}}}(\boldsymbol{x}|\boldsymbol{u})\,\frac{d^N\boldsymbol{x}}{(2\pi i)^N},
\end{equation}
where the weight function $\omega_{\vec{\boldsymbol{\lambda}}}(\boldsymbol{x}|\boldsymbol{u})$ is given by \eqref{off-shell-projected-computed}. At the same time using \eqref{off-shell-bethe-vector-action-on-KZ-projected} one finds
\begin{equation}\label{KZ-equation-rhs}
 _{\boldsymbol{u}}\hspace*{-2pt}\langle\boldsymbol{\varnothing}|a^{(1)}_{\boldsymbol{\lambda}^{(1)}}\dots a^{(n)}_{\boldsymbol{\lambda}^{(n)}}|T_1|\psi(\boldsymbol{u})\rangle=\oint q^{\sum_a\frac{x_a}{\hbar}} \prod_{a=1}^N\prod_{j=1}^{n}V^{(\hbar)}(u_j-x_a)\prod\limits_{a\neq b}\Phi^{(\hbar)}(x_a-x_b)q^{|\boldsymbol{\lambda}^{|1|}|}\omega_{\vec{\boldsymbol{\lambda'}}}(\boldsymbol{x}|\boldsymbol{u'})\,\frac{d^N\boldsymbol{x}}{(2\pi i)^N},
\end{equation}
where $\boldsymbol{u'}=(u_2,\dots,u_n,u_1)$ and $\vec{\boldsymbol{\lambda'}}=\{\boldsymbol{\lambda}^{(2)},...,\boldsymbol{\lambda}^{(n)},\boldsymbol{\lambda}^{(1)}\}$. As the integration kernel in \eqref{KZ-equation-lhs} and \eqref{KZ-equation-rhs} is $\boldsymbol{x}$ symmetric, one can replace (see \eqref{Integrand-function-transformed})
\begin{equation}
 \omega_{\vec{\boldsymbol{\lambda}}}(\boldsymbol{x}|\boldsymbol{u})\rightarrow F_{\vec{\boldsymbol{\lambda}}}^{\boldsymbol{u}}(\boldsymbol{x})\prod_{a<b}S(x_a-x_b),\qquad \omega_{\vec{\boldsymbol{\lambda}'}}(\boldsymbol{x}|\boldsymbol{u'})\rightarrow F_{\vec{\boldsymbol{\lambda}'}}^{\boldsymbol{u}'}(\boldsymbol{x})\prod_{a<b}S(x_a-x_b).
\end{equation}
Then it is immediately to see that the shift 
\begin{equation}
 u_1\rightarrow u_1+\hbar,\qquad \left(x_1,\dots,x_{|\boldsymbol{\lambda}^{(1)}|}\right)\rightarrow\left(x_1+\hbar,\dots,x_{|\boldsymbol{\lambda}^{(1)}|}+\hbar\right)
\end{equation}
reduces \eqref{KZ-equation-lhs} to \eqref{KZ-equation-rhs} after relabeling of integration variables. The statement \eqref{eqKZ}  follows.
	
Of course these considerations are correct modulo choice of integration contour. Integrals of the form  \eqref{WaveFunction} have been discussed in details in the literature \cite{Aganagic:2013tta,Zenkevich:2017tnb}. Following these approaches, we treat the integral \eqref{WaveFunction} as a sum over residues, the poles contributing to the integral are in one to one correspondence with a collection of $n$ 3D partitions, with fixed floor shape $\boldsymbol{\lambda}$:
\begin{equation}
 x_{\boldsymbol{I}}=v_{k}-(i_k-1)\epsilon_1-(j_k-1)\epsilon_2+\hbar n^{(k)}_{i,j}
\end{equation}
Here we treat the 3D partition as 2D Young diagram filled with integer numbers $n^{(k)}_{i,j}$ such that:
\begin{equation}
 n^{(k)}_{i,j}\ge n^{(k)}_{i+1,j} \quad  n^{(k)}_{i,j}\ge n^{(k)}_{i,j+1}
\end{equation}
\subsection{Diagonalization of \texorpdfstring{$\mathbf{I}_{2}$}{I2} Integral} \label{LocalDiag}
The diagonalization problem of KZ integral given above does not work for $n=1$. Specially for this case and also for academic purposes we consider diagonalization problem for $\mathbf{I}_{2}(q)$ IM \eqref{ILW-local-integrals}. We have to remember that we have changed normalization in \eqref{epsilonNotation}-\eqref{uepsilonNotation}. It is also convenient to subtract the vacuum eigenvalue and $\frac{\epsilon_3}{2}\mathbf{I}_1(q)$ from $\mathbf{I}_{2}(q)$. Altogether, one has       
\begin{equation}
  \tilde{\mathbf{I}}_2(q)=-\epsilon_3\int\left[ \frac{1}{3}\sqrt{\frac{\sigma_3}{\epsilon_3}}\sum\limits_i(\partial \phi_i)^3-\epsilon_3\left(\frac{1}{2}\sum\limits_{i,j}\partial\phi_iD(q)\partial\phi_j+\sum\limits_{i<j}\partial\phi_i\partial^2\phi_j\right) \right] \frac{dx}{2\pi}-\frac{\epsilon_3\mathbf{I}_1(q)}{2}- \frac{\epsilon_3}{3}\sqrt{\frac{\sigma_3}{\epsilon_3}}\sum\limits_iu_i^3
\end{equation}
where the last two terms are added for a convenience, in order to have more simple formula for the eigenvalues \eqref{Eigenvalues}. 

Since our total Fock space splits into quantum and auxiliary parts \eqref{total-Fock}, it will be convenient to add an upper index to $\tilde{\mathbf{I}}_2(q)$, either $\boldsymbol{x}$, $\boldsymbol{u}$ or $(\boldsymbol{x},\boldsymbol{u})$ referring to auxiliary, quantum or total spaces respectively. The key observation is that the Integral of Motion $\tilde{\mathbf{I}}_2^{(\boldsymbol{x},\boldsymbol{u})}(q)$ is almost the sum of terms acting separately on spaces $\mathcal{F}_{\boldsymbol{u}}$ and $\mathcal{F}_{\boldsymbol{x}}$ plus a cross term
\begin{eqnarray}\label{J-complition}
\tilde{\mathbf{I}}_2^{(\boldsymbol{x},\boldsymbol{u})}(q)=\tilde{\mathbf{I}}_2^{\boldsymbol{x}}(q)+\tilde{\mathbf{I}}_2^{(\boldsymbol{u}}(q)-\frac{1}{2\pi}\int J^{\boldsymbol{u}}(\xi)(D(q)+\partial)J^{\boldsymbol{x}}(\xi)d\xi,
\end{eqnarray}
where $J^{\boldsymbol{x}/\boldsymbol{v}}(\xi)$ is the $U(1)$ mode \eqref{Jn-}, in our particular representation $J^{\boldsymbol{x}/\boldsymbol{v}}(\xi)=\sqrt{\epsilon_3}\sum\limits_{i\in \mathcal{F}_{\boldsymbol{x}/\boldsymbol{u}}} \partial \phi_i(\xi)$.

We will show that on-shell Bethe vector, is an eigenvector of $\tilde{\mathbf{I}}_2^{u}(q)$
\begin{equation}
 \tilde{\mathbf{I}}_2^{\boldsymbol{u}}(q)\,_{\boldsymbol{x}}\hspace*{-1pt}\langle\boldsymbol{\varnothing}|\mathcal{R}(\boldsymbol{x},\boldsymbol{u})
 |\chi\rangle_{\boldsymbol{x}}\otimes|\boldsymbol{\varnothing}\rangle_{\boldsymbol{u}}=\left(\sum_{k=1}^Nx_k\right)\,_{\boldsymbol{x}}\hspace*{-1pt}\langle\boldsymbol{\varnothing}|\mathcal{R}(\boldsymbol{x},\boldsymbol{u})
 |\chi\rangle_{\boldsymbol{x}}\otimes|\boldsymbol{\varnothing}\rangle_{\boldsymbol{u}}
\end{equation}
where $x_k$ obeys Bethe ansatz equations \eqref{Bethe-ansatz-equations}. We start with off-shell Bethe vector \eqref{off-shell-bethe-vector-definition} and insert Integral of Motion for a system with zero twist $q=0$ acting on $\boldsymbol{x}$ space 
\begin{equation}
|B(\boldsymbol{x})\rangle_{\boldsymbol{u}}=_{\boldsymbol{x}}\hspace*{-4pt}\langle\boldsymbol{\varnothing}|\mathcal{R}(\boldsymbol{x},\boldsymbol{u})
|\chi\rangle_{\boldsymbol{x}}\otimes|\boldsymbol{\varnothing}\rangle_{\boldsymbol{u}}\longrightarrow
_{\boldsymbol{x}}\hspace*{-1pt}\langle\boldsymbol{\varnothing}|\mathcal{R}(\boldsymbol{x},\boldsymbol{u})
\tilde{\boldsymbol{I}}_2^{\boldsymbol{x}}(0)|\chi\rangle_{\boldsymbol{x}}\otimes|\boldsymbol{\varnothing}\rangle_{\boldsymbol{u}}
\end{equation}
We have the following chain of arguments
\begin{enumerate}
 \item Since $|\chi\rangle_{\boldsymbol{x}}$ is an eigenvector of zero twist integrable system \eqref{chi-eigenvalue}, it is also an eigenvector for $\tilde{\boldsymbol{I}}_2^{\boldsymbol{x}}(0)$ with eigenvalue $\sum_kx_k$. It implies
  \begin{equation}\label{I2-BA-1}
	 _{\boldsymbol{x}}\hspace*{-1pt}\langle\boldsymbol{\varnothing}|\mathcal{R}(\boldsymbol{x},\boldsymbol{u})
	 \tilde{\boldsymbol{I}}_2^{\boldsymbol{x}}(0)|\chi\rangle_{\boldsymbol{x}}\otimes|\boldsymbol{\varnothing}\rangle_{\boldsymbol{u}}=\sum_{k=1}^Nx_k|B(\boldsymbol{x})\rangle_{\boldsymbol{u}}
  \end{equation}
\item On the other hand, the Integral $\tilde{\boldsymbol{I}}_2^{\boldsymbol{x}}(0)$ can be completed by \eqref{J-complition} to   $\tilde{\boldsymbol{I}}_2^{(\boldsymbol{x},\boldsymbol{u})}(0)$ which acts on the whole $(\boldsymbol{x},\boldsymbol{u})$ space
\begin{equation}
 \tilde{\boldsymbol{I}}_2^{(\boldsymbol{x},\boldsymbol{u})}(0)=\tilde{\boldsymbol{I}}_2^{\boldsymbol{x}}(0)+\tilde{\boldsymbol{I}}_2^{\boldsymbol{u}}(0)-i\sum\limits_{k\in\mathbb{Z}}(|k|+k)J_{k}^{\boldsymbol{u}}J_{-k}^{\boldsymbol{x}},
\end{equation}
because the last two terms vanish on $|\boldsymbol{\varnothing}\rangle_{\boldsymbol{u}}$. It implies
\begin{equation}\label{I2-BA-2}
_{\boldsymbol{x}}\hspace*{-1pt}\langle\boldsymbol{\varnothing}|\mathcal{R}(\boldsymbol{x},\boldsymbol{u})
\tilde{\boldsymbol{I}}_2^{\boldsymbol{x}}(0)|\chi\rangle_{\boldsymbol{x}}\otimes|\boldsymbol{\varnothing}\rangle_{\boldsymbol{u}}=_{\boldsymbol{x}}\hspace*{-4pt}\langle\boldsymbol{\varnothing}|\mathcal{R}(\boldsymbol{x},\boldsymbol{u})
\tilde{\boldsymbol{I}}_2^{\boldsymbol{x},\boldsymbol{u}}(0)|\chi\rangle_{\boldsymbol{x}}\otimes|\boldsymbol{\varnothing}\rangle_{\boldsymbol{u}}
\end{equation}
\item From the definition of $\mathcal{R}(\boldsymbol{x},\boldsymbol{u})$ (see \eqref{off-shell-bethe-vector-definition}) we have 
\begin{equation}
 \mathcal{R}(\boldsymbol{x},\boldsymbol{u})\tilde{\boldsymbol{I}}_2^{(\boldsymbol{x},\boldsymbol{u})}(0)=\tilde{\boldsymbol{I}}_2^{(\boldsymbol{u},\boldsymbol{x})}(0)\mathcal{R}(\boldsymbol{x},\boldsymbol{u})\quad\text{where}\quad \boldsymbol{I}_2^{(\boldsymbol{u},\boldsymbol{x})}(0)=\tilde{\boldsymbol{I}}_2^{\boldsymbol{x}}(0)+\tilde{\boldsymbol{I}}_2^{\boldsymbol{u}}(0)-i\sum\limits_{k\in\mathbb{Z}}(|k|+k)J_{k}^{\boldsymbol{x}}J_{-k}^{\boldsymbol{u}},
\end{equation}
and hence
\begin{equation}\label{I2-BA-3}
_{\boldsymbol{x}}\hspace*{-1pt}\langle\boldsymbol{\varnothing}|\mathcal{R}(\boldsymbol{x},\boldsymbol{u})
\tilde{\boldsymbol{I}}_2^{\boldsymbol{x},\boldsymbol{u}}(0)|\chi\rangle_{\boldsymbol{x}}\otimes|\boldsymbol{\varnothing}\rangle_{\boldsymbol{u}}=_{\boldsymbol{x}}\hspace*{-4pt}\langle\boldsymbol{\varnothing}|\left(\tilde{\boldsymbol{I}}_2^{\boldsymbol{u}}(0)-2i\sum_{k>0}kJ_{-k}^{\boldsymbol{u}}J_{k}^{\boldsymbol{x}}\right)\mathcal{R}(\boldsymbol{x},\boldsymbol{u})
|\chi\rangle_{\boldsymbol{x}}\otimes|\boldsymbol{\varnothing}\rangle_{\boldsymbol{u}}
\end{equation} 
\item One has a remarkable property
\begin{equation}\label{I2-BA-4}
_{\boldsymbol{x}}\hspace*{-1pt}\langle\boldsymbol{\varnothing}|J_{k}^{\boldsymbol{x}}\,\mathcal{R}(\boldsymbol{x},\boldsymbol{u})
|\chi\rangle_{\boldsymbol{x}}\otimes|\boldsymbol{\varnothing}\rangle_{\boldsymbol{u}}=q^k\,_{\boldsymbol{x}}\hspace*{-1pt}\langle\boldsymbol{\varnothing}|\mathcal{R}(\boldsymbol{x},\boldsymbol{u})J_{k}^{\boldsymbol{x}}
|\chi\rangle_{\boldsymbol{x}}\otimes|\boldsymbol{\varnothing}\rangle_{\boldsymbol{u}},
\end{equation}
which holds provided that $\boldsymbol{x}$ satisfy  \eqref{Bethe-ansatz-equations}.
\item One can replace $J_{k}^{\boldsymbol{x}}\rightarrow J_{k}^{\boldsymbol{x}}+J_{k}^{\boldsymbol{u}}$ in the r.h.s. of \eqref{I2-BA-4} and use the property
\begin{equation}
[\mathcal{R}(\boldsymbol{x},\boldsymbol{u}),J_{k}^{\boldsymbol{x}}+J_{k}^{\boldsymbol{u}}]=0,
\end{equation}
to obtain
\begin{equation}\label{I2-BA-5}
 _{\boldsymbol{x}}\hspace*{-1pt}\langle\boldsymbol{\varnothing}|J_{k}^{\boldsymbol{x}}\,\mathcal{R}(\boldsymbol{x},\boldsymbol{u})
 |\chi\rangle_{\boldsymbol{x}}\otimes|\boldsymbol{\varnothing}\rangle_{\boldsymbol{u}}=\frac{q^k}{1-q^k}\,_{\boldsymbol{x}}\hspace*{-1pt}\langle\boldsymbol{\varnothing}|\,J_{k}^{\boldsymbol{u}}\,\mathcal{R}(\boldsymbol{x},\boldsymbol{u})
 |\chi\rangle_{\boldsymbol{x}}\otimes|\boldsymbol{\varnothing}\rangle_{\boldsymbol{u}}.
\end{equation}
\item Equations \eqref{I2-BA-1}, \eqref{I2-BA-2}, \eqref{I2-BA-3}, \eqref{I2-BA-4} and \eqref{I2-BA-5} imply that
\begin{equation} \label{Eigenvalues}
\tilde{I}_2^{\boldsymbol{u}}(q)|B(\boldsymbol{x})\rangle_{\boldsymbol{u}}=\sum_{k=1}^Nx_k|B(\boldsymbol{x})\rangle_{\boldsymbol{u}}. 
\end{equation}
on Bethe ansatz equations \eqref{Bethe-ansatz-equations}.
\end{enumerate}
	
In the above reasoning \eqref{I2-BA-4} requires explanation. In Appendix \ref{VectorChi} we have shown that
\begin{equation}
J_{k}^{\boldsymbol{x}}=\textrm{Ad}_{f_1}^{k-1}f_0
\end{equation}
Using this formula, one finds explicitly
\begin{equation}\label{Jk-integral-representation}
J_{k}^{\boldsymbol{x}}=\oint g_k(\boldsymbol{\xi})f(\xi_1)...f(\xi_k)d\boldsymbol{\xi}\quad\text{with}\quad
g_n(\vec{\xi})=\prod\limits_i \xi_i\left(\sum (-1)^iC_n^i \xi^{-1}_i\right)
\end{equation}
where $C_n^i$ are the binomial coefficients. 
	
Consider a matrix element of the l.h.s. of \eqref{I2-BA-4} with generic state
\begin{equation}\label{I2-BA-4-projected}
\,_{\boldsymbol{u}}\hspace*{-1pt}\langle\boldsymbol{\varnothing}|a^{(1)}_{\boldsymbol{\lambda}^{(1)}}\dots a^{(n)}_{\boldsymbol{\lambda}^{(n)}}\otimes\hspace*{.3pt}_{\boldsymbol{x}}\hspace*{-1pt}\langle\boldsymbol{\varnothing}|\,
J_{k}^{\boldsymbol{x}}\,\mathcal{R}(\boldsymbol{x},\boldsymbol{u})
|\chi\rangle_{\boldsymbol{x}}\otimes|\boldsymbol{\varnothing}\rangle_{\boldsymbol{u}},
\end{equation}
with
\begin{equation}
 \sum_{j=1}^n|\boldsymbol{\lambda}^{(j)}|+k=N.
\end{equation}
It can be rewritten as
\begin{multline}\label{I2-BA-4-projected-integral}
\,_{\boldsymbol{u}}\hspace*{-1pt}\langle\boldsymbol{\varnothing}|a^{(1)}_{\boldsymbol{\lambda}^{(1)}}\dots a^{(n)}_{\boldsymbol{\lambda}^{(n)}}\otimes\hspace*{.3pt}_{\boldsymbol{x}}\hspace*{-1pt}\langle\boldsymbol{\varnothing}|\,
J_{k}^{\boldsymbol{x}}\,\mathcal{R}(\boldsymbol{x},\boldsymbol{u})
|\chi\rangle_{\boldsymbol{x}}\otimes|\boldsymbol{\varnothing}\rangle_{\boldsymbol{u}}=_{\boldsymbol{x}}\hspace*{-4pt}\langle\boldsymbol{\varnothing}|
J_k^{\boldsymbol{x}}\mathcal{L}_{\boldsymbol{\lambda}^{(1)},\varnothing}(u_{1})\dots\mathcal{L}_{\boldsymbol{\lambda}^{(n)},\varnothing}(u_{n})|\chi\rangle_{\boldsymbol{x}}=\\=\oint
g_k(\boldsymbol{\xi})F_{\vec{\boldsymbol{\lambda}}}(\vec{\boldsymbol{z}}|\boldsymbol{u})\,
_{\boldsymbol{x}}\hspace*{-1pt}\langle\boldsymbol{\varnothing}|f(\xi_1)...f(\xi_k)\,h(u_1)\underbrace{f(z_1^{(1)})\dots}_{|\boldsymbol{\lambda}^{(1)}|}\,
h(u_2)\underbrace{f(z_1^{(2)})\dots}_{|\boldsymbol{\lambda}^{(2)}|}\,\,\dots\,\, h(u_n)\underbrace{f(z_1^{(n)})\dots}_{|\boldsymbol{\lambda}^{(n)}|}|\chi\rangle_{\boldsymbol{x}}\,d\boldsymbol{\xi}\,d\vec{\boldsymbol{z}}=\\=q^k\oint
g_k(\boldsymbol{\xi})F_{\vec{\boldsymbol{\lambda}}}(\vec{\boldsymbol{z}}|\boldsymbol{u})\,
_{\boldsymbol{x}}\hspace*{-1pt}\langle\boldsymbol{\varnothing}|h(u_1)\underbrace{f(z_1^{(1)})\dots}_{|\boldsymbol{\lambda}^{(1)}|}\,
h(u_2)\underbrace{f(z_1^{(2)})\dots}_{|\boldsymbol{\lambda}^{(2)}|}\,\,\dots\,\, h(u_n)\underbrace{f(z_1^{(n)})\dots}_{|\boldsymbol{\lambda}^{(n)}|}f(\xi_1)...f(\xi_k)|\chi\rangle_{\boldsymbol{x}}\,d\boldsymbol{\xi}\,d\vec{\boldsymbol{z}},
\end{multline}
which is equivalent to \eqref{I2-BA-4}. In the first line in \eqref{I2-BA-4-projected-integral} we have used definition of $\mathcal{L}_{\boldsymbol{\lambda},\varnothing}(u)$, in the second line \eqref{Jk-integral-representation}, \eqref{matrix-element-Lax-integral-representaion} and definition \eqref{F-lambda-vec}. While in the third line we have used argument similar to the one in \eqref{eigen-equation-on-off-shell-bethe-vector-integral-2} that is dragging  all $f(\xi_j)$'s to the right, abandoning local terms in commutation relations \eqref{hf-relation} and \eqref{ff-exact-relation} and using the fact that all the factors (here $\boldsymbol{z}$ denotes the set of all $\xi_j$ and $z_i^{(k)}$)
\begin{equation}
 \prod_{z_j\neq z}\prod_{\alpha=1}^3\frac{z-z_j+\epsilon_{\alpha}}{z-z_j-\epsilon_{\alpha}}\prod_{k=1}^n\frac{z-u_k}{z-u_k+\epsilon_3}
\end{equation}
are equal to $q$ on Bethe ansatz equations \eqref{Bethe-ansatz-equations}.
\subsection{Okounkov-Pandharipande equation}\label{OPe}
We saw that there are two related problems: diagonalization of KZ integral and solution of KZ difference equation.  Both problems can be solved in terms of Bethe vector. Similarly to KZ case, both counterparts exists for local Integrals of Motion. Let us consider the following equation \cite{okounkov2010quantum}:
\begin{equation}
  \tilde{\boldsymbol{I}}_2(q) |\psi\rangle=\hbar q\frac{d}{dq} |\psi\rangle \label{OP}
\end{equation}
We will show now that this equation is solved by the same function \eqref{WaveFunction}. In order to do that let us notice that while acting  \eqref{WaveFunction} differential operator $\hbar q \frac{d}{dq}$ is equal to multiplication on $\sum x_k$. The later can be expressed as an action of Integral of Motion in auxiliary space:
\begin{equation}
\sum_{k=1}^N x_k |B(\boldsymbol{x})\rangle_{\boldsymbol{u}}=\sum _{k=1}^Nx_k\,  {}_{\boldsymbol{x}}\hspace*{-1pt}\langle\boldsymbol{\varnothing}|\mathcal{R}(\boldsymbol{x},\boldsymbol{u})
|\chi\rangle_{\boldsymbol{x}}\otimes|\boldsymbol{\varnothing}\rangle_{\boldsymbol{u}}=
_{\boldsymbol{x}}\hspace*{-1pt}\langle\boldsymbol{\varnothing}|\mathcal{R}(\boldsymbol{x},\boldsymbol{u})
\tilde{\boldsymbol{I}}_2^{\boldsymbol{x}}(0)|\chi\rangle_{\boldsymbol{x}}\otimes|\boldsymbol{\varnothing}\rangle_{\boldsymbol{u}}
\end{equation}
Then we may repeat all the steps from previous section and found that under the integral over Bethe roots: action of $I_2^{\boldsymbol{x}}(0)$ on auxiliary space is equal to the action of $I_2^{\boldsymbol{v}}(q)$  on the quantum space. The only problematic point is number 4, let us explain it more details.
	
Let us consider the matrix element of the wave function:
\begin{multline}
\,_{\boldsymbol{u}}\hspace*{-1pt}\langle\boldsymbol{\varnothing}|a^{(1)}_{\boldsymbol{\lambda}^{(1)}}\dots a^{(n)}_{\boldsymbol{\lambda}^{(n)}}\otimes\hspace*{.3pt}_{\boldsymbol{x}}\hspace*{-1pt}\langle\boldsymbol{\varnothing}|\,
	\,\mathcal{R}(\boldsymbol{x},\boldsymbol{u})J_{k}^{\boldsymbol{x}}
	|\chi\rangle_{\boldsymbol{x}}\otimes|\boldsymbol{\varnothing}\rangle_{\boldsymbol{u}}=_{\boldsymbol{x}}\hspace*{-4pt}\langle\boldsymbol{\varnothing}|
	\mathcal{L}_{\boldsymbol{\lambda}^{(1)},\varnothing}(u_{1})\dots\mathcal{L}_{\boldsymbol{\lambda}^{(n)},\varnothing}(u_{n})J_k^{\boldsymbol{x}}|\chi\rangle_{\boldsymbol{x}}=\\=\textrm{Sym}_{\boldsymbol{x}}\Big( \Omega_{\vec{\boldsymbol{\lambda}}}(x_1,...,x_{N-k}|\boldsymbol{u})g_{{k}}(x_{N-k+1},...,x_N) \prod\limits_{i<j}S(x_i-x_j) \Big)
\end{multline}
On the other hand, if we insert $J_k^x$ from the left we will have:
\begin{multline}
\,_{\boldsymbol{u}}\hspace*{-1pt}\langle\boldsymbol{\varnothing}|a^{(1)}_{\boldsymbol{\lambda}^{(1)}}\dots a^{(n)}_{\boldsymbol{\lambda}^{(n)}}\otimes\hspace*{.3pt}_{\boldsymbol{x}}\hspace*{-1pt}\langle\boldsymbol{\varnothing}|\,
\,J_{k}^{\boldsymbol{x}}\mathcal{R}(\boldsymbol{x},\boldsymbol{u})
|\chi\rangle_{\boldsymbol{x}}\otimes|\boldsymbol{\varnothing}\rangle_{\boldsymbol{u}}=_{\boldsymbol{x}}\hspace*{-4pt}\langle\boldsymbol{\varnothing}|
J_{k}^{\boldsymbol{x}}\mathcal{L}_{\boldsymbol{\lambda}^{(1)},\varnothing}(u_{1})\dots\mathcal{L}_{\boldsymbol{\lambda}^{(n)},\varnothing}(u_{n})|\chi\rangle_{\boldsymbol{x}}=\\=\textrm{Sym}_{\boldsymbol{x}} \Big(\mathcal{D}(x_1,...,x_k|x_{k+1},...x_N) \Omega_{\vec{\boldsymbol{\lambda}}}(x_1,...,x_{N-k}|\boldsymbol{u})g_{{k}}(x_{N-k+1},...,x_N) \prod\limits_{i<j}S(x_i-x_j)\Big),
\end{multline}
where
\begin{equation}
\mathcal{D}(\boldsymbol{x_i}|\boldsymbol{y_j})=\prod_{i,j}\prod_{\alpha=1}^3\frac{x_i-y_j-\epsilon_{\alpha}}{x_i-y_j+\epsilon_{\alpha}}\prod_i\prod_{k=1}^n\frac{x_i-u_k+\epsilon_3}{x_i-u_k}.
\end{equation}
As we explain in Appendix \ref {VectorChi} (see \eqref{kerS}), the function $g_k(\boldsymbol{x_i})$ is transnational  invariant under the matrix element. And under the integral \eqref{WaveFunction} we can freely perform simultaneous shift of all $x_i$ for $i > N-k$. After this shift factor $\mathcal{D}(\boldsymbol{x_i}|\boldsymbol{y_j})$ will be canceled, and we arrive to the desired identity
\begin{eqnarray}
\,_{\boldsymbol{u}}\hspace*{-1pt}\langle\boldsymbol{\varnothing}|a^{(1)}_{\boldsymbol{\lambda}^{(1)}}\dots a^{(n)}_{\boldsymbol{\lambda}^{(n)}}\otimes\hspace*{.3pt}_{\boldsymbol{x}}\hspace*{-1pt}\langle\boldsymbol{\varnothing}|\,
\,J_{k}^{\boldsymbol{x}}\mathcal{R}(\boldsymbol{x},\boldsymbol{u})
|\chi\rangle_{\boldsymbol{x}}\otimes|\boldsymbol{\varnothing}\rangle_{\boldsymbol{u}}\sim q^k \,_{\boldsymbol{u}}\hspace*{-1pt}\langle\boldsymbol{\varnothing}|a^{(1)}_{\boldsymbol{\lambda}^{(1)}}\dots a^{(n)}_{\boldsymbol{\lambda}^{(n)}}\otimes\hspace*{.3pt}_{\boldsymbol{x}}\hspace*{-1pt}\langle\boldsymbol{\varnothing}| \,\mathcal{R}(\boldsymbol{x},\boldsymbol{u})J_{k}^{\boldsymbol{x}}
|\chi\rangle_{\boldsymbol{x}}\otimes|\boldsymbol{\varnothing}\rangle_{\boldsymbol{u}},
\end{eqnarray}
where the $"\sim"$ means equivalence under the integral \eqref{WaveFunction}. Other steps are completely similar to the ones in section \ref{LocalDiag}. And we recover the equation \eqref{OP}.
\subsection{Difference equations and norms of Bethe eigenvectors}
In this section we will treat $\hbar$ to be purely imaginary. As we already discussed in section \eqref{Section KZ}, solutions of KZ-OP \eqref{eqKZ}, \eqref{OP} equations on level $N$ are labeled by collections of Young diagrams with $N$ boxes. Let us compute the scalar product of two different solutions
\begin{equation}
 \langle\psi_{\lambda}(\boldsymbol{u})|\psi_{\mu}(\boldsymbol{u})\rangle
\end{equation}
As a consequence of KZ and OP equations this scalar product obey.
\begin{equation}
D_{\hbar}^{v_i}\langle\psi_{\lambda}(\boldsymbol{u})|\psi_{\mu}(\boldsymbol{u})\rangle=q\frac{d}{dq}\langle\psi_{\lambda}(\boldsymbol{u})|\psi_{\mu}(\boldsymbol{u})\rangle=0,
\end{equation}
and can be computed at the point $q=0$:
\begin{equation}
\langle\psi_{\lambda}(\boldsymbol{u})|\psi_{\mu}(\boldsymbol{u})\rangle\Bigl|_{q=0}=\delta_{\lambda,\mu} C(N,\hbar). \label{q0}
\end{equation}
We note that the constant $C(N,\hbar)$ is independent on the spectral parameters $u_i$ and the twist $q$. Let us compute the integral \eqref{WaveFunction} for general $q$ , but in the limit $\hbar \to 0$. In this limit \eqref{WaveFunction} can be taken by the saddle point method.  Using
\begin{equation}
\Gamma\left(\frac{x}{\hbar}\right)=\sqrt{\frac{2\pi \hbar}{x}}e^{\frac{x}{\hbar}\left(\log(x)-1 \right)-\frac{x}{\hbar}log(\hbar)}+o(\hbar),
\end{equation}
we find that the integration kernel in \eqref{WaveFunction} turns to the exponent of the Yang-Yang function:
\begin{eqnarray}
|\psi_{\lambda}(\boldsymbol{u})\rangle=\oint\limits_{C_{\lambda}} \frac{1}{\mathcal{F}(\boldsymbol{x})}e^{\frac{\mathcal{Y}(\boldsymbol{x})}{\hbar}}|B(\boldsymbol{x})\rangle_{\boldsymbol{u}}\,\frac{d^N\boldsymbol{x}}{(2\pi i)^N}, \label{oint}
\end{eqnarray}
where
\begin{multline}
\mathcal{Y}(\boldsymbol{x})=\sum\limits_{i\ne j}\Big[\mathfrak{\omega}(x_i-x_j-\epsilon_1)+\mathfrak{\omega}(x_i-x_j-\epsilon_2)-\mathfrak{\omega}(x_i-x_j+\epsilon_3)-\mathfrak{\omega}(x_i-x_j)\Big]+\\+\sum\limits_{i,k}\left(\mathfrak{\omega}(v_k-x_i)-\mathfrak{\omega}(v_k-x_i-\epsilon_3)\right)+\sum\limits_{i} x_i \log(q),\quad\text{with}\quad 	\mathfrak{\omega}(x)=x(\log(x)-1).
\end{multline}
and
\begin{eqnarray}
\mathcal{F}(\boldsymbol{x})=\prod\limits_{i\ne j} S(x_i-x_j) \prod\limits_{i,k}\frac{u_k-x_i}{u_k-x_i-\epsilon_3}
\end{eqnarray}
Computing the integral \eqref{oint} by saddle point, we found:
\begin{eqnarray}
|\psi_{\lambda}(\boldsymbol{u})\rangle= e^{\frac{\mathcal{Y}_{crit}(\boldsymbol{x})}{\hbar}}\frac{\hbar^{\frac{N}{2}}}{\sqrt{\mathcal{F}(\boldsymbol{x})H(\mathcal{Y})}}|B_{\lambda}(\boldsymbol{x})\rangle_{\boldsymbol{u}},
\end{eqnarray}
where $H(\mathcal{Y})$ is a Hessian:
\begin{eqnarray}
H(\mathcal{Y})=\textrm{det}\left(\frac{\partial^2 \mathcal{Y}}{\partial x_i \partial x_j}\right)
\end{eqnarray}
Comparing to the \eqref{q0}, we immediately recover the Slavnov's determinant formula for the norms of on-shell Bethe vectors \cite{Slavnov:1989kx}
\begin{eqnarray}
\langle B_{\lambda}|B_{\mu}\rangle=C(N) \delta_{\lambda,\mu}H(\mathcal{Y}) \mathcal{F}(\boldsymbol{x}) \label{norms}
\end{eqnarray}
Where $C(N)$ is the limit $C(N)=\lim\limits_{\hbar \to 0} C(N,\hbar)\hbar^{-N}$. The formula $\eqref{norms}$ can be rewritten in a different way:
\begin{eqnarray}
\frac{\langle B_{\lambda}|B_{\lambda}\rangle_q}{H(\mathcal{Y}) \mathcal{F}(\boldsymbol{x})}=\frac{\langle B_{\lambda}|B_{\lambda}\rangle_q}{H(\mathcal{Y}) \mathcal{F}(\boldsymbol{x})} \Big|_{q=0}
\end{eqnarray}
\section{Concluding remarks}\label{Conclusions}
This paper represents our efforts to understand the affine Yangian of $\mathfrak{gl}(1)$ and its role in integrability of conformal field theory. Many aspects have not been touched. Below we present some open problems and preliminary results that will be left for future work.
\paragraph{Other representations of the Yangian.} As we have seen, the commutation relations of $\textrm{YB}(\widehat{\mathfrak{gl}}(1))$ \eqref{Yangian-relation-main} are symmetric with respect to permutations of $\epsilon_k$. It implies that the algebra $\textrm{YB}(\widehat{\mathfrak{gl}}(1))$ admits three types of Fock modules $\mathcal{F}_u^{(k)}$ with $k=1,2,3$. Taking a representation of  generic type
\begin{equation}
   \mathcal{F}_{u_1}^{(k_1)}\otimes\mathcal{F}_{u_2}^{(k_2)}\otimes\dots\otimes\mathcal{F}_{u_n}^{(k_n)},
\end{equation}
will lead to ILW type integrable system corresponding to more general $W$ algebras introduced in \cite{2015arXiv151208779B,Litvinov:2016mgi}. The corresponding Miura transformation is explicitly known \cite{Prochazka:2018tlo,Prochazka:2019dvu}. All the results obtained in our paper can be generalized with a mild modification to this case. We collect some details in appendix \ref{SUSY-ILW}.
\paragraph{Massive deformation of $\textrm{ILW}_n$ integrable system.} The twist deformation of CFT integrable system \eqref{ILW-local-integrals} leads to certain $\tau-$deformation of Toda action \eqref{Toda-action-gln}. Namely, for our choice of twist deformation \eqref{Transfer-matrix-q-deformed}, one exponent in \eqref{Toda-action-gln} gets replaced by its non-local counterpart
\begin{equation}
e^{b(\varphi_2(x,t)-\varphi_1(x,t))}\xrightarrow{\text{twist deformation}}e^{b(\varphi_2(x,t)-\varphi_1(x+\pi\tau,t))},\quad\text{where}\quad q=e^{i\pi\tau}
\end{equation}
The corresponding classical field theory called \emph{non-local} $\mathfrak{gl}(n)$ Toda field theory  is known to be integrable in a Lax sense \cite{Degasperis,LEBEDEV1991166}. Its quantization has not been studied in the literature so far.

The simplest model of this kind is a free boson perturbed by a single exponent
\begin{equation}\label{NLToda}
S=\int \left( \frac{1}{8\pi} \partial_{\mu}\varphi\partial_{\mu}\varphi+\Lambda e^{b(\varphi(x,t)-\varphi(x+\pi\tau,t))} \right)\,d^{2}x, 
\end{equation}
This model has an interesting feature, in a finite volume of circumference $L=\pi n\tau$: ($x\sim x+L$),  relabeling the fields: $\phi(x+\pi k\tau) \overset{def}{=} \phi_k(x)$ we found that non local theory $\eqref{NLToda}$ in a volume $L=\pi n\tau$ is mapped to a local affine $A_n$ Toda in volume $\tau$. So it will be interesting to study the $S$ matrix and the spectrum of theory $\eqref{NLToda}$ in finite volume.
\paragraph{Relation to Sklyanin's results.}
Bethe anzatz equations similar to the ones studied in this paper were recently obtained by Sklyanin in a slightly different context of quantization of the first Hamiltonian structure of KP equation \cite{Kozlowski:2016too}. We believe that our results may be relevant in this context.   
\paragraph{Integrable systems corresponding to different root systems.} It is interesting to discover analytic continuation of general $\mathcal{W}(\mathfrak{g})$ algebra, associated to lie algebra $\mathfrak{g}$, and their relation to Yangian structures. Some results in the $\mathfrak{q}-$deformed case has been obtained in a recent paper \cite{feigin2020deformations}. Namely, there were proposed a new algebra $\mathcal{K}$ which is an analytic continuation of $\mathcal{W}(\mathcal{B}_n,\mathcal{D}_n,\mathcal{D}_n)$. As $\mathcal{W}_{1+\infty}$ algebra is related to Affine Yangian of $\widehat{\mathfrak{gl}}(1)$, it turns out that $\mathcal{K}$ algebra is related to Sklyanin boundary algebra \cite{Sklyanin:1988yz} associated to Affine Yangian. However authors were unable to find Bethe eigenvectors and Bethe equations in this case, which is an interesting open question. 
\paragraph{ODE/IM correspondence.} The spectrum of untwisted integrable systems (i.e. at $q=1$) can be studied by means of ODE/IM correspondence (see eg \cite{Dorey:2007zx} for review). We do not known any transparent relation between these two approaches. In particular the transfer matrices are quite different and we were unable to relate them.  We note also that algebraic equations for the spectrum are rather different in two approaches. In Yangian approach one has BA equations \eqref{Bethe-ansatz-equations}, while on ODE/IM side the spectrum is given by Gaudin-like equations (see for example \cite{Bazhanov:2004fk,Lukyanov:2013wra,Kotousov:2019nvt}).  It looks similar to the known duality between trigonometric
Gaudin and rational XXX models \cite{Mukhin:2006fk}, however it has not been clarified yet (see discussions in  \cite{Feigin:2017gcv}).
\paragraph{Yangian Double.} 
The algebra called Yangian Double has been introduced in \cite{Khoroshkin:1994uk} following  Drinfeld's quantum double construction \cite{Drinfeld-second}. The  Yangian Double seems to be more appropriate for construction of Bethe vectors  by the so called "method of projections" developed in \cite{enriquez2006weight,Khoroshkin_2007}  (see \cite{Hutsalyuk_2017,Liashyk:2019owy} for latest results). We introduce Yangian Double of $YB(\widehat{\mathfrak{gl}}(1))$ in  Appendix \ref{Yangian-double} and discuss some its properties. Unfortunately, we were unable to repeat the procedure executed in \cite{Hutsalyuk_2017,Liashyk:2019owy} and define the off-shell Bethe vector as a projection of a state build of "total" currents. This is an interesting open problem.
\section*{Acknowledgments}
Our special gratitude is to Borya Feigin for constantly guiding us through the subject. Also this work would not be possible without numerous conversations with Misha Bershtein, Serezha Lukyanov, Andrii Liashyk and Tom\'a\v{s} Proch\'azka.   
	
A.L. has been supported by the Russian Science Foundation under the grant 18-12-00439. I.V. has been Supported in part by Young Russian Mathematics award.
\Appendix
\section{Large \texorpdfstring{$u$}{u} expansion of the operator \texorpdfstring{$\mathcal{R}(u)$}{R(u)}}\label{R-expan}
We look for the solution to the equation
\begin{equation}\label{R-equation}
\mathcal{R}(u)\Bigl(-\bigl(u+J(x)\bigr)^{2}+QJ'(x)\Bigr)=\Bigl(-\bigl(u+J(x)\bigr)^{2}-QJ'(x)\Bigr)\mathcal{R}(u)
\end{equation}
where $J(x)=\sum_{k\neq0}a_{k}e^{-ikx}$ in the form \cite{Maulik:2012wi}
\begin{equation}\label{R-expansion}
\mathcal{R}(u)=\exp\left(iQ\Bigl(2u\log u+\sum_{k=1}^{\infty}(-1)^{k-1}\frac{r_{k}}{u^{k}}\Bigr)\right),\quad\text{where}\quad r_{k}=\frac{1}{2\pi}\int_{0}^{2\pi}g_{k+1}(x)dx,
\end{equation}
Solving \eqref{R-equation} one can find first few densities $g_{k}(x)$ explicitly:
\begin{equation}\label{first-local-densities-S-matrix}
\begin{gathered}
g_{2}=J^{2},\quad
g_{3}=\frac{J^{3}}{3},\quad
g_{4}=\frac{J^{4}}{6}+\frac{1-2Q^{2}}{24}J_{x}^{2},\quad
g_{5}=\frac{J^{5}}{10}+\frac{1-2Q^{2}}{8}JJ_{x}^{2},\\
g_{6}=\frac{J^{6}}{15}+\frac{1-2Q^{2}}{4}J^{2}J_{x}^{2}+\frac{2-9Q^{2}+6Q^{4}}{480}J_{xx}^{2},\quad
g_{7}=\frac{J^{7}}{21}+\frac{5(1-2Q^{2})}{12}J^{3}J_{x}^{2}+\frac{2-9Q^{2}+6Q^{4}}{96}JJ_{xx}^{2},
\end{gathered}
\end{equation}
and more disgusting expression for $g_{8}$
\begin{multline}
g_{8}=\frac{1}{28}J^{8}+\frac{5}{8}(1-2Q^{2})J^{4}J_{x}^{2}+\frac{1}{32}(2-9Q^{2}+6Q^{4})J^{2}J_{xx}^{2}
+\frac{1}{576}\left(-9+41Q^2-26Q^4\right)J_{x}^{4}+\\+
\frac{1}{161280}\left(90-671Q^2+998Q^4-360Q^6\right)J_{xxx}^{2}.
\end{multline}
For all densities $g_{k}(x)$ in \eqref{first-local-densities-S-matrix} we used zeta-function regularization. 
For example
\begin{equation*}
\int J^{2}=\int :J^{2}:-\frac{1}{24},\quad 
\int J^{4}=\int :J^{4}:-\int\frac{1}{4}:J^{2}:+\frac{1}{192},\quad
\int J_{x}^{2}=\int:J_{x}^{2}:+\frac{1}{240},
\end{equation*}
where $::$ stands for the Wick ordering. 
	
Explicit formula \eqref{R-expansion} is useful for us, because it provides a relation between Yangian currents \eqref{efh-def} and $W^{n}(z)$ currents. For example for the first few modes:
\begin{align}
f_0=Q a_{1} \ , \quad f_1=Q \sum_n a_{n+1}a_{-n} \\
e_0=Q a_{-1} \ , \quad e_1=Q \sum_n a_{n-1}a_{-n}.
\end{align}
Of course these formulas are only true in a bosonic representation, however it is easy to analytically continue them to the arbitrary number of bosons:
\begin{align} \label{modes}
f_0=Q J_{1} \ , \quad f_1=Q L_{1} \\ 
e_0=Q J_{-1} \ , \quad e_1=Q L_{-1}.
\end{align}
Where $L_{\pm 1}$ is a special $\mathcal{W}^{(2)}$ current, such that:
\begin{eqnarray}
[L_{\pm 1},J_n]=J_{n\pm 1}.
\end{eqnarray}
	
Let us also note that there is easily established pattern in the densities \eqref{first-local-densities-S-matrix}. Namely, the first terms in \eqref{first-local-densities-S-matrix} can be written as
\begin{multline}
g_{n}=\frac{2}{n(n-1)}J^{n}+\frac{(n-2)(n-3)}{48}(1-2Q^{2})J^{n-4}J_{x}^{2}+\\+
\frac{(n-2)(n-3)(n-4)(n-5)}{11520}(2-9Q^{2}+6Q^{4})J^{n-6}J_{xx}^{2}+\dots
\end{multline}
Using this observation one can formally do the resumation in \eqref{R-expansion}.
\begin{equation}
G(x)\overset{\text{def}}{=}2u\log u+\sum_{k=1}^{\infty}(-1)^{k-1}g_{k+1}(x)/u^{k},
\end{equation}
which admits the derivative expansion  
\begin{equation}\label{Phi-expansion-resummed}
G(x)=2(u+J)\log\left(u+J\right)+\frac{1-2Q^{2}}{24}\frac{J_{x}^{2}}{(u+J)^{3}}
+\frac{2-9Q^{2}+6Q^{4}}{480}\frac{J_{xx}^{2}}{(u+J)^{5}}+\dots
\end{equation}
The expansion \eqref{Phi-expansion-resummed} suggests the following general form 
\begin{equation}\label{Phi-expansion-resummed-II}
G(x)=2(u+J)\log\left(u+J\right)+\sum_{k=1}^{\infty}\frac{U_{2k+2}(J_{x},J_{xx},\dots)}{(u+J)^{2k+1}},
\end{equation}
where $U_{2k+2}(J_{x},J_{xx},\dots)$ is a  homogeneous and even with respect to the  transformation $J\rightarrow-J$  density of degree $2k+2$. It would be interesting to find the densities $U_{2k+2}(J_{x},J_{xx},\dots)$ exactly.
	
One can also compute the $\mathcal{R}(u)$ operator in the ``free fermion'' point $c=-2$. Namely, take $Q=-\frac{i}{\sqrt{2}}$ in \eqref{R-expansion} and represent the current $J(x)$ by the complex fermion $\psi(x)$ as
\begin{equation}
J(x)=\frac{1}{\sqrt{2}}:\psi^{+}(x)\psi(x):.
\end{equation}
Then one can check that (see also appendix \ref{SUSY-ILW})
\begin{equation}
\mathcal{R}(u)\Bigl|_{c=-2}\sim
\exp\left(\frac{1}{2\pi}\int_{0}^{2\pi}\psi^{+}(x)\log\left(1+\frac{i}{u\sqrt{2}}\partial\right)\psi(x)\,dx\right)
\end{equation}
\section{Affine Yangian commutation relations}\label{Yangian-relations}
Here we consider commutation relations of the Yangian algebra \eqref{YB-algebra} in components. Similar analysis has been performed in \cite{Prochazka:2019dvu}. We use the following notations
\begin{equation}
\begin{picture}(300,40)(410,5)
\Thicklines
\unitlength 7pt 	
\put(50,0){\line(0,1){5}}
\put(50.5,0){\line(0,1){5}}
\put(51,0){\line(0,1){5}}
\put(51.5,0){\line(0,1){5}}
\put(52,0){\line(0,1){5}}
\put(48,2.5){\line(1,0){6}}
\put(46.4,2.25){$\scriptstyle{\langle u|}$}
\put(54.1,2.25){$\scriptstyle{|u\rangle}$}
\put(56,2.13){$=\mathcal{L}_{\scriptscriptstyle{\varnothing,\varnothing}}(u)\overset{\text{def}}{=}h(u)$}
\put(77,0){\line(0,1){5}}
\put(77.5,0){\line(0,1){5}}
\put(78,0){\line(0,1){5}}
\put(78.5,0){\line(0,1){5}}
\put(79,0){\line(0,1){5}}
\put(75,2.5){\line(1,0){6}}
\put(73.4,2.25){$\scriptstyle{\langle u|}$}
\put(81.3,2.25){$\scriptstyle{a_{-1}|u\rangle}$}
\put(85,2.13){$=\mathcal{L}_{\scriptscriptstyle{\varnothing,\Box}}(u)$}
\put(102,0){\line(0,1){5}}
\put(102.5,0){\line(0,1){5}}
\put(103,0){\line(0,1){5}}
\put(103.5,0){\line(0,1){5}}
\put(104,0){\line(0,1){5}}
\put(100,2.5){\line(1,0){6}}
\put(97.3,2.25){$\scriptstyle{\langle u|a_{1}}$}
\put(106.1,2.25){$\scriptstyle{|u\rangle}$}
\put(108,2.13){$=\mathcal{L}_{\scriptscriptstyle{\Box,\varnothing}}(u)$}
\end{picture}
\end{equation}
and admit the convention that the operators acts in ``quantum'' space from up to down. It is also convenient  to define according to \eqref{efh-def}
\begin{equation}
h(u)\overset{\text{def}}{=}\mathcal{L}_{\scriptscriptstyle{\varnothing,\varnothing}}(u),\qquad
e(u)\overset{\text{def}}{=}h^{-1}(u)\cdot\mathcal{L}_{\scriptscriptstyle{\varnothing},\includegraphics[scale=0.04]{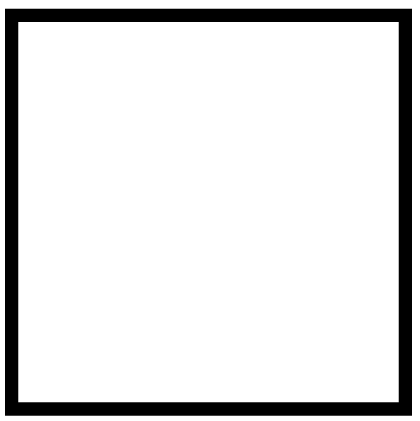}}(u)\quad\text{and}\quad
f(u)\overset{\text{def}}{=}\mathcal{L}_{\includegraphics[scale=0.04]{Jack0.eps},\scriptscriptstyle{\varnothing}}(u)\cdot h^{-1}(u).
\end{equation} 
We will introduce currents $e_{\scriptscriptstyle{\boldsymbol{\lambda}}}(u)$ and $f_{\scriptscriptstyle{\boldsymbol{\lambda}}}(u)$ associated to $3D$ partitions. There are $3$ currents on level $2$ \eqref{level2-currents}, $6$ currents on level $3$ (see \eqref{level3-currents}) etc. Similar expressions one has for $f_{\scriptscriptstyle{\boldsymbol{\lambda}}}(u)$.

All these and other generators of the Yang-Baxter  \eqref{YB-algebra} admit large $u$ expansion which is inherited from the large $u$ expansion of the $R-$matrix \eqref{R-expansion}. In particular, 
\begin{equation}
h(u)=1+\frac{h_{0}}{u}+\frac{h_{1}}{u^{2}}+\dots,\quad
e(u)=\frac{e_{0}}{u}+\frac{e_{1}}{u^{2}}+\dots,\quad
f(u)=\frac{f_{0}}{u}+\frac{f_{1}}{u^{2}}+\dots,
\end{equation}
while the higher currents are expected to behave at $u\rightarrow\infty$ as
\begin{equation}
e_{\boldsymbol{\lambda}}(u)\sim\frac{1}{u^{\boldsymbol{\lambda}}},\quad
f_{\boldsymbol{\lambda}}(u)\sim\frac{1}{u^{\boldsymbol{\lambda}}}.
\end{equation}
	
The relations of the Yang-Baxter algebra \eqref{YB-algebra} appear from the tensor product of two Fock spaces $\mathcal{F}_{u}$ and $\mathcal{F}_{v}$. We will use the following notations for the bra and ket highest weight states in  $\mathcal{F}_{u}\otimes\mathcal{F}_{v}$
\begin{equation}
\langle\textrm{vac}|\overset{\text{def}}{=}\langle u|\otimes\langle v|,\qquad
|\textrm{vac}\rangle\overset{\text{def}}{=}|u\rangle\otimes|v\rangle.
\end{equation}
The action of the zero-mode $a_0$ on the vacuum state $|u\rangle$ is 
\begin{equation}
  a_0|u\rangle=-iu|u\rangle.
\end{equation} 
\subsection*{$he$ and $hf$ relations:}
Then on level $0$ we  have
\begin{equation}\label{level-0-YB-algebra}
\begin{picture}(300,70)(400,5)
\Thicklines
\unitlength 7pt 
\put(57,0){\line(0,1){10}}
\put(57.5,0){\line(0,1){10}}
\put(58,0){\line(0,1){10}}
\put(58.5,0){\line(0,1){10}}
\put(59,0){\line(0,1){10}}
\put(55,8){\line(1,0){6}}
\put(55,2){\line(1,0){6}}
\put(61.15,1.73){$\scriptstyle{|u\rangle}$}
\put(61.15,7.73){$\scriptstyle{|v\rangle}$}
\put(55,2){\line(-2,3){4}}
\put(55,8){\line(-2,-3){4}}
\put(51,2){\line(-1,0){2}}
\put(51,8){\line(-1,0){2}}
\put(47.5,1.73){$\scriptstyle{\langle v|}$}
\put(47.4,7.73){$\scriptstyle{\langle u|}$}
\put(65,4.55){$=$}
\put(72,0){\line(0,1){10}}
\put(72.5,0){\line(0,1){10}}
\put(73,0){\line(0,1){10}}
\put(73.5,0){\line(0,1){10}}
\put(74,0){\line(0,1){10}}
\put(70,8){\line(1,0){6}}
\put(70,2){\line(1,0){6}}
\put(68.5,1.73){$\scriptstyle{\langle v|}$}
\put(68.4,7.73){$\scriptstyle{\langle u|}$}
\put(76,2){\line(2,3){4}}
\put(76,8){\line(2,-3){4}}
\put(80,2){\line(1,0){2}}
\put(80,8){\line(1,0){2}}
\put(82.15,1.73){$\scriptstyle{|u\rangle}$}
\put(82.15,7.73){$\scriptstyle{|v\rangle}$}
\put(87,4.55){$\implies$}
\put(92,4.55){$h(u)h(v)=h(v)h(u)$}
\end{picture}
\end{equation}
On level $1$ one has two relations
\begin{equation}\label{level-1-YB-algebra}
\begin{picture}(300,60)(350,0)
\Thicklines
\unitlength 5pt 
\hspace*{.5cm}
\put(57,0){\line(0,1){10}}
\put(57.5,0){\line(0,1){10}}
\put(58,0){\line(0,1){10}}
\put(58.5,0){\line(0,1){10}}
\put(59,0){\line(0,1){10}}
\put(55,8){\line(1,0){6}}
\put(55,2){\line(1,0){6}}
\put(61.15,1.73){$\scriptstyle{|u\rangle}$}
\put(61.15,7.73){$\scriptstyle{a_{-1}|v\rangle}$}
\put(55,2){\line(-2,3){4}}
\put(55,8){\line(-2,-3){4}}
\put(51,2){\line(-1,0){2}}
\put(51,8){\line(-1,0){2}}
\put(47,1.73){$\scriptstyle{\langle v|}$}
\put(46.9,7.73){$\scriptstyle{\langle u|}$}
\put(65,4.55){$=$}
\put(72,0){\line(0,1){10}}
\put(72.5,0){\line(0,1){10}}
\put(73,0){\line(0,1){10}}
\put(73.5,0){\line(0,1){10}}
\put(74,0){\line(0,1){10}}
\put(70,8){\line(1,0){6}}
\put(70,2){\line(1,0){6}}
\put(68,1.73){$\scriptstyle{\langle v|}$}
\put(67.9,7.73){$\scriptstyle{\langle u|}$}
\put(76,2){\line(2,3){4}}
\put(76,8){\line(2,-3){4}}
\put(80,2){\line(1,0){2}}
\put(80,8){\line(1,0){2}}
\put(82.15,1.73){$\scriptstyle{|u\rangle}$}
\put(82.15,7.73){$\scriptstyle{a_{-1}|v\rangle}$}
\put(87,4.55){$\implies$}
\put(93,4.55){$h(u)\mathcal{L}_{\scriptscriptstyle{\varnothing,\Box}}(v)=
\frac{u-v}{u-v+Q}\mathcal{L}_{\scriptscriptstyle{\varnothing,\Box}}(v)h(u)+\frac{Q}{u-v+Q}h(v)\mathcal{L}_{\scriptscriptstyle{\varnothing,\Box}}(u)$}
\end{picture}
\end{equation}
and 
\begin{equation}\label{level-1-YB-algebra-II}
\begin{picture}(300,60)(350,0)
\Thicklines
\unitlength 5pt 
\hspace*{.5cm}
\put(57,0){\line(0,1){10}}
\put(57.5,0){\line(0,1){10}}
\put(58,0){\line(0,1){10}}
\put(58.5,0){\line(0,1){10}}
\put(59,0){\line(0,1){10}}
\put(55,8){\line(1,0){6}}
\put(55,2){\line(1,0){6}}
\put(61.15,1.73){$\scriptstyle{a_{-1}|u\rangle}$}
\put(61.15,7.73){$\scriptstyle{|v\rangle}$}
\put(55,2){\line(-2,3){4}}
\put(55,8){\line(-2,-3){4}}
\put(51,2){\line(-1,0){2}}
\put(51,8){\line(-1,0){2}}
\put(47,1.73){$\scriptstyle{\langle v|}$}
\put(46.9,7.73){$\scriptstyle{\langle u|}$}
\put(65,4.55){$=$}
\put(72,0){\line(0,1){10}}
\put(72.5,0){\line(0,1){10}}
\put(73,0){\line(0,1){10}}
\put(73.5,0){\line(0,1){10}}
\put(74,0){\line(0,1){10}}
\put(70,8){\line(1,0){6}}
\put(70,2){\line(1,0){6}}
\put(68,1.73){$\scriptstyle{\langle v|}$}
\put(67.9,7.73){$\scriptstyle{\langle u|}$}
\put(76,2){\line(2,3){4}}
\put(76,8){\line(2,-3){4}}
\put(80,2){\line(1,0){2}}
\put(80,8){\line(1,0){2}}
\put(82.15,1.73){$\scriptstyle{a_{-1}|u\rangle}$}
\put(82.15,7.73){$\scriptstyle{|v\rangle}$}
\put(87,4.55){$\implies$}
\put(93,4.55){$\mathcal{L}_{\scriptscriptstyle{\varnothing,\Box}}(u)h(v)=
\frac{Q}{u-v+Q}\mathcal{L}_{\scriptscriptstyle{\varnothing,\Box}}(v)h(u)+\frac{u-v}{u-v+Q}h(v)\mathcal{L}_{\scriptscriptstyle{\varnothing,\Box}}(u)$}
\end{picture}
\end{equation}
In fact \eqref{level-1-YB-algebra} and \eqref{level-1-YB-algebra-II} are not independent. Taking the linear combination $Q\times\eqref{level-1-YB-algebra}-(u-v)\times\eqref{level-1-YB-algebra-II}$ one arrives to the equation  \eqref{level-1-YB-algebra} with $u\leftrightarrow v$. 
	
Now, multiplying \eqref{level-1-YB-algebra} by $(u-v+Q)h^{-1}(v)$ from the left and using \eqref{level-0-YB-algebra} one get the relation
\begin{equation}\label{Yangian-relation-1}
(u-v+Q)h(u)e(v)=(u-v)e(v)h(u)+Q\mathcal{L}_{\scriptscriptstyle{\varnothing,\Box}}(u)\implies
(u-v+Q)e(v)=(u-v)h^{-1}(u)e(v)h(u)+Qe(u).
\end{equation}
In the leading order in large $v$ expansion one obtains (compare to \eqref{first-local-relation})
\begin{equation}
[e_{0},h(u)]=Q\mathcal{L}_{\scriptscriptstyle{\varnothing,\Box}}(u).
\end{equation}
	
We note that the relation \eqref{level-0-YB-algebra} implies that the product $h(u)e(v)$ could not have poles, which implies that
\begin{equation}
\mathcal{L}_{\scriptscriptstyle{\varnothing,\Box}}(u)=h(u)e(u)=e(u+Q)h(u).
\end{equation}
Formula \eqref{Yangian-relation-1} allows one to rewrite 
\begin{equation}
e_{j_{1}}\dots e_{j_{n}}h(u)e_{j_{n+1}}\dots e_{j_{n+m}}=\oint\limits_{\mathcal{C}_{n+m}}\dots\oint\limits_{\mathcal{C}_{1}}\prod_{k=1}^{n+m}z_{k}^{j_{k}}\cdot\prod_{k=n+1}^{n+m}\frac{u-z_{k}}{u-z_{k}+Q}\,e(z_{n+m})\dots e(z_{1})h(u)d\vec{z},
\end{equation}
where the contours $\mathcal{C}_{\infty}$ go concentrically around $\infty$ in such a way that all singularities of the function $F(\boldsymbol{z})$ are kept \emph{inside} of these contours.

Similarly to \eqref{Yangian-relation-1} one also obtains the relation
\begin{equation}\label{Yangian-relation-2}
(u-v+Q)f(v)h(u)=(u-v)h(u)f(v)+Q\mathcal{L}_{\scriptscriptstyle{\Box,\varnothing}}(u),
\end{equation}
which enables one to rewrite
\begin{equation}\label{fh-under-intagral-identity}
f_{j_{n+m}}\dots f_{j_{n+1}}h(u)f_{j_{n}}\dots f_{k_{1}}=\oint\limits_{\mathcal{C}_{n+m}}\dots\oint\limits_{\mathcal{C}_{1}}\prod_{k=1}^{n+m}z_{k}^{j_{k}}\cdot\prod_{k=1}^{n}\frac{u-z_{k}}{u-z_{k}-Q}\,h(u)f(z_{n+m})\dots f(z_{1})d\vec{z}.
\end{equation}
\subsection*{$ee$ and $ff$ relations:}
On level $2$ we have three independent equations 
\begin{equation}\label{level-2-YB-algebra}
\begin{picture}(300,160)(310,-100)
\Thicklines
\unitlength 5pt 
\hspace*{-.5cm}
\put(57,0){\line(0,1){10}}
\put(57.5,0){\line(0,1){10}}
\put(58,0){\line(0,1){10}}
\put(58.5,0){\line(0,1){10}}
\put(59,0){\line(0,1){10}}
\put(55,8){\line(1,0){6}}
\put(55,2){\line(1,0){6}}
\put(61.15,1.73){$\scriptstyle{|u\rangle}$}
\put(61.15,7.73){$\scriptstyle{a_{-1}^{2}|v\rangle}$}
\put(55,2){\line(-2,3){4}}
\put(55,8){\line(-2,-3){4}}
\put(51,2){\line(-1,0){2}}
\put(51,8){\line(-1,0){2}}
\put(47,1.73){$\scriptstyle{\langle v|}$}
\put(46.9,7.73){$\scriptstyle{\langle u|}$}
\hspace*{.3cm}
\put(65,4.55){$=$}
\hspace*{.3cm}
\put(72,0){\line(0,1){10}}
\put(72.5,0){\line(0,1){10}}
\put(73,0){\line(0,1){10}}
\put(73.5,0){\line(0,1){10}}
\put(74,0){\line(0,1){10}}
\put(70,8){\line(1,0){6}}
\put(70,2){\line(1,0){6}}
\put(68,1.73){$\scriptstyle{\langle v|}$}
\put(67.9,7.73){$\scriptstyle{\langle u|}$}
\put(76,2){\line(2,3){4}}
\put(76,8){\line(2,-3){4}}
\put(80,2){\line(1,0){2}}
\put(80,8){\line(1,0){2}}
\put(82.15,1.73){$\scriptstyle{|u\rangle}$}
\put(82.15,7.73){$\scriptstyle{a_{-1}^{2}|v\rangle}$}
\put(107,0){\line(0,1){10}}
\put(107.5,0){\line(0,1){10}}
\put(108,0){\line(0,1){10}}
\put(108.5,0){\line(0,1){10}}
\put(109,0){\line(0,1){10}}
\put(105,8){\line(1,0){6}}
\put(105,2){\line(1,0){6}}
\put(111.15,1.73){$\scriptstyle{|u\rangle}$}
\put(111.15,7.73){$\scriptstyle{a_{-2}|v\rangle}$}
\put(105,2){\line(-2,3){4}}
\put(105,8){\line(-2,-3){4}}
\put(101,2){\line(-1,0){2}}
\put(101,8){\line(-1,0){2}}
\put(97,1.73){$\scriptstyle{\langle v|}$}
\put(96.9,7.73){$\scriptstyle{\langle u|}$}
\hspace*{.3cm}
\put(115,4.55){$=$}
\hspace*{.3cm}
\put(122,0){\line(0,1){10}}
\put(122.5,0){\line(0,1){10}}
\put(123,0){\line(0,1){10}}
\put(123.5,0){\line(0,1){10}}
\put(124,0){\line(0,1){10}}
\put(120,8){\line(1,0){6}}
\put(120,2){\line(1,0){6}}
\put(118,1.73){$\scriptstyle{\langle v|}$}
\put(117.9,7.73){$\scriptstyle{\langle u|}$}
\put(126,2){\line(2,3){4}}
\put(126,8){\line(2,-3){4}}
\put(130,2){\line(1,0){2}}
\put(130,8){\line(1,0){2}}
\put(132.15,1.73){$\scriptstyle{|u\rangle}$}
\put(132.15,7.73){$\scriptstyle{a_{-2}|v\rangle}$}
\hspace*{-2cm}
\put(87,-20){\line(0,1){10}}
\put(87.5,-20){\line(0,1){10}}
\put(88,-20){\line(0,1){10}}
\put(88.5,-20){\line(0,1){10}}
\put(89,-20){\line(0,1){10}}
\put(85,-18){\line(1,0){6}}
\put(85,-12){\line(1,0){6}}
\put(91.15,-18.27){$\scriptstyle{a_{-1}|u\rangle}$}
\put(91.15,-12.27){$\scriptstyle{a_{-1}|v\rangle}$}
\put(85,-18){\line(-2,3){4}}
\put(85,-12){\line(-2,-3){4}}
\put(81,-18){\line(-1,0){2}}
\put(81,-12){\line(-1,0){2}}
\put(77,-18.27){$\scriptstyle{\langle v|}$}
\put(76.9,-12.27){$\scriptstyle{\langle u|}$}
\hspace*{.3cm}
\put(95,-15.45){$=$}
\hspace*{.3cm}
\put(102,-20){\line(0,1){10}}
\put(102.5,-20){\line(0,1){10}}
\put(103,-20){\line(0,1){10}}
\put(103.5,-20){\line(0,1){10}}
\put(104,-20){\line(0,1){10}}
\put(100,-12){\line(1,0){6}}
\put(100,-18){\line(1,0){6}}
\put(98,-18.27){$\scriptstyle{\langle v|}$}
\put(97.9,-12.27){$\scriptstyle{\langle u|}$}
\put(106,-18){\line(2,3){4}}
\put(106,-12){\line(2,-3){4}}
\put(110,-18){\line(1,0){2}}
\put(110,-12){\line(1,0){2}}
\put(112.15,-18.27){$\scriptstyle{a_{-1}|u\rangle}$}
\put(112.15,-12.27){$\scriptstyle{a_{-1}|v\rangle}$}
\end{picture}
\end{equation}
In the r.h.s of any of these relation one has a linear combination of five terms
\begin{equation}
h(v)\mathcal{L}_{\scriptscriptstyle{\varnothing},\zerounderset{\scriptscriptstyle{\Box}}{\vspace*{-1.55pt}\scriptscriptstyle{\Box}}}(u),\quad
h(v)\mathcal{L}_{\scriptscriptstyle{\varnothing,\Box\hspace*{-1.2pt}\vspace*{2pt}\Box}}(u),\quad
\mathcal{L}_{\scriptscriptstyle{\varnothing},\scriptscriptstyle{\Box}}(v)\mathcal{L}_{\scriptscriptstyle{\varnothing},\scriptscriptstyle{\Box}}(u),\quad
\mathcal{L}_{\scriptscriptstyle{\varnothing},\zerounderset{\scriptscriptstyle{\Box}}{\vspace*{-1.55pt}\scriptscriptstyle{\Box}}}(v)h(u),\quad
\mathcal{L}_{\scriptscriptstyle{\varnothing,\Box\hspace*{-1.2pt}\vspace*{2pt}\Box}}(v)h(u).
\end{equation}
One can always find special linear combination of three equations \eqref{level-2-YB-algebra} which kills contributions of last two terms. Explicitly, one has
\begin{multline}\label{level-2-special-combination}
A_{1}(u-v)h(u)\mathcal{L}_{\scriptscriptstyle{\varnothing},\zerounderset{\scriptscriptstyle{\Box}}{\vspace*{-1.55pt}\scriptscriptstyle{\Box}}}(v)+
A_{2}(u-v)h(u)\mathcal{L}_{\scriptscriptstyle{\varnothing,\Box\hspace*{-1.2pt}\vspace*{2pt}\Box}}(v)+
A_{3}(u-v)\mathcal{L}_{\scriptscriptstyle{\varnothing},\scriptscriptstyle{\Box}}(u)\mathcal{L}_{\scriptscriptstyle{\varnothing},\scriptscriptstyle{\Box}}(v)=\\=
A_{1}(v-u)h(v)\mathcal{L}_{\scriptscriptstyle{\varnothing},\zerounderset{\scriptscriptstyle{\Box}}{\vspace*{-1.55pt}\scriptscriptstyle{\Box}}}(u)+
A_{2}(v-u)h(v)\mathcal{L}_{\scriptscriptstyle{\varnothing,\Box\hspace*{-1.2pt}\vspace*{2pt}\Box}}(u)+
A_{3}(v-u)\mathcal{L}_{\scriptscriptstyle{\varnothing},\scriptscriptstyle{\Box}}(v)\mathcal{L}_{\scriptscriptstyle{\varnothing},\scriptscriptstyle{\Box}}(u),
\end{multline}
where
\begin{equation}
A_{1}(u)=-Q(u+Q),\quad A_{2}(u)=-iQ,\quad A_{3}(u)=(u+b)(u+b^{-1})
\end{equation}
We note that using \eqref{Yangian-relation-1} one can express\footnote{We also have more general relation  
\begin{equation}
\mathcal{L}_{\scriptscriptstyle{\varnothing},\scriptscriptstyle{\Box}}(u)\mathcal{L}_{\scriptscriptstyle{\varnothing},\scriptscriptstyle{\boldsymbol{\lambda}}}(v)=
\frac{h(u)h(v)}{u-v}\Bigl[(u-v-Q)e(u)e_{\scriptscriptstyle{\boldsymbol{\lambda}}}(v)+Qe(v)e_{\scriptscriptstyle{\boldsymbol{\lambda}}}(v)\Bigr].
\end{equation}}
\begin{equation}
\mathcal{L}_{\scriptscriptstyle{\varnothing},\scriptscriptstyle{\Box}}(u)\mathcal{L}_{\scriptscriptstyle{\varnothing},\scriptscriptstyle{\Box}}(v)=
\frac{h(u)h(v)}{u-v}\left[(u-v-Q)e(u)e(v)+Qe^{2}(v)\right],
\end{equation}
and hence the relation \eqref{level-2-special-combination} can be rewritten as
\begin{multline}\label{ee-Yangian-relation-3D-basis}
(u-v+b)(u-v+b^{-1})(u-v-Q)e(u)e(v)=(u-v-b)(u-v-b^{-1})(u-v+Q)e(v)e(u)+\\+
\Bigl((u-v-b^{-1})(u-v+Q)e_{\includegraphics[scale=0.035]{partition2.eps}}(u)+(u-v-b)(u-v+Q)e_{\includegraphics[scale=0.035]{partition3.eps}}(u)+
(u-v-b)(u-v-b^{-1})e_{\includegraphics[scale=0.035]{partition1.eps}}(u)+(u\longleftrightarrow v)\Bigr),
\end{multline}
where the higher currents $e_{\boldsymbol{\lambda}}(u)$ are given by \eqref{level2-currents}.

Other two relations from \eqref{level-2-YB-algebra} are equivalent to commutation relations between $h(u)$ end $e_{\scriptscriptstyle{\boldsymbol{\lambda}}}(v)$ (similar to \eqref{Yangian-relation-1})
\begin{multline}
(u-v+Q)\Bigl(u-v+Q+\frac{1}{b}\Bigr)e_{\includegraphics[scale=0.035]{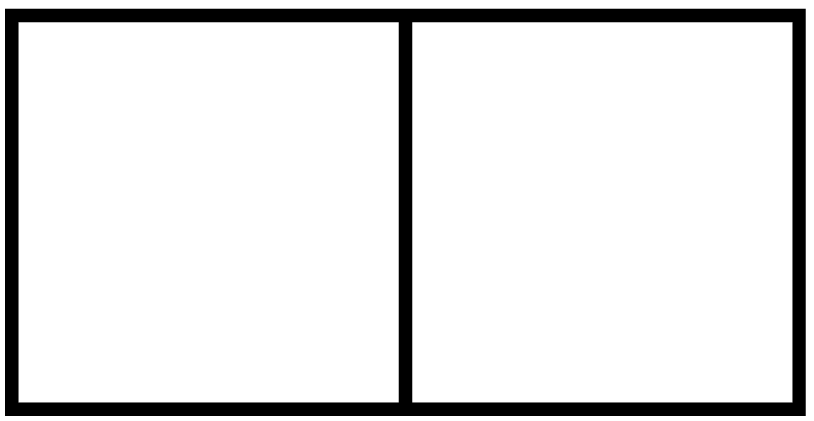}}(v)=
(u-v)\Bigl(u-v+\frac{1}{b}\Bigr)h^{-1}(u)e_{\includegraphics[scale=0.035]{Jack1.eps}}(v)h(u)+\\+
\left((u-v+Q)\Bigl(u-v+Q+\frac{1}{b}\Bigr)-(u-v)\Bigl(u-v+\frac{1}{b}\Bigr)\right)e_{\includegraphics[scale=0.035]{Jack1.eps}}(u)-
\frac{2bQ(u-v)}{b-b^{-1}}\Bigl(e_{\includegraphics[scale=0.035]{Jack1.eps}}(u)-
e_{\includegraphics[scale=0.035]{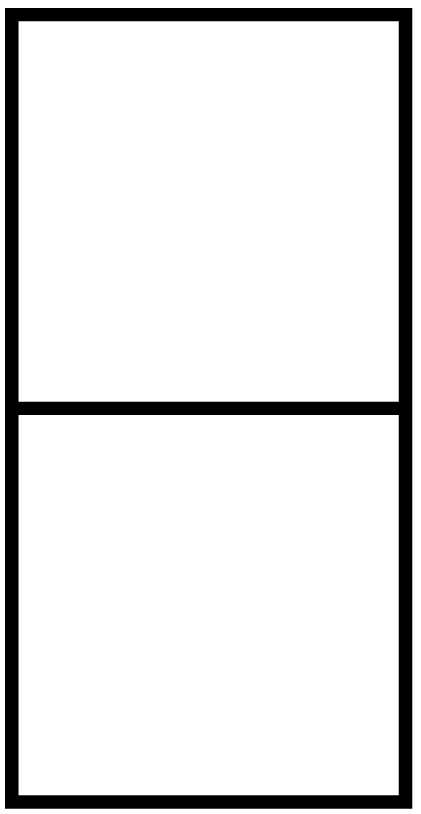}}(u)\Bigr)-\\-
2ibQ\Bigl((u-v+Q)e(v)e(u)-Qe^{2}(u)\Bigr)
\end{multline}
and similar for $e_{\includegraphics[scale=0.035]{Jack2.eps}}(v)$ 
\begin{multline}
(u-v+Q)(u-v+Q+b)e_{\includegraphics[scale=0.035]{Jack2.eps}}(v)=
(u-v)(u-v+b)h^{-1}(u)e_{\includegraphics[scale=0.035]{Jack2.eps}}(v)h(u)+\\+
\bigl((u-v+Q)(u-v+Q+b)-(u-v)(u-v+b)\bigr)e_{\includegraphics[scale=0.035]{Jack2.eps}}(u)-
\frac{2b^{-1}Q(u-v)}{b-b^{-1}}\Bigl(e_{\includegraphics[scale=0.035]{Jack1.eps}}(u)-
e_{\includegraphics[scale=0.035]{Jack2.eps}}(u)\Bigr)-\\-
2ib^{-1}Q\Bigl((u-v+Q)e(v)e(u)-Qe^{2}(u)\Bigr)
\end{multline}
\subsection*{$ef$ relation} 
In order to obtain the relation \eqref{ef-relation}, we consider matrix element 
\begin{equation}\label{level-1-1-YB-algebra}
\begin{picture}(150,70)(400,15)
\Thicklines
\unitlength 7pt 
\put(57,0){\line(0,1){10}}
\put(57.5,0){\line(0,1){10}}
\put(58,0){\line(0,1){10}}
\put(58.5,0){\line(0,1){10}}
\put(59,0){\line(0,1){10}}
\put(55,8){\line(1,0){6}}
\put(55,2){\line(1,0){6}}
\put(61.15,1.73){$\scriptstyle{a_{-1}|u\rangle}$}
\put(61.15,7.73){$\scriptstyle{|v\rangle}$}
\put(55,2){\line(-2,3){4}}
\put(55,8){\line(-2,-3){4}}
\put(51,2){\line(-1,0){2}}
\put(51,8){\line(-1,0){2}}
\put(46.3,1.73){$\scriptstyle{\langle v|a_{1}}$}
\put(47.4,7.73){$\scriptstyle{\langle u|}$}
\put(65,4.55){$=$}
\put(72,0){\line(0,1){10}}
\put(72.5,0){\line(0,1){10}}
\put(73,0){\line(0,1){10}}
\put(73.5,0){\line(0,1){10}}
\put(74,0){\line(0,1){10}}
\put(70,8){\line(1,0){6}}
\put(70,2){\line(1,0){6}}
\put(67.3,1.73){$\scriptstyle{\langle v|a_{1}}$}
\put(68.4,7.73){$\scriptstyle{\langle u|}$}
\put(76,2){\line(2,3){4}}
\put(76,8){\line(2,-3){4}}
\put(80,2){\line(1,0){2}}
\put(80,8){\line(1,0){2}}
\put(82.15,1.73){$\scriptstyle{a_{-1}|u\rangle}$}
\put(82.15,7.73){$\scriptstyle{|v\rangle}$}
\end{picture}
\end{equation}
which reads explicitly as
\begin{equation}
\frac{u-v}{u-v+Q}\mathcal{L}_{\scriptscriptstyle{\varnothing,\Box}}(u)\mathcal{L}_{\scriptscriptstyle{\Box,\varnothing}}(v)+
\frac{Q}{u-v+Q}\mathcal{L}_{\scriptscriptstyle{\Box,\Box}}(u)h(v)=
\frac{u-v}{u-v+Q}\mathcal{L}_{\scriptscriptstyle{\Box,\varnothing}}(v)\mathcal{L}_{\scriptscriptstyle{\varnothing,\Box}}(u)+
\frac{Q}{u-v+Q}\mathcal{L}_{\scriptscriptstyle{\Box,\Box}}(v)h(u).\label{Lboxbox}
\end{equation}
Using \eqref{level-1-YB-algebra-II} and similar relation
\begin{equation}
\frac{u-v}{u-v+Q}h(u)\mathcal{L}_{\scriptscriptstyle{\Box,\varnothing}}(v)+
\frac{Q}{u-v+Q}\mathcal{L}_{\scriptscriptstyle{\Box,\varnothing}}(u)h(v)=\mathcal{L}_{\scriptscriptstyle{\Box,\varnothing}}(v)h(u) 
\end{equation}
nothing that
\begin{equation}
e(u+Q)=\mathcal{L}_{\scriptscriptstyle{\varnothing,\Box}}(u)h^{-1}(u),
\end{equation}
and multiplying by $h^{-1}(u)h^{-1}(v)$ from the right one obtains
\begin{multline}\label{ef-preliminary-equation}
e(u+Q)f(v)+\frac{Q}{u-v+Q}\left(\mathcal{L}_{\scriptscriptstyle{\Box,\Box}}(u)h^{-1}(u)-\mathcal{L}_{\scriptscriptstyle{\varnothing,\Box}}(u)h^{-1}(u)
\mathcal{L}_{\scriptscriptstyle{\Box,\varnothing}}(u)h^{-1}(u)\right)=\\=
f(v)e(u+Q)+\frac{Q}{u-v+Q}\left(\mathcal{L}_{\scriptscriptstyle{\Box,\Box}}(v)h^{-1}(v)-\mathcal{L}_{\scriptscriptstyle{\Box,\varnothing}}(v)h^{-1}(v)
\mathcal{L}_{\scriptscriptstyle{\varnothing,\Box}}(v)h^{-1}(v)\right).
\end{multline}
After shifting $u\rightarrow u-Q$, equation  \eqref{ef-preliminary-equation} reads 
\begin{equation}
 [e(u),f(v)]=-Q\frac{\psi(u)-\psi(v)}{u-v}\quad\text{where}\quad
 \psi(u+Q)=\mathcal{L}_{\scriptscriptstyle{\Box,\Box}}(u)h^{-1}(u)-\mathcal{L}_{\scriptscriptstyle{\varnothing,\Box}}(u)h^{-1}(u)
 \mathcal{L}_{\scriptscriptstyle{\Box,\varnothing}}(u)h^{-1}(u).
\end{equation}
\subsection*{Serre relations}
Formulas at level $3$ becomes tough, however they are straightforward. For example for $e(u)e_{\includegraphics[scale=0.035]{partition2.eps}}(v)$ we will have:
\begin{multline}\label{eEp}
e(u)e_{\includegraphics[scale=0.035]{partition2.eps}}(v)=\frac{\bar{g}(u-v)\bar{g}(u-v+b)}{g(u-v)g(u-v+b)}
e_{\includegraphics[scale=0.035]{partition2.eps}}(v)e(u)+
\frac{2Q}{(b-b^{-1})(Q+b)}\frac{\bar{g}(u-v)\bar{g}(u-v+b)}{g(u-v)g(u-v+b)}\times\\\times
e(v)\Big(\frac{1}{v-u-Q-b}e_{\includegraphics[scale=0.035]{partition1.eps}}(u)+\frac{1}{v-u}e_{\includegraphics[scale=0.035]{partition2.eps}}(u)+\frac{1}{v-u-b-b^{-1}}e_{\includegraphics[scale=0.035]{partition3.eps}}(u)\Big)+\text{locals}
\end{multline}
We will specify "locals" terms later (see \eqref{Level3-rels}), here we just want to point out that l.h.s by definition doesn't have any poles and so does the r.h.s. This condition will imply some additional relations, most of them will be non local, and we are not gonna discuss them. However we note that the multiplier
\begin{equation}
\frac{\bar{g}(u-v)\bar{g}(u-v+Q)}{g(u-v)g(u-v+Q)}
\end{equation}
doesn't have pole at $u=v+Q$.  Surprisingly, there is a "local" term with pole at this point.  Setting the residue to zero, one finds a relation
\begin{align}
&(1+b^2)(1+2b^2)e(u)e_{\includegraphics[scale=0.035]{partition2.eps}}(u)+Q(1+2b^2) e(u)e_{\includegraphics[scale=0.035]{partition3.eps}}(u) -b (b^4-b^2-4) e_{\includegraphics[scale=0.035]{partition2.eps}}(u)e(u)+\\
&+ 2b (1 +b^2) e_{\includegraphics[scale=0.035]{partition3.eps}}(u)e(u)+2b(1+b^2)e_{\includegraphics[scale=0.035]{partition1.eps}}(u)e(u)-(b^2+1)^2e_{\includegraphics[scale=0.035]{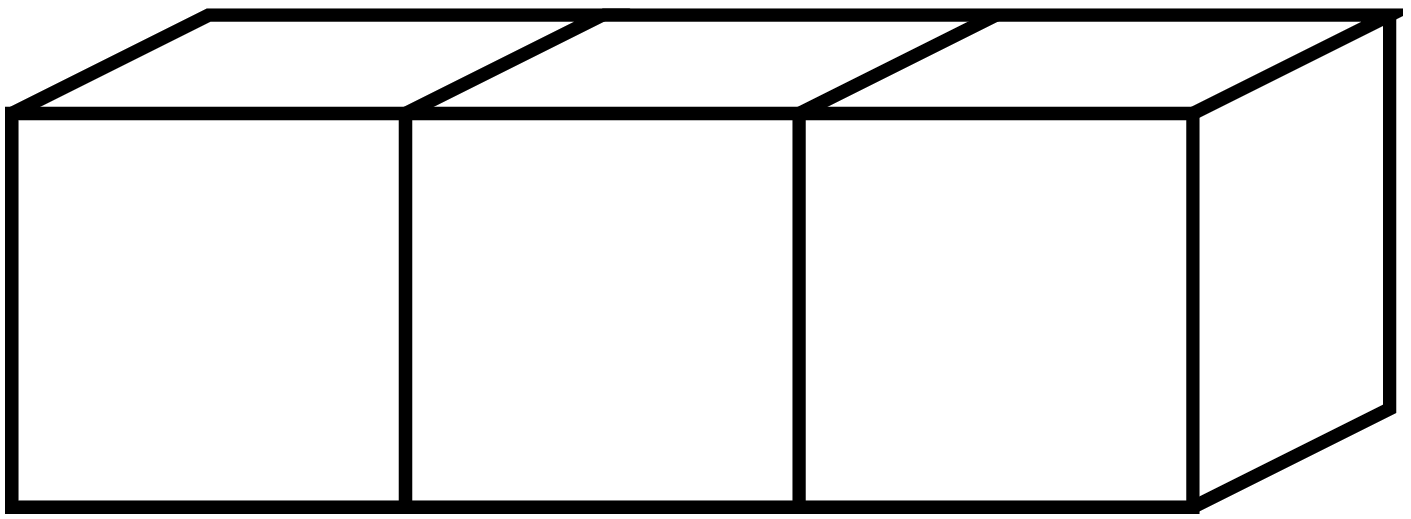}}(u))\overset{\text{def}}{=}e_{\includegraphics[scale=0.035]{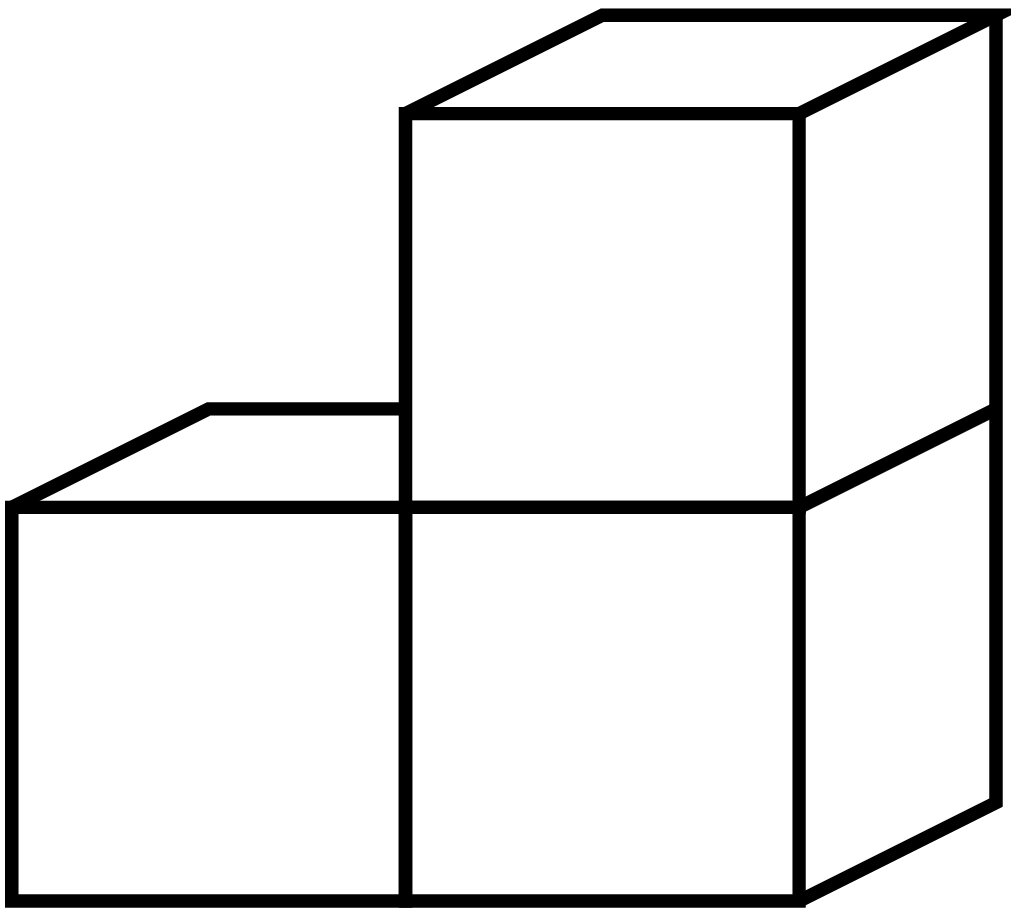}}(u)=0 \label{rels}
\end{align}
We will have similar relation for $e(u)e_{\includegraphics[scale=0.035]{partition3.eps}}(v)$. And also one trivial relation which follows from the fact that $e(u)e(u)^2=e(u)^2e(u)=e(u)^3$. As a result we will have 6 independent currents at level 3, which is equal to the number of 3d young diagrams with 3 boxes. In practice we used this three relations in order to exclude composite currents ($e_{\includegraphics[scale=0.035]{partition1.eps}}(u)e(u) , e_{\includegraphics[scale=0.035]{partition2.eps}}(u)e(u), e_{\includegraphics[scale=0.035]{partition3.eps}}(u)e(u)$). This three relations may look mysterious, however, after explicit calculation we found that this three relations are equivalent to Serre relations, the later could be written down in terms of currents as
\begin{equation}
\sum_{\sigma\in\mathbb{S}_{3}}(x_{\sigma_{1}}-2x_{\sigma_{2}}+x_{\sigma_{3}})e(x_{\sigma_{1}})e(x_{\sigma_{2}})e(x_{\sigma_{3}})+\sum_{\sigma\in\mathbb{S}_{3}}
[e(x_{\sigma_1}),e_{\includegraphics[scale=0.035]{partition1.eps}}(x_{\sigma_2})+e_{\includegraphics[scale=0.035]{partition2.eps}}(x_{\sigma_2})+e_{\includegraphics[scale=0.035]{partition3.eps}}(x_{\sigma_2})]=0  \label{Serre}
\end{equation}
Namely, using the quadratic relations, we may reorder any polynomial in $e(u_i)$ in a way that it will contain only monomials with ordered arguments: $e(u_{i_{1}})e(u_{i_{2}})...e(u_{i_{n}})$ , $i_1<i_2...<i_n$ , this could be done explicitly with the formulas \eqref{Level3-rels}. After doing this procedure with Serre relations we found that they proportional to linear combination of three currents, and so equal to zero in Yangian algebra.

Finally, after imposing three relations \eqref{rels} we will have explicitly: \begin{multline}\label{Level3-rels}
g(u-v)g(u-v+b)\Big[e(u)e_{\includegraphics[scale=0.035]{partition2.eps}}(v)\Big]_d=\bar{g}(u-v)\bar{g}(u-v+b)
\Big[e_{\includegraphics[scale=0.035]{partition2.eps}}(v)e(u)\Big]_d+\\+
\frac{2Q}{(b-b^{-1})(Q+b)}\left(\frac{\bar{g}(u-v)\bar{g}(u-v+b)}{v-u-Q-b}\Big[e(v)e_{\includegraphics[scale=0.035]{partition1.eps}}(u)\Big]_d+\frac{\bar{g}(u-v)\bar{g}(u-v+b)}{v-u}\Big[e(v)e_{\includegraphics[scale=0.035]{partition2.eps}}(u)\Big]_d+\right.\\+
\left.\frac{\bar{g}(u-v)\bar{g}(u-v+b)}{v-u-b-b^{-1}}\Big[e(v)e_{\includegraphics[scale=0.035]{partition3.eps}}(u)\Big]_d\right),
\end{multline}
where
\begin{align}
&\Big[e(u)e_{\includegraphics[scale=0.035]{partition2.eps}}(v)\Big]_d\overset{\text{def}}{=}e(u)e_{\includegraphics[scale=0.035]{partition2.eps}}(v)-\frac{1}{u-v+2b }e_{\includegraphics[scale=0.035]{partition7.eps}}(v)-\frac{b-2b^{-1}}{u-v+b^{-1} }e_{\includegraphics[scale=0.035]{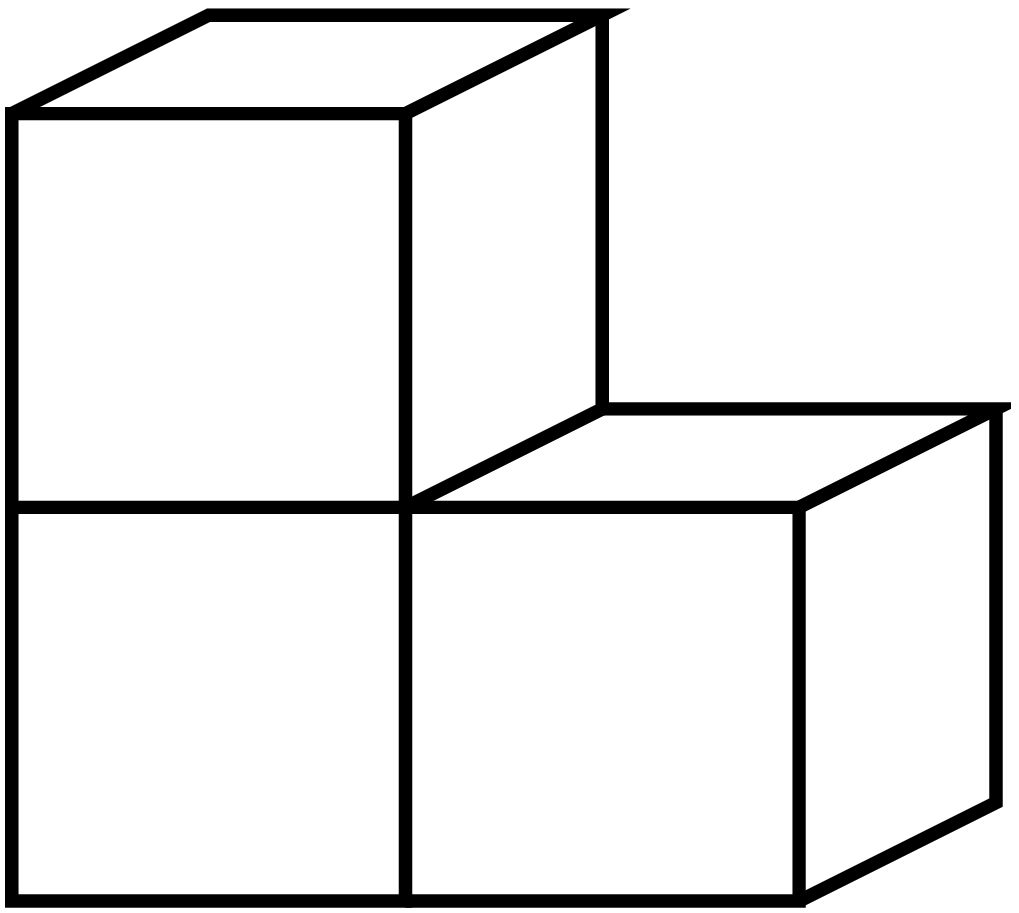}}(v)-\frac{b+2Q}{u-v-Q }e_{\includegraphics[scale=0.035]{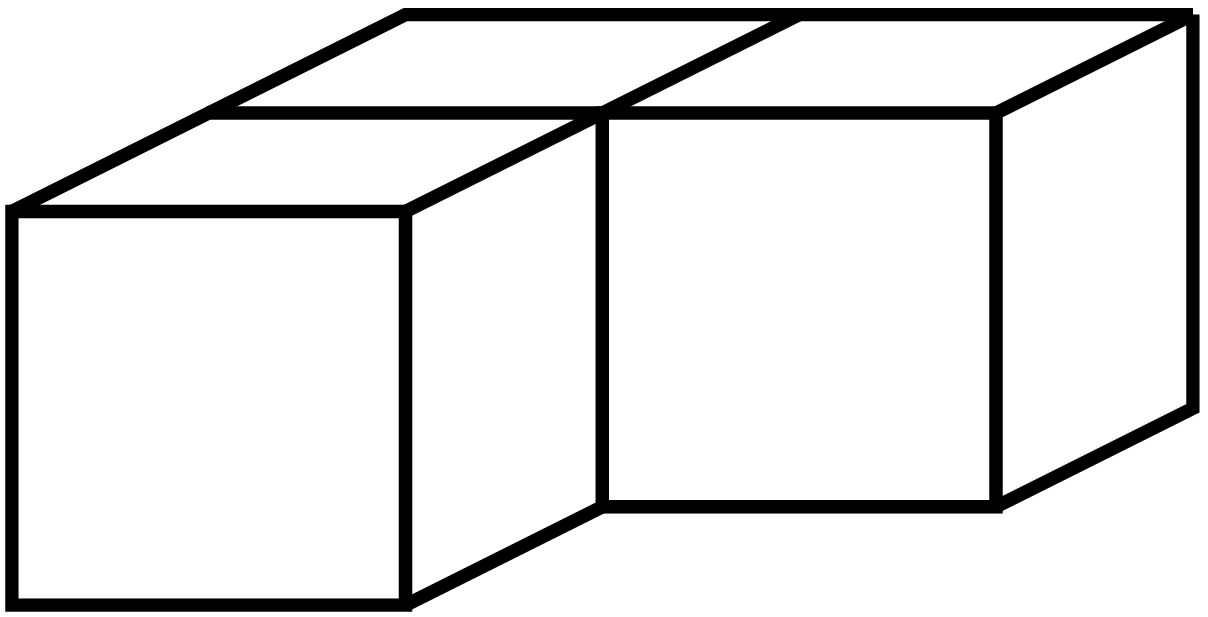}}(v),\\
&\Big[e(u)e_{\includegraphics[scale=0.035]{partition3.eps}}(v)\Big]_d\overset{\text{def}}{=}e(u)e_{\includegraphics[scale=0.035]{partition3.eps}}(v)-\frac{1}{u-v+2b^{-1} }e_{\includegraphics[scale=0.035]{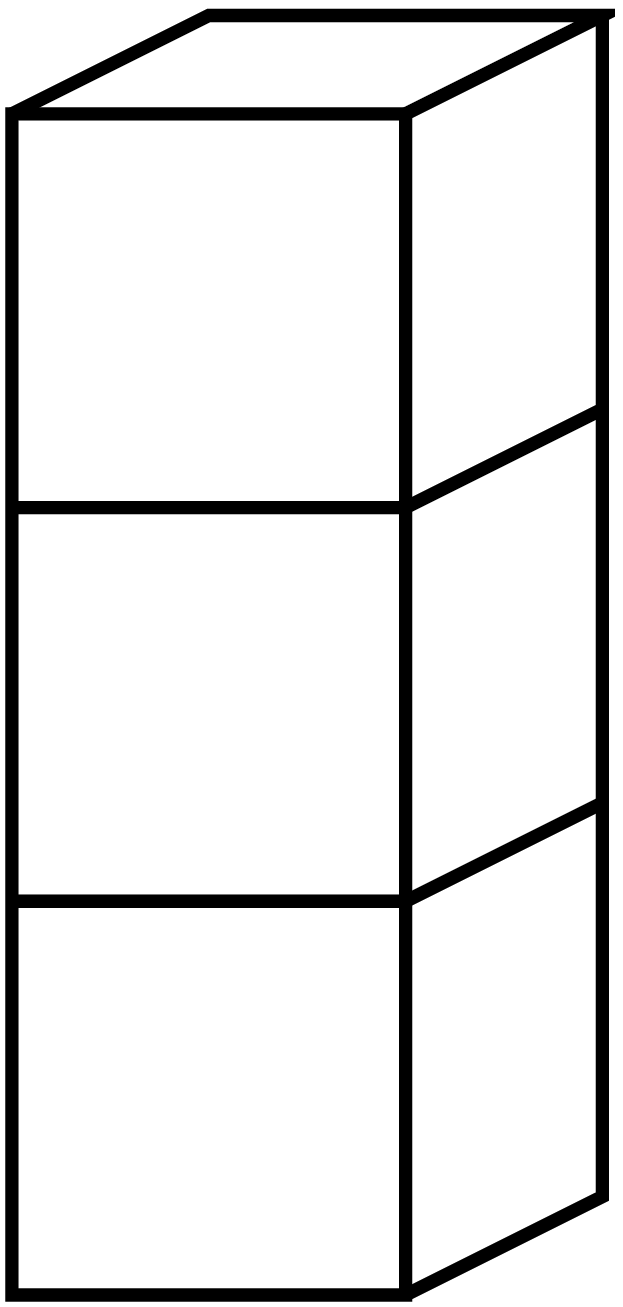}}(v)-\frac{2b-b^{-1}}{u-v+b }e_{\includegraphics[scale=0.035]{partition6.eps}}(v)-\frac{2b+b^{-1}}{u-v-Q }e_{\includegraphics[scale=0.035]{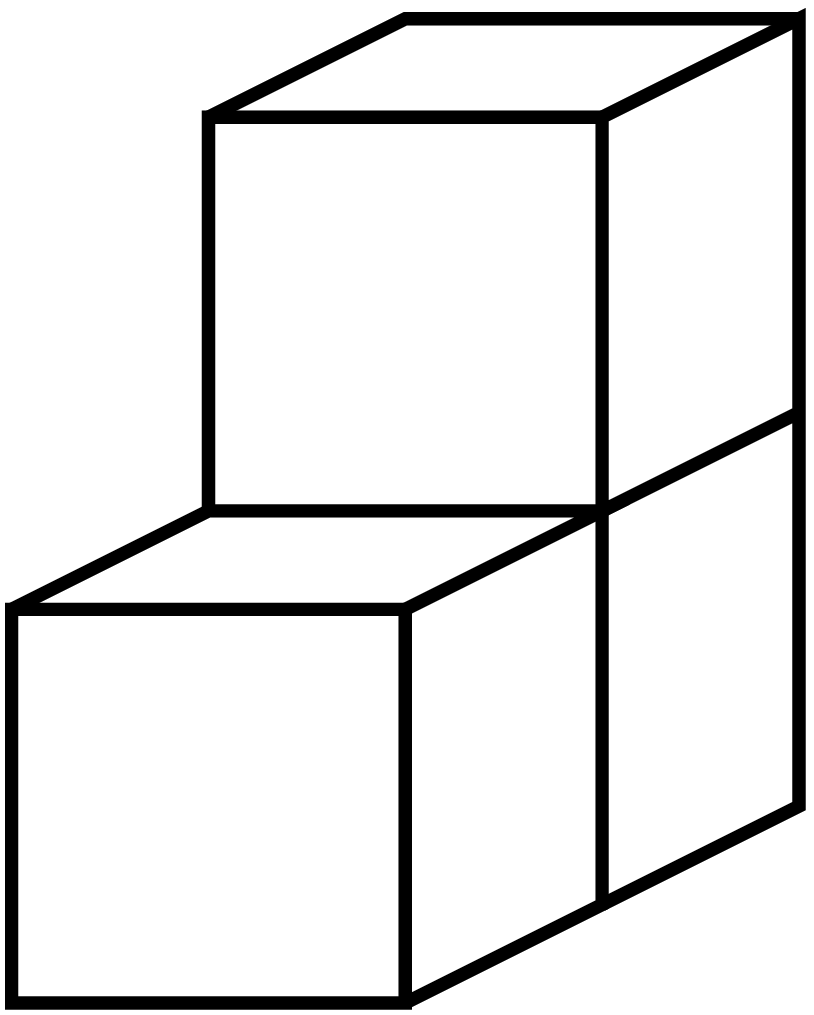}}(v),\\
&\Big[e(u)e_{\includegraphics[scale=0.035]{partition1.eps}}(v)\Big]_d\overset{\text{def}}{=}e(u)e_{\includegraphics[scale=0.035]{partition1.eps}}(v)-\frac{1}{u-v-2Q }e_{\includegraphics[scale=0.035]{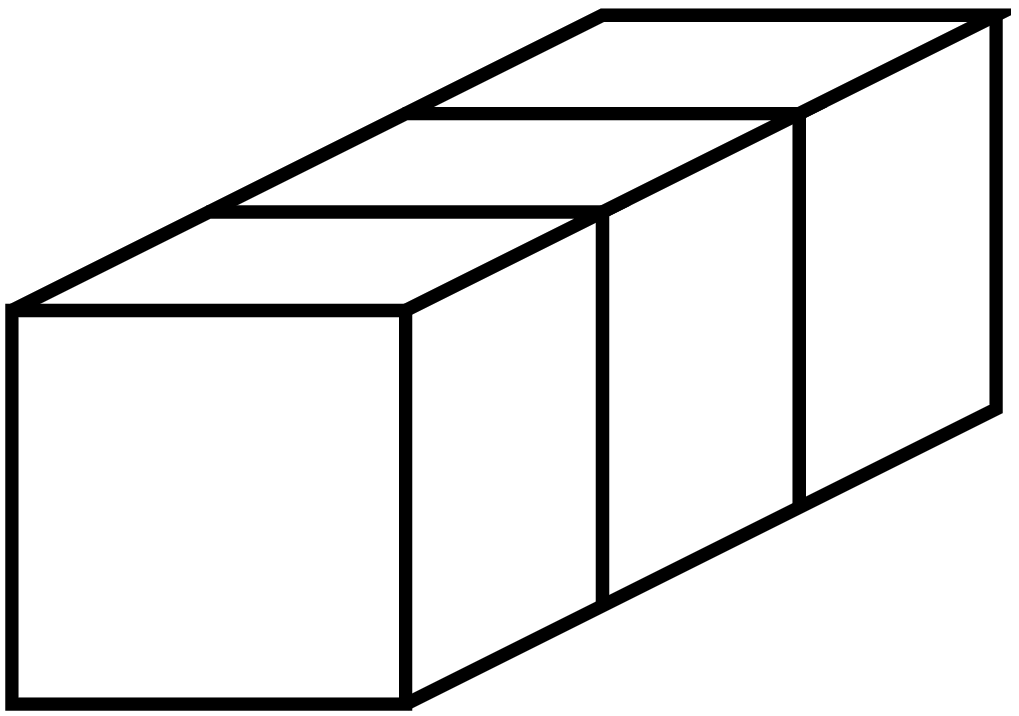}}(v)-\frac{Q+2b}{u-v+b }e_{\includegraphics[scale=0.035]{partition4.eps}}(v)-\frac{(Q+2b^{-1})(Q+b)}{u-v+b^{-1} }\frac{e_{\includegraphics[scale=0.035]{partition5.eps}}(v)}{2Q+b^{-1}},
\end{align}
and
\begin{multline}
\hspace*{-5pt}\Bigl[e_{\includegraphics[scale=0.035]{partition2.eps}}(v)e(u)\Bigr]_d\overset{\text{def}}{=}e_{\includegraphics[scale=0.035]{partition2.eps}}(v)e(u)-\frac{2}{(b-b^{-1})(b+Q)}
\frac{1}{u-v-2b^{-1}}e_{\includegraphics[scale=0.035]{partition8.eps}}(u)-\frac{2bQ}{Q+b}\frac{u-v-b+b^{-1}}{(u-v-b)(u-v-b^{-1})}e_{\includegraphics[scale=0.035]{partition6.eps}}(u)-\\
-\frac{2}{b-b^{-1}}\frac{u-v-Q-b}{(u-v-b)(u-v-Q)}e_{\includegraphics[scale=0.035]{partition4.eps}}(u)-\frac{2Q}{(b-b^{-1})(Q+b)}\frac{1}{(u-v-2 Q)}e_{\includegraphics[scale=0.035]{partition9.eps}}(u)+\\
+\frac{b^{2}}{(b-b^{-1})(b+Q)}\frac{2v-2u-4b+b^{-1}+b^{-3}}{(u-v-b)(u-v-2b^)}e_{\includegraphics[scale=0.035]{partition7.eps}}(u)-\frac{(Q+b^{-1})b^{-1}}{(2Q+b^{-1})(b-b^{-1})}\frac{2v-2u-3Qb^{-2}}{(u-v-b^{-1})(u-v+Q)}e_{\includegraphics[scale=0.035]{partition5.eps}}(u)
\end{multline}
with the higher currents given by
\begin{subequations}\label{level3-currents}
	\begin{equation}\label{level3-currents-1}	
	e_{\includegraphics[scale=0.035]{partition7.eps}}(v)=\frac{2Q^2}{(b-b^{-1})(2b-b^{-1})(Q+b)(Q+2b)}h^{-1}(v)\big(\mathcal{L}_{\scriptscriptstyle{\varnothing},\zerounderset{\scriptscriptstyle{\Box}}{\vspace*{-1.55pt}\zerounderset{\scriptscriptstyle{\Box}}{\vspace*{-1.55pt}\scriptscriptstyle{\Box}}}}(v)+3 i b \mathcal{L}_{\scriptscriptstyle{\varnothing},\hspace*{4pt}\zerounderset{\scriptscriptstyle{\Box}\hspace*{-1.3pt}\scriptscriptstyle{\Box}}{\vspace*{-1.5pt}\hspace*{-4.07pt}\scriptscriptstyle{\Box}}}\hspace*{3pt}(v)-2b^2 \mathcal{L}_{\scriptscriptstyle{\varnothing,\Box\hspace*{-1.2pt}\vspace*{2pt}\Box\hspace*{-1.2pt}\vspace*{2pt}\Box}}(v)\big),
	\end{equation}
	\begin{equation}\label{level3-currents-2}
	e_{\includegraphics[scale=0.035]{partition8.eps}}=\frac{2Q^2}{(b-b^{-1})(b-2b^{-1})(2b^{-1}+b)(3b^{-1}+b)}h^{-1}(v)\big(\mathcal{L}_{\scriptscriptstyle{\varnothing},\zerounderset{\scriptscriptstyle{\Box}}{\vspace*{-1.55pt}\zerounderset{\scriptscriptstyle{\Box}}{\vspace*{-1.55pt}\scriptscriptstyle{\Box}}}}(v)+3 i b^{-1} \mathcal{L}_{\scriptscriptstyle{\varnothing},\hspace*{4pt}\zerounderset{\scriptscriptstyle{\Box}\hspace*{-1.3pt}\scriptscriptstyle{\Box}}{\vspace*{-1.5pt}\hspace*{-4.07pt}\scriptscriptstyle{\Box}}}\hspace*{3pt}(v)-2b^{-2} \mathcal{L}_{\scriptscriptstyle{\varnothing,\Box\hspace*{-1.2pt}\vspace*{2pt}\Box\hspace*{-1.2pt}\vspace*{2pt}\Box}}(v)\big),
	\end{equation}
	\begin{multline}\label{level3-currents-3}
	e_{\includegraphics[scale=0.035]{partition9.eps}}=\frac{2Q^4}{(Q+b^{-1})(Q+b)}h^{-1}(v)\big(\mathcal{L}_{\scriptscriptstyle{\varnothing},\zerounderset{\scriptscriptstyle{\Box}}{\vspace*{-1.55pt}\zerounderset{\scriptscriptstyle{\Box}}{\vspace*{-1.55pt}\scriptscriptstyle{\Box}}}}(v)+\frac{12 i Q}{(2Q+b)(2Q+b^{-1})} \mathcal{L}_{\scriptscriptstyle{\varnothing},\hspace*{4pt}\zerounderset{\scriptscriptstyle{\Box}\hspace*{-1.3pt}\scriptscriptstyle{\Box}}{\vspace*{-1.5pt}\hspace*{-4.07pt}\scriptscriptstyle{\Box}}}\hspace*{3pt}(v)-\frac{4}{(2Q+b)(2Q+b^{-1})} \mathcal{L}_{\scriptscriptstyle{\varnothing,\Box\hspace*{-1.2pt}\vspace*{2pt}\Box\hspace*{-1.2pt}\vspace*{2pt}\Box}}(v)\big)-\\-\frac{2Q^2(Q+2b^{-1})}{2Q+b^{-1}}e(v)e_{\includegraphics[scale=0.035]{partition3.eps}}(v)-\frac{2Q^2(Q+2b)}{2Q+b}e(v)e_{\includegraphics[scale=0.035]{partition2.eps}}(v)-2Qe(v)e_{\includegraphics[scale=0.035]{partition1.eps}}(v),
	\end{multline}
	\begin{equation}\label{level3-currents-4}
	e_{\includegraphics[scale=0.035]{partition6.eps}}=-\frac{2Q^2}{(b-b^{-1})(b-2b^{-1})(Q+b^{-1})(2b-b^{-1})(Q+b)}h^{-1}(v)\big(\mathcal{L}_{\scriptscriptstyle{\varnothing},\zerounderset{\scriptscriptstyle{\Box}}{\vspace*{-1.55pt}\zerounderset{\scriptscriptstyle{\Box}}{\vspace*{-1.55pt}\scriptscriptstyle{\Box}}}}(v)+ i Q \mathcal{L}_{\scriptscriptstyle{\varnothing},\hspace*{4pt}\zerounderset{\scriptscriptstyle{\Box}\hspace*{-1.3pt}\scriptscriptstyle{\Box}}{\vspace*{-1.5pt}\hspace*{-4.07pt}\scriptscriptstyle{\Box}}}\hspace*{3pt}(v)- \mathcal{L}_{\scriptscriptstyle{\varnothing,\Box\hspace*{-1.2pt}\vspace*{2pt}\Box\hspace*{-1.2pt}\vspace*{2pt}\Box}}(v)\big),
	\end{equation}
	\begin{multline}\label{level3-currents-5}
	e_{\includegraphics[scale=0.035]{partition5.eps}}=\frac{Q^3b^{-1}}{(b-b^{-1})(Q+b)^2(Q+b^{-1})(Q+2b^{-1})}\times\\\times
	h^{-1}(v)\Big((2Q+b^{-1})\mathcal{L}_{\scriptscriptstyle{\varnothing},\zerounderset{\scriptscriptstyle{\Box}}{\vspace*{-1.55pt}\zerounderset{\scriptscriptstyle{\Box}}{\vspace*{-1.55pt}\scriptscriptstyle{\Box}}}}(v)+ i (3Q+b)b^{-1} \mathcal{L}_{\scriptscriptstyle{\varnothing},\hspace*{4pt}\zerounderset{\scriptscriptstyle{\Box}\hspace*{-1.3pt}\scriptscriptstyle{\Box}}{\vspace*{-1.5pt}\hspace*{-4.07pt}\scriptscriptstyle{\Box}}}\hspace*{3pt}(v)-2b^{-1} \mathcal{L}_{\scriptscriptstyle{\varnothing,\Box\hspace*{-1.2pt}\vspace*{2pt}\Box\hspace*{-1.2pt}\vspace*{2pt}\Box}}(v)\Big)-\frac{Q}{Q+b^{-1}}e(v)e_{\includegraphics[scale=0.035]{partition3.eps}}(v),
	\end{multline}
	and
	\begin{multline}\label{level3-currents-6}
	e_{\includegraphics[scale=0.035]{partition4.eps}}=\frac{Q^3b}{(b-b^{-1})(Q+b)(Q+b^{-1})(Q+2b)(2Q+b)}\times\\\times
	h^{-1}(v)\big((2Q+b)\mathcal{L}_{\scriptscriptstyle{\varnothing},\zerounderset{\scriptscriptstyle{\Box}}{\vspace*{-1.55pt}\zerounderset{\scriptscriptstyle{\Box}}{\vspace*{-1.55pt}\scriptscriptstyle{\Box}}}}(v)+ i (3Q+b^{-1})b \mathcal{L}_{\scriptscriptstyle{\varnothing},\hspace*{4pt}\zerounderset{\scriptscriptstyle{\Box}\hspace*{-1.3pt}\scriptscriptstyle{\Box}}{\vspace*{-1.5pt}\hspace*{-4.07pt}\scriptscriptstyle{\Box}}}\hspace*{3pt}(v)-2b \mathcal{L}_{\scriptscriptstyle{\varnothing,\Box\hspace*{-1.2pt}\vspace*{2pt}\Box\hspace*{-1.2pt}\vspace*{2pt}\Box}}(v)\big)-\frac{Q}{2Q+b}e(v)e_{\includegraphics[scale=0.035]{partition2.eps}}(v),
	\end{multline}
\end{subequations}
In principle we may go further, and calculate quadratic relations at next levels, however as we already have shown, the algebra is generated by the $h(u)$, $e(u)$ and $f(u)$ currents, so in principle we don't need to use higher currents. The only problem is to prove that quadratic and Serre relations are the only ones which currents $e(u)$ obeys.
\subsection*{Relations in $\epsilon$ notations}
We see that there is an $S_{3}$ symmetry associated to permutation of the triple $(b,b^{-1},-Q)$. In fact it is more convenient to go to epsilon notations:
\begin{equation}\label{epsilonNotation-appendix}
b= \frac{\epsilon_{1}}{\sqrt{\epsilon_1 \epsilon_2}},\quad b^{-1}= \frac{\epsilon_{2}}{\sqrt{\epsilon_1 \epsilon_2}},\quad Q=-\frac{\epsilon_{3}}{\sqrt{\epsilon_1 \epsilon_2}}\implies\epsilon_{1}+\epsilon_{2}+\epsilon_{3}=0.
\end{equation}
It is also convenient to change a normalization of the highest weight/spectral parameters, together with the whole normalization of basic fields:
\begin{equation}\label{uepsilonNotation-appendix}
 \varphi(x)\to \phi(x)= -i \frac{\varphi(x)}{\sqrt{\epsilon_1\epsilon_2}}. 
\end{equation}
Then the relation \eqref{ee-Yangian-relation-3D-basis} takes apparently symmetric form
\begin{multline}
g(u-v)\Bigl[e(u)e(v)+\frac{e_{\includegraphics[scale=0.035]{partition2.eps}}(v)}{u-v+\epsilon_{1}}+
\frac{e_{\includegraphics[scale=0.035]{partition3.eps}}(v)}{u-v+\epsilon_{2}}+\frac{e_{\includegraphics[scale=0.035]{partition1.eps}}(v)}{u-v+\epsilon_{3}}\Bigr]=\\=
\bar{g}(u-v)\Bigl[e(v)e(u)+\frac{e_{\includegraphics[scale=0.035]{partition2.eps}}(u)}{u-v-\epsilon_{1}}+
\frac{e_{\includegraphics[scale=0.035]{partition3.eps}}(u)}{u-v-\epsilon_{2}}+\frac{e_{\includegraphics[scale=0.035]{partition1.eps}}(u)}{u-v-\epsilon_{3}}\Bigr],
\end{multline}
where 
\begin{equation}
g(x)=(x+\epsilon_{1})(x+\epsilon_{2})(x+\epsilon_{3}),\quad
\bar{g}(x)=(x-\epsilon_{1})(x-\epsilon_{2})(x-\epsilon_{3}).
\end{equation}

Our conventions about relation between $\epsilon$ and $b,Q$ notations are summarized in a table
\begin{center}
 \begin{tabular}{c|c|c|c}
	 & \text{fields normalisation} & \text{Current} $e$ & commutator $[e,f]$\\
	 $b$ \text{notations} & $\partial\varphi(x)\partial\varphi(y)=-\frac{1}{\sin^2(x-y)}+\text{reg}$ & $e(u)=h^{-1}(u)\mathcal{L}_{\scriptscriptstyle{\varnothing,\Box}}(u)$ &  $[e(u_1),f(u_2)]=-Q\frac{\psi(u_1)-\psi(u_2)}{u_1-u_2}$ \\
	 $\epsilon$ \text{notations} &  $\partial\phi(x)\partial\phi(y)= \frac{\epsilon_1\epsilon_2}{\sin^2(x-y)}+\text{reg}$& $e(v)=\sqrt{\epsilon_3}h^{-1}(v)\mathcal{L}_{\scriptscriptstyle{\varnothing,\Box}}(v)$ & $[e(v_1),f(v_2)]=\frac{\psi(v_1)-\psi(v_2)}{v_1-v_2}$
 \end{tabular}
\end{center}
In definition of matrix elements $\mathcal{L}_{\boldsymbol{\lambda},\boldsymbol{\mu}}(u)$ we define the state $|\square\rangle$, as well as any state $|\boldsymbol{\lambda}\rangle$ to be normalized as $\langle\boldsymbol{\lambda}|\boldsymbol{\lambda}\rangle=1$ in any notation.

Note that here we used a Maulik-Okounkov $\mathcal{R}$ matrix, which breaks the symmetry between $\epsilon_1,\epsilon_2,\epsilon_3$, so that we have a selected  $\epsilon_3$. In fact, there exist additional $\mathcal{R}_{f}^{(1/2)}$ matrices with either $\epsilon_1$ or $\epsilon_2$ selected (see appendix \ref{SUSY-ILW}).
\section{Yangian Double and higher relations}\label{Yangian-double}
In section \ref{comrel} we have seen that $e(z)$ currents obeys simple commutation relations \eqref{ee-exact-relation}, except some local terms
\begin{equation} \label{exc}
g(u-v) e(u)e(v) \sim \bar{g}(u-v)e(v)e(u)
\end{equation}
Where $\sim$ means equality up to "local" terms, see discussion in section \ref{comrel}. 

It may be convenient to enlarge the algebra of currents, by introducing negative modes, so that relation \eqref{exc} will hold exactly. Formally this may be achieved by introducing Yangian Double $\textrm{DY}(\widehat{\mathfrak{gl}}(1))$.

Let us introduce $RLL$ realization of the Yangian Double, we have two types of generators
\begin{equation}
  \mathcal{L}^{\pm}_{\boldsymbol{\lambda},\boldsymbol{\mu}}(u),
\end{equation}
which are considered as a series at $u=\infty$ or $u=0$ correspondingly. These operators satisfy $RLL=LLR$ quadratic relations
\begin{eqnarray}
\mathcal{R}_{ij}(u-v)\mathcal{L}_i^{\pm}(u)\mathcal{L}_j^{\pm}(v)=\mathcal{L}_i^{\pm}(v)\mathcal{L}_j^{\pm}(u)\mathcal{R}_{ij}(u-v), \label{RLL1}\\
\mathcal{R}_{ij}(u-v)\mathcal{L}_i^{+}(u)\mathcal{L}_j^{-}(v)=\mathcal{L}_i^{-}(v)\mathcal{L}_j^{+}(u)\mathcal{R}_{ij}(u-v), \label{RLL2}
\end{eqnarray}
where in the second equation the $R-$matrix $\mathcal{R}_{ij}(u-v)$ is understood as a series in $\frac{u}{v}$\footnote{This definition is inspired by the one from $\mathfrak{q}$-deformed algebra (see \cite{ding1993})}.

Let us define \emph{total} current $e(u), f(u)$ currents as
\begin{equation}
  e(z)=e^{+}(z)-e^{-}(z),\qquad   f(z)=f^{+}(z)-f^{-}(z),
\end{equation}
where
\begin{equation}
 h^{\pm}(u)=\mathcal{L}^{\pm}_{\scriptscriptstyle{\varnothing,\varnothing}}(u),\quad
 e^{\pm}(u)=(h^{\pm}(u))^{-1}\mathcal{L}^{\pm}_{\scriptscriptstyle{\varnothing,\Box}}(u),\quad
 f^{\pm}(u)=\mathcal{L}^{\pm}_{\scriptscriptstyle{\Box,\varnothing}}(u)(h^{\pm}(u))^{-1}.
\end{equation}
It is easy to check that the local terms cancel in commutation relations for higher currents
\begin{equation}
\begin{aligned}
&[h^{\pm}(u),h^{\pm}(v)]=[h^{\pm}(u),\psi^{\pm}(v)]=[\psi^{\pm}(u),\psi^{\pm}(v)]=0
\\
&(u-v+\epsilon_3)h^{\pm}(u)e(v)=(u-v)e(v)h^{\pm}(u)\\
&(u-v+\epsilon_3)f(v)h^{\pm}(u)=(u-v)h^{\pm}(u)f(v)h^{\pm}(u)\\
&[e^{\pm}(u),f^{\pm}(v)]=\frac{\psi^{\pm}(u)-\psi^{\pm}(v)}{u-v}\\
&\prod\limits_{\alpha=1}^{3}(u-v+\epsilon_\alpha)\psi^+(u)e(v)=\prod\limits_{\alpha=1}^{3}(u-v-\epsilon_\alpha)e(v)\psi^+(u)\\
&\prod\limits_{\alpha=1}^{3}(u-v+\epsilon_\alpha)e(v)\psi^+(u)=\prod\limits_{\alpha=1}^{3}(u-v-\epsilon_\alpha)\psi^+(u)e(v)  
\end{aligned}
\end{equation}
and
\begin{equation}
\begin{aligned}
&[e^{\pm}(u),f^{\pm}(v)]=\frac{\psi^{\pm}(u)-\psi^{\pm}(v)}{u-v}\\
&e(u)e(v)=\frac{(u-v-\epsilon_1)(u-v-\epsilon_2)(u-v-\epsilon_3)}{(u-v+\epsilon_1)(u-v+\epsilon_2)(u-v+\epsilon_3)}e(v)e(u),\\
&f(u)f(v)=\frac{(u-v+\epsilon_1)(u-v+\epsilon_2)(u-v+\epsilon_3)}{(u-v-\epsilon_1)(u-v-\epsilon_2)(u-v-\epsilon_3)}f(v)f(u)
\end{aligned}
\end{equation}
The price for this relations is that product of two currents $e(u)e(v)$ has poles at points $v=u+\epsilon_{\alpha}$ and similar for $f(u)f(v)$.

Yangian Double may be useful for understanding the structure of the relations \eqref{Level3-rels} between higher currents.
Let us, for example, derive the relation between  $e(u)$ and the higher current $e_{\includegraphics[scale=0.035]{partition2.eps}}(v)$.  First of all we have
\begin{equation}
g(u-v)\Bigl[e^+(u)e(v)-\frac{e_{\includegraphics[scale=0.035]{partition2.eps}}(v)}{u-v+\epsilon_1}-
\frac{e_{\includegraphics[scale=0.035]{partition3.eps}}(v)}{u-v+\epsilon_2}-\frac{e_{\includegraphics[scale=0.035]{partition1.eps}}(v)}{u-v+\epsilon_3}\Bigr]=
\bar{g}(u-v)e(v)e^+(u). \label{first}
\end{equation}
So that
\begin{equation}\label{e2}
e_{\includegraphics[scale=0.035]{partition2.eps}}(v)=\textrm{res}_{u=v-\epsilon_1}\left(e^+(u)e(v)\right)-\frac{2\epsilon_1\epsilon_2\epsilon_3}{(\epsilon_1-\epsilon_2)(\epsilon_1-\epsilon_3)}e(v)e^+(v-\epsilon_1)
\end{equation}
Now let us find an exchange relation between $e(\xi)$ and $e_{\includegraphics[scale=0.035]{partition2.eps}}(v)$. Using \eqref{first}, one finds
\begin{multline}
e(\xi)e(v)e^+(v-\epsilon_1)=\frac{\bar{g}(\xi-v)}{g(\xi-v)}e(v)e(\xi)e^+(v-\epsilon_1)=\frac{\bar{g}(\xi-v)\bar{g}(\xi-v+\epsilon_1)}{g(u-v)g(u-v+\epsilon_1)}e(v)\times\\\times
\Big(e^+(v-\epsilon_1)e(\xi)-\frac{1}{v-\xi}e_{\includegraphics[scale=0.035]{partition2.eps}}(\xi)-\frac{1}{v-\xi+\epsilon_2-\epsilon_1}e_{\includegraphics[scale=0.035]{partition3.eps}}(\xi)-\frac{1}{v-\xi+\epsilon_3-\epsilon_1}e_{\includegraphics[scale=0.035]{partition1.eps}}(\xi)\Big)
\end{multline}
Exchange relation with the first term of \eqref{e2} is simple, because local terms doesn't contribute to the residue
\begin{eqnarray}
e(\xi)\textrm{res}_{u=v-b}\left(e^+(u)e(v)\right)=\textrm{res}_{u=v-b}\left(e^+(u)e(v)\right)e(\xi)\frac{\bar{g}(\xi-v)\bar{g}(\xi-v+\epsilon_3)}{g(\xi-v)g(\xi-v+\epsilon_3)}
\end{eqnarray}
Combining this together, we find the relation
\begin{multline}\label{Level3-rels-total-currents}
 e(\xi)e_{\includegraphics[scale=0.035]{partition2.eps}}(v)=\frac{\bar{g}(\xi-v)\bar{g}(\xi-v+\epsilon_1)}{g(\xi-v)g(\xi-v+\epsilon_1)}
 e_{\includegraphics[scale=0.035]{partition2.eps}}(v)e(\xi)+\frac{2\epsilon_1\epsilon_2\epsilon_3}{(\epsilon_1-\epsilon_2)(\epsilon_1-\epsilon_3)}\frac{\bar{g}(\xi-v)\bar{g}(\xi-v+\epsilon_{1})}{g(\xi-v)g(\xi-v+\epsilon_{1})}e(v)\times\\\times
 \left(\frac{1}{v-\xi}e_{\includegraphics[scale=0.035]{partition2.eps}}(\xi)+\frac{1}{v-\xi+\epsilon_2-\epsilon_1}e_{\includegraphics[scale=0.035]{partition3.eps}}(\xi)+\frac{1}{v-u+\epsilon_3-\epsilon_1}e_{\includegraphics[scale=0.035]{partition1.eps}}(\xi)\right),
\end{multline}
which reproduces non-local part of the relation between the half currents \eqref{Level3-rels}. We note that \eqref{Level3-rels}  contains more information. In particular they contain cubic Serre relation \eqref{Serre-relations}. 
\section{Special vector \texorpdfstring{$|\chi\rangle$}{chi} and shuffle functions}\label{VectorChi}
In the later we will need a more detailed description of a subalgebra  $\mathfrak{n}^{+}$ generated by currents $f(z)$. Easy to understand that the subspace of the form $\mathcal{L}_{\mu,\varnothing_1}(v_1)\dots \mathcal{L}_{\mu,\varnothing_n}(v_n)$ may be identified with the subspace :$\mathfrak{n}^+(\boldsymbol{v})=h(v_1)\dots h(v_n) \mathfrak{n}^+$. A particular result of this section is an explicit realization of this mapping.
	
First of all let us note that both spaces are graded by the number of $f(\xi_i)$ currents in the monomial, let us note each graded component of corresponding algebras by $\mathfrak{n}^+_N$, $\mathfrak{n}^+_N(\boldsymbol{v})$.
	
It is a natural idea to identify elements of $\mathfrak{n}^+$ and $\mathfrak{n}^+(\boldsymbol{v})$ by their matrix elements in some representation:
\begin{eqnarray}
\mathfrak{n}^+_N \to \langle \varnothing|\mathfrak{n}^+_N|\chi\rangle \label{shuffle}
\end{eqnarray}
In order to unambiguously characterize the elements of $\mathfrak{n}^+_N$, $\mathfrak{n}^+_N(\boldsymbol{v})$ we need a big enough set of representations and vectors $|\chi\rangle$. Our choice is the following: let us pick an $N$ Fock spaces: $\mathcal{F}_{x_1}\otimes\dots \mathcal{F}_{x_N}$, and consider simplest vector of grade $N$ : 
\begin{equation}
|\chi\rangle_{\scriptscriptstyle{\boldsymbol{x}}}\overset{\text{def}}{=}|\underbrace{\Box,\dots,\Box}_N\rangle =\lim\limits_{\xi_i\to x_i}\prod\limits_{i,k} \frac{\xi_i-x_k}{\xi_i-x_k-\epsilon_3} \prod\limits_{i<j}S(\xi_i-\xi_j)e(\xi_N)...e(\xi_1)|0\rangle 
\end{equation}
Then, our mapping \eqref{shuffle} maps an element of $\mathfrak{n}^+_N$, $\mathfrak{n}^+_N(\boldsymbol{v})$ to a rational function of $N$ variables $f(x_1,\dots, x_N)$ obeying the so called "wheel" condition \cite{Feigin:2015raa}:
\begin{eqnarray}
f(x_1,x_1+\epsilon_i,x_1+\epsilon_i+\epsilon_j,x_4,\dots)=0
\end{eqnarray}
For $\mathfrak{n}_N^+$ and additional condition:
\begin{eqnarray}
f(v,v+\epsilon_3,x_3,\dots)=0
\end{eqnarray}
For $\mathfrak{n}^+_N(\boldsymbol{v})$. This functions is a rational limits of $Sh_0$ and $Sh_1$ functions from \cite{Feigin:2015raa}. The multiplication in algebra, implies the multiplication of Shuffle functions
\begin{align}\label{ShuffleMap}
S_0: \mathfrak{n}^+_N &\times \mathfrak{n}^+_M \to \mathfrak{n}^+_{N+M}\\
f(\boldsymbol{x})\star g(\boldsymbol{y})&\equiv \textrm{Sym}_{x,y}\Bigl(f(\boldsymbol{x})g(\boldsymbol{y})\prod\limits_{i,j}S(x_i-y_j)\Bigr) 
\end{align}
For $\mathfrak{n}^+$, And
\begin{align}
S_1: \mathfrak{n}^+_N(\boldsymbol{v}) &\times \mathfrak{n}^+_M(\boldsymbol{u}) \to \mathfrak{n}^+_{N+M}(\boldsymbol{u,v})\\
f(\boldsymbol{x})\star g(\boldsymbol{y}) &\equiv \textrm{Sym}_{x,y}\Big(f(\boldsymbol{x})g(\boldsymbol{y})\prod\limits_{n,i} \frac{u_n-x_i}{u_n-x_i-\epsilon_3} \prod_{i,j} S(x_{i}-y_j) \Big) 
\end{align}
For $\mathfrak{n}^+(\boldsymbol{v})$.

Let us introduce, $W^{(1)}(z)$ current $J_n$
\begin{align}
 &\langle\varnothing|\mathcal{L}(u) \  a^{(0)}_{-n}|\varnothing\rangle=\frac{J_n}{u}+o\left(\frac{1}{u^2}\right) \quad , n>0 \label{Jn+}\\
 &\langle\varnothing|a^{(0)}_{n}\ \mathcal{L}(u) |\varnothing\rangle=\frac{J_{-n}}{u}+o\left(\frac{1}{u^2}\right) \quad , n>0 \label{Jn-}
\end{align}
It is clear from the $\mathcal{R}\mathcal{L}\mathcal{L}$ relation that $R(u)$ matrix commute with $W^{(1)}$ current:
\begin{eqnarray}
(a_n^{(0)}+J_n)R^{0,v}=R^{0,v}(a_n^{(0)}+J_n)
\end{eqnarray}
Taking the matrix element over the auxiliary space $\langle\varnothing|\dots |\mu \rangle$  for positive $n$ we will get:
\begin{eqnarray}\label{Ad}
[\mathcal{L}_{\boldsymbol{\mu},\varnothing}(u),J_n]=\mathcal{L}_{\boldsymbol{\mu}+n,\varnothing}(u), 
\end{eqnarray}
where $\langle\boldsymbol{\mu}+n| \overset{\text{def}}{=} \langle \boldsymbol{\mu}|a_n$.
	
It is also clear, that $J_n$ for $n>0$ belongs to the sub-algebra $\mathfrak{n}^+$. Indeed, explicit calculation of the large $u$ limit of $R(u)$ matrix \eqref{R-expansion} shows that:
\begin{align} \label{Jef}
 J_1=f_0 &\quad J_{-1}=e_0,\\
	J_{n+1}=[f_1,J_n] &\quad J_{-n-1}=[e_1,J_{-n}].
\end{align}
Then we get:
\begin{align}
 J_{k}^{\boldsymbol{x}}=\oint\dots\oint g_k(\boldsymbol{\xi})f(\xi_1)...f(\xi_k)d\boldsymbol{\xi}\quad\text{with} \label{JSh},\\
 g_{n+1}(\vec{\xi})=\xi_1g_n(x_2\dots \xi_{n+1})-g_n(\xi_1\dots \xi_n)\xi_{n+1},
\end{align}
and
\begin{eqnarray}
g_{n}(\boldsymbol{\xi})=\prod\limits_i \xi_i\left(\sum (-1)^iC_n^i \xi^{-1}_i\right),
\end{eqnarray}
where $C_n^i$ are the binomial coefficients. 
	
Note that the function $g(\boldsymbol{\xi})$ defined ambiguously, indeed algebra $Y(\widehat{\mathfrak{gl}}(1))$ enjoys Serre relations \eqref{Serre}
\begin{eqnarray}
 \textrm{Sym}_{i,j,k}[f_i,[f_j[f_{k+1}]]]=0
\end{eqnarray}
Indeed such an element lies in the kernel of the Shuffle map \eqref{ShuffleMap}
\begin{eqnarray}
 \textrm{Sym}_{i,j,k} \Big(\xi_1^i\xi_2^k\xi_3^k(\xi_1-2\xi_2+\xi_3) S(\xi_1-\xi_2)S(\xi_1-\xi_3)S(\xi_2-\xi_3)\Big)=0
\end{eqnarray}
In particular, commutativity of $J_n$ may be thought as a consequence of Serre relation, for example choosing $i=j=k=0$
\begin{eqnarray}
 [J_1,J_2]=[f_0,[f_1,f_0]]\overset{\text{Serre}}{=}0
\end{eqnarray}
We should consider functions $g_n(\boldsymbol{\xi})$ modulo equivalence:
\begin{eqnarray} \label{kerS} 
 g^{(1)}_n(\boldsymbol{\xi}) \sim g^{(2)}_n(\boldsymbol{\xi})+\text{Ker}(S_0) 
\end{eqnarray}
It is easy to understand that modulo this equivalence function $g_n(\boldsymbol{\xi})$ is invariant under the simultaneous shift of all variables $\xi\to \xi+\hbar$ we will use this fact in section \ref{OPe}.
	
As we announced, operators  $\mathcal{L}(u)_{\mu,\varnothing}$ belongs to the subspace $\mathfrak{n}^+(u)_{|\mu|}$:
\begin{equation}\label{LAsSh}
 \mathcal{L}_{\boldsymbol{\lambda},\varnothing}(u)=\frac{1}{(2\pi i)^{|\boldsymbol{\lambda}|}}\oint\dots\oint F_{{\lambda}}(\boldsymbol{z})\,h(u)f(z_{|\boldsymbol{\lambda}|})\dots f(z_1)dz_1\dots dz_{|\boldsymbol{\lambda}|}, g
\end{equation} 
where contours go clockwise around $\infty$ and all poles of $F_{{\lambda}}(\boldsymbol{z})$. 
	
Let us prove this statement, and find recurrence relations for the rational function $F_{\boldsymbol{\lambda}}(\boldsymbol{z})$. Now in order to recover formula \eqref{LAsSh} we have to use relation \eqref{Ad} together with the formula \eqref{JSh}. In order to reproduce \eqref{LAsSh} we have to reorder $h$ and $f$ current, in order to move $h$ to the left, this can be done with the simple fact
\begin{align}
 \oint\limits_{\infty}\xi^nf(\xi)h(u)\frac{d\xi}{2\pi i}=\oint\limits_{\infty}\Big[ \frac{(u-\xi)}{(u-\xi-\epsilon_{3})}h(u)f(\xi)-\frac{\epsilon_3}{(u-\xi-\epsilon_{3})} f(u)h(u)\Big]\xi^n\frac{d\xi}{2\pi i}=\\=\oint\limits_{\infty+\{u-\epsilon_3\}} \frac{(u-\xi)}{(u-\xi-\epsilon_{3})}h(u)f(\xi)\xi^n\frac{d\xi}{2\pi i} \label{fh}
\end{align}
Here in the first equality we used equation \eqref{hf-relation}, while in the second we used a simple fact that l.h.s of \eqref{hf-relation} doesn't have pole at $u=v+\epsilon_3$, and so r.h.s does ($h(u)f(u+\epsilon_3)=f(u)h(u)$), thus we may deform integration contour.
	
Equation \eqref{Ad} together with \eqref{fh} implies integral  formula \eqref{LAsSh} together with recurrence representation for $F_{\boldsymbol{\lambda}}(\boldsymbol{z}|u)$
\begin{eqnarray}
 F_{\boldsymbol{\lambda}+n}(\boldsymbol{z},\boldsymbol{w}|u)=F_{\boldsymbol{\lambda}}(\boldsymbol{z})g_n(\boldsymbol{w})\left(1-\prod\limits_{i,j} G(z_i-w_j)\prod_i\frac{u-w_i}{u-w_i-\epsilon_3}\right)
\end{eqnarray}
\section{Other representations of \texorpdfstring{$\textrm{YB}\bigl(\widehat{\mathfrak{gl}}(1)\bigr)$}{YB(gl(1))}}\label{SUSY-ILW}
In this paper we were concentrated on an examples of "spin chain" with $n$ sites and periodic boundary conditions, this setup corresponds to an affine $A_n$  Toda field theory. At the each site of our "spin chain" we should place a representation of $RLL$ algebra. The generating function of IM's is equal to
\begin{equation}
T(u)=\textrm{Tr}_{\mathcal{F}_0} \left(q^{\sum\limits_n a^{(0)}_{-n}{a^{(0)}_n}}\mathcal{R}_{0,1}(u-u_1)\dots\mathcal{R}_{0,n}(u-u_n) \right)
\end{equation}
One possibility is to choose $\mathcal{R}_{0,k}(u-u_k)$ to be the Maulik-Okounkov $R-$matrix. However we have already seen that $RLL$ algebra in current realization is symmetric under permutation of three parameters $\epsilon_{\alpha}$, in terms of usual parameters $b,Q,b^{-1}$ this means a symmetry between $b$ and $Q=b+\frac{1}{b}$ where $b=\sqrt{\frac{\epsilon_1}{\epsilon_2}}$.

In order to see two additional representations of $RLL$ algebra let us realize  representation of $W$ algebra in the space of two bosons as commutant of Screening charge, according to \cite{2015arXiv151208779B,Litvinov:2016mgi} there are three choices of screening currents. Our notation is that there exist three different types of representation of $YB(\widehat{\mathfrak{gl}}(1)$: we call them $\mathcal{F}_{u}^{(1)}$, $\mathcal{F}_{u}^{(2)}$ and $\mathcal{F}_{u}^{(3)}$. We assign the screening charge $S_k$ to a tensor product of two Fock spaces of the same type $\mathcal{F}_{u}^{(k)}\otimes\mathcal{F}_{v}^{(k)}$, and we assign "fermionic" screening charge $S_{i,f}$ to the tensor product of different Fock spaces $\mathcal{F}_{u}^{(j)}\otimes\mathcal{F}_{v}^{(k)}$ with $(i,j,k)=\textrm{cycl}(1,2,3)$. Fixing  one Fock space to be of type $3$, we will have three options for the other one
\begin{equation}
 S_{f,1}=\oint e^{b \phi_0(z)+i\beta \phi_1(z)}dz,\quad
 S_{f,2}=\oint e^{b^{-1} \phi_0(z)+i\bar{\beta} \phi_1(z)}dz,\quad
 S_3^{\pm}=\oint e^{b^{\pm 1}(\phi_0(z)-\phi_1(z))}dz,
\end{equation}
where $\beta=i\sqrt{b^2+1}$ and  $\bar{\beta}=i\sqrt{1+b^{-2}}$.

While the third screening charge $S_3$ leads to the MO $R-$matrix
\begin{equation}
\mathcal{R}^{(3)}_{0,1}=\mathcal{R}^{MO}_{0,1}=e^{iQ\int\limits_{x=0}^{2\pi} \bigl[\frac{1}{2u}(\partial\phi_0(x)-\partial\phi_1(x))^2-\frac{1}{3u^2}(\partial\phi_0(x)-\partial\phi_1(x))^3\bigr]+o(\frac{1}{u^2})]\frac{dx}{2\pi}},
\end{equation}
the first and the second screenings have dimension $\frac{1}{2}$ and the  corresponding $W$ algebra admits free fermion representation. For  example for the first screening, let us introduce two fermionic currents
\begin{align}
 &\psi(x)=e^{-ibux}e^{b \phi_0(x)+i\beta \phi_1(x)} \label{psi}\\
 &\psi^{\dagger}(x)=e^{ibux}e^{-b \phi_0(x)-i\beta \phi_1(x)}\label{psid}, 
\end{align}
where $iu$ is the zero mode of $ \phi_0(x)$.
It is easy to check that they obeys free fermionic OPE's
\begin{equation}
 \psi(x)\psi^{\dagger}(y)=\frac{1}{\sin(x-y)}+\text{reg},\quad
 \psi(x)\psi(y)=\text{reg},\quad \psi^{\dagger}(x)\psi^{\dagger}(y)=\text{reg}.
\end{equation}
Correspondingly, $W^{(2)}(z)$ current which commutes with $S_1$ is simply
\begin{equation}
W^{(2)}(z)=\psi^{\dagger}(z)(i\partial+ub)\psi(z)
\end{equation}
Intertwining relation implies
\begin{equation}
\mathcal{R}_f^{(1)}\psi^{\dagger}(z)(i\partial+ub)\psi(z)=\psi(z)(i\partial-ub)\psi^{\dagger}(z)\mathcal{R}_f^{(1)}\label{SuperIntertwining}
\end{equation}

One can find that the $\mathcal{R}_f^{(1)}$ matrix is given by the explicit formula
\begin{equation}\label{Rf} 
\mathcal{R}_f^{(1)}(u)=\exp\Big[\frac{1}{2\pi}\int_{0}^{2\pi}\psi^{\dagger}(x)\log\left(1+\frac{i \partial}{ub}\right)\psi(x)dx\Big] 
\end{equation}
Indeed under the adjoint action of $\mathcal{R}$ matrix fermions transform as:
\begin{gather}
\mathcal{R}_f^{(1)}\psi(z)\Big(\mathcal{R}_f^{(1)}\Big)^{-1}=\frac{1}{1+\frac{i \partial}{ub}}\ \psi(z) \\
\mathcal{R}_f^{(1)}\psi^{\dagger}(z)\Big(\mathcal{R}_f^{(1)}\Big)^{-1}=\big({1-\frac{i \partial}{ub}}\big)\psi^{\dagger}(z)
\end{gather}
Such that \eqref{SuperIntertwining} holds. Although formula for $\mathcal{R}$ matrix looks pretty simple, it's structure is quite complicated because one should remember that $\psi(z)$ operator is nontrivial in terms of individual bosons \eqref{psi} and \eqref{psid}.

In order to find local integrals of motion we have to expand $\mathcal{R}$ matrix in powers of $\frac{1}{u}$. Let us introduce a shorthand notation
\begin{equation}
\Phi(x)=ib\phi_0(x)+i\beta\phi_1(x)
\end{equation}
It is easy to find that
\begin{gather}
:\psi^{\dagger}(x)\psi(x):=\partial \Phi(x) \label{psipsi}\\
:\psi^{\dagger}(x)\partial\psi(x):= \frac{1}{2}\big(\partial\Phi(x)\big)^2+\frac{1}{2}\partial^2\Phi(x) \label{psidpsi}\\
:\psi^{\dagger}(x)\partial^2\psi(x):=\frac{1}{3}(\partial\Phi(x))^3+\partial\Phi(x)\partial^2\Phi(x)+\frac{1}{3}\partial^3\Phi(x) \label{psipddsi}\\
\dots
\end{gather}
Using \eqref{Rf}, \eqref{psipsi} and \eqref{psidpsi},  it is easy to find first non trivial integral of motion in the space of one boson $F_2$:
\begin{equation}
T(u)=\textrm{Tr}'_{\textrm{aux}}(q^{\sum_na_{-n}a_n}R^{(1)}_f)=e^{\frac{I_1}{u}+\frac{I_2}{u^2}+\dots},
\end{equation}
where
\begin{equation}
I_1=\frac{iQ}{2\pi}\int_{x=0}^{2\pi}\partial\phi^2 dx,\qquad
I_2=-\frac{iQ}{b}\int_{x=0}^{2\pi}\Big[ \frac{1}{3}\beta(\partial \phi)^3-\frac{1}{2}b^2 \partial\phi D\partial\phi \Big] \frac{dx}{2\pi}
\end{equation}

In general, representation may contain Fock modules of different types. Let us consider the following one
\begin{equation}
\mathcal{F}_q=\left(\mathcal{F}_u^{(1)}\right)^{\otimes n_1}\left(\mathcal{F}_u^{(2)}\right)^{\otimes n_2}\left(\mathcal{F}_u^{(3)}\right)^{\otimes n_3} \ , \quad \mathcal{F}_{\text{aux}}=F_3
\end{equation}
Where $\mathcal{F}_q$ is our quantum space, and $\mathcal{F}_{\textrm{aux}}$ is an auxiliary space.  As usual the generating function of Integrals of Motion is
\begin{equation}
T(u)=\textrm{Tr}_{\text{aux}}\Big(q^{\sum_na_{n}a_{-n}}R_{aux,q}\Big)=tr_{aux}\Big(q^{\sum_na_{n}a_{-n}}\prod\limits_{j=1}^{n_1}R_{2,f}(u-v_j)\prod\limits_{i=n_1+1}^{n_1+n_2}R_{1,f}(u-v_{i})\prod\limits_{k=n_1+n_2}^{n_1+n_2+n_3}R^{MO}(u-v_k)\Big)
\end{equation}
Expanding at large spectral parameter, easy to find first non trivial integral of motion:
\begin{equation}
I_2=iQ\int\limits_{x=0}^{2\pi}\Big[ \frac{\bar{\beta}}{3}\sum\limits_{i=1}^{n_1}(\partial \phi_i)^3+\frac{\beta}{3}\sum\limits_{i=n_1+1}^{n_1+n_2}(\partial \phi_i)^3-\frac{1}{3}\sum\limits_{i=n_1+n_2+1}^{n_1+n_2+n_3}(\partial \phi_i)^3-\left(\frac{1}{2}\sum\limits_{i,j} B_{i,j}\partial\phi_i D\partial\phi_j +\sum\limits_{i<j} B_{i,j}\partial\phi_i \partial^2\phi_j\right)\Big] \frac{dx}{2\pi}
\end{equation}
Where $B$ is a $n_1\times n_2\times n_3$ block matrix:
\begin{equation}
B=\begin{pmatrix} b && 1 && \beta  \\ 1 && b^{-1} && \bar{\beta}\\
\beta && \bar{\beta}  && -Q
\end{pmatrix}
\end{equation}
Alternatively switching to epsilon notation:
\begin{multline}
I_2=-\epsilon_3\int\limits_{x=0}^{2\pi}\Big[ \frac{1}{3}\frac{\epsilon_1}{\sigma_3}\sum\limits_{i=1}^{n_1}(\partial \phi_i)^3+\frac{1}{3}\frac{\epsilon_2}{\sigma_3}\sum\limits_{i=n_1+1}^{n_1+n_2}(\partial \phi_i)^3+\frac{1}{3}\frac{\epsilon_3}{\sigma_3}\sum\limits_{i=n_1+n_2+1}^{n_1+n_2+n_3}(\partial \phi_i)^3-\\-\left(\frac{1}{2}\sum\limits_{i,j} \frac{\epsilon_i \epsilon_j}{\sigma_3}\partial\phi_i D\partial\phi_j +\sum\limits_{i<j} \frac{\epsilon_i \epsilon_j}{\sigma_3}\partial\phi_i \partial^2\phi_j\right)\Big] \frac{dx}{2\pi} \label{I2OM}
\end{multline}
And basic fields normalized as follows:
\begin{eqnarray}
\partial_i\phi(x)\partial_j\phi(y)=-\delta_{i,j}\frac{\sigma_3}{\epsilon_i}\frac{1}{\sin^2(x-y)}
\end{eqnarray}
Where $\sigma_3=\epsilon_1\epsilon_2\epsilon_3$.

Bethe ansatz for the models considered in this section could be derived along the same lines, the only difference is in the action of $\psi(u)$ generators on vacuum. For a Fock space representation $\mathcal{F}=\prod_k \otimes\mathcal{F}_{u_k}^{(\alpha_k)}$ we have:
\begin{eqnarray}
\psi(u)|\varnothing\rangle =\prod\limits_{k=1}^n\frac{u-u_k-\epsilon_{\alpha_k}}{u-u_k}|\varnothing\rangle
\end{eqnarray}
So that we will have the same Bethe equations as in \eqref{Bethe-ansatz-equations}, but with different source function
\begin{eqnarray}
A(u)=\prod\limits_{k=1}^n\frac{u-u_k-\epsilon_{\alpha_k}}{u-u_k}.
\end{eqnarray}

\bibliographystyle{\string~/BibTeX/mybst/MyStyle}
\bibliography{\string~/BibTeX/mybib/MyBib} 
	
\end{document}